\documentclass[aps,pra]{revtex4-2}
\usepackage{amsfonts}
\usepackage{amsmath}
\usepackage{amsthm}
\usepackage{amssymb}
\usepackage{graphicx}
\usepackage{epstopdf}
\usepackage{appendix}
\usepackage{hyperref}
\usepackage{textcomp}
\usepackage{multirow}
\usepackage{color}
\usepackage{epsfig}
\usepackage{bm}
\usepackage{braket}
\usepackage{subfigure}

\begin{document}

\title{Multidimensional self-trapping in linear and nonlinear potentials}
\author{Boris A. Malomed$^{1,2}$}
\affiliation{$^{1}$Department of Physical Electronics, School of Electrical Engineering,
Faculty of Engineering, and Center for Light-Matter Interaction, Tel Aviv
University, P.O. Box 39040 Tel Aviv, Israel\\
$^{2}$Instituto de Alta Investigaci\'{o}n, Universidad de Tarapac\'{a},
Casilla 7D, Arica, Chile}

\begin{abstract}
Solitons are typically stable objects in diverse one-dimensional (1D)
models, but their straightforward extensions to 2D and 3D settings tend to
be unstable. In particular, the ubiquitous nonlinear Schr\"{o}dinger (NLS)
equation with the cubic self-focusing, which is also widely known as the
Gross-Pitaevskii (GP) equation in the theory of Bose-Einstein condensates
(BECs), creates only unstable 2D and 3D solitons, because the same equation
gives rise to destructive effects in the form of the critical and
supercritical wave collapse in the 2D and 3D cases, respectively. This
chapter offers, first, a review of physically relevant settings which,
nevertheless, make it possible to create stable 2D and 3D solitons,
including ones with embedded vorticity. The main stabilization schemes
considered here are: (i) competing (e.g., cubic-quintic) and saturable
nonlinearities; (2) linear and nonlinear trapping potentials; (3) the
Lee-Huang-Yang correction to the mean-field BEC dynamics, leading to the
formation of robust \textit{quantum droplets}; (4) spin-orbit-coupling (SOC)
effects in binary BEC; (5) emulation of SOC in nonlinear optical waveguides,
including $\mathcal{PT}$-symmetric ones. Further, the chapter presents a
detailed summary of results which demonstrate the creation of stable 2D and
3D solitons by the schemes based on the usual linear trapping potentials or
effective nonlinear ones, which may be induced by means of spatial
modulation of the local nonlinearity strength. The latter setting is
especially promising, making it possible to use \emph{self-defocusing}
media, with the local nonlinearity strength growing fast enough from the
center to periphery, for the creation of a great variety of stable
multidimensional modes. In addition to fundamental states and vortex rings,
the respective 3D modes may be \textit{hopfions}, i.e., twisted vortex rings
which carry two indpednent topological charges. Many results for the
multidimensional solitons have been obtained, in such settings, not only in
a numerical form, but also by means of analytical methods, such as the
variational and Thomas-Fermi approximation.
\end{abstract}

\maketitle

\noindent \textbf{Acronyms}

\noindent 1D -- one-dimensional

\noindent 2D -- two-dimensional

\noindent 3D -- three-dimensional

\noindent \textrm{a.r.} -- aspect ratio

\noindent b.c. -- boundary condition(s)

\noindent BdG -- Bogoliubov -- de Gennes (linearized equations for
perturbations around stationary solutions of GP/NLS equations)

\noindent BEC -- Bose-Einstein condensate

\noindent CQ -- cubic-quintic (nonlinearity))

\noindent CW -- continuous wave

\noindent FR -- Feshbach resonance

\noindent GP -- Gross-Pitaevskii (equation)

\noindent GVD -- group-velocity dispersion

\noindent HO -- harmonic-oscillator (potential)

\noindent HV -- hidden vorticity

\noindent IST inverse-scattering transform (the method for solving
integrable nonlinear partial differential equations)

\noindent LHY -- Lee-Huang-Yang (correction to the MF theory)

\noindent MF -- mean-field (approximation)

\noindent NLS -- nonlinear Schr\"{o}dinger (equation))

\noindent OL -- optical lattice

\noindent PDE -- partial differential equation

\noindent PhR -- photorefractive (optical material)

\noindent $\mathcal{PT}$ -- parity-time (symmetry)

\noindent QD -- quantum droplet

\noindent TF -- Thomas-Fermi (approximation)

\noindent TS -- Townes soliton

\noindent VA -- variational approximation

\noindent VK -- Vakhitov-Kolokolov (stability criterion)

\noindent XPM -- cross-phase modulation

\section{Introduction: The objective of this chapter}


Solitons, alias solitary waves, are self-trapped (localized) objects
existing in a great variety of physical media, due to the interplay of basic
linear properties, such as dispersion and/or diffraction, and nonlinearity
which represents self-attraction of matter or fields that fill the media
(Kivshar and Agrawal, 2003; Dauxois and Peyrard, 2006). Parallel to the
development of experimental research of solitons in a large number of
physical realizations, a great deal of work has been performed on
theoretical models producing solitons (as it usually happens, the progress
in the theoretical work was much faster). The theory has been developing in
two related but distinct directions: on the one hand, elaboration of
mathematical models of diverse physical setups, in which the concept of
solitons is relevant, and, on the other hand, mathematical investigation of
these and many other models (Zakharov \textit{et al}., 1980; Ablowitz and
Segur, 1981; Calogero and Degasperis, 1982; Newell, 1985; Takhtadjian and
Faddeev, 1986; Yang, 2010). Actually, some models were introduced on the
basis of their mathematical interest, rather than being directly suggested
by physical realizations.

The concept of solitons and self-trapping had appeared in one-dimensional
(1D) settings. Up to this day, an absolute majority of experimental and
theoretical/mathematical studies of solitons have been performed in
effectively 1D setups, and in the framework of 1D nonlinear partial
differential equations (PDEs). The development of the studies for two- and
three-dimensional (2D and 3D) systems, aimed at prediction and experimental
creation of multidimensional solitons, is a fascinating possibility.
However, a fundamental obstacle which strongly impedes the progress in this
direction is the problem of stability of 2D and 3D solitons (Malomed \textit{%
et al}., 2005 and 2016; Malomed, 2016; Mihalache, 2017; Kartashov \textit{et
al}., 2019; Malomed, 2019). In most cases, 1D solitons appear as fully
stable solutions of the underlying PDEs, and they are readily observed as
stable objects in the experiment. On the other hand, the ubiquitous NLS
equation with the cubic self-focusing nonlinearity creates only unstable
solitons in 2D and 3D spaces, because precisely the same equations gives
rise to the phenomena of the\textit{\ wave collapse} (alias \textit{blowup}%
), i.e., spontaneous formation of singularities in finite times, starting
from regular localized (soliton-like) inputs. The collapse governed by the
cubic NLS equation is \textit{critical} in the 2D geometry, i.e., it sets in
if the integral norm of the input exceeds a certain critical value;
otherwise, the input spreads out. In 3D, the same equation gives rise to the
\textit{supercritical collapse}, for which the threshold value of the norm
is zero, i.e., the formation of the singularity may be initiated by the
input with an arbitrarily small norm. In either case, the possibility of the
collapse makes the formally existing 2D and 3D soliton solutions of the NLS
equation completely unstable.

For this reason, the cardinal direction in the work on the vast area of
self-trapping in the multidimensional geometry has been elaboration of
physically relevant setups in which 3D and 3D solitons may be stabilized
(Malomed \textit{et al}., 2005 and 2016; Malomed, 2016; Mihalache, 2017;
Kartashov \textit{et al}., 2019; Malomed, 2019). One of promising directions
is the use of trapping potentials. First of all, this may be a
straightforward linear potential, such as the parabolic (alias
harmonic-oscillator (OH) one). A more sophisticated option is the use of
self-repulsive nonlinearity with a spatially modulate strength of the local
interaction. While, in the uniform space, the self-repulsion obviously
cannot give rise to self-trapping, it can readily support a remarkable
variety of stable 1D, 2D, and 3D localized states (quasi-solitons), provided
that the local strength of the self-repulsion in the space of dimension $D$
grows fast enough from the center to periphery (faster than $r^{D}$, where $r
$ is the radial coordinate), as was first proposed by Borovkova \textit{et al%
}. (2011a,2011b) and later developed in many other works, see below. The
latter setup may be considered as one with an effective \textit{nonlinear
potential}.

The objective of this chapter is to provide a summary of theoretical results
obtained on this topic, including a review of diverse methods elaborated for
the stabilization of multidimensional solitons in systems of the NLS type,
and a detailed account of theoretical results predicting stable 2D and 3D
solitons in the framework of the NLS equation including linear or nonlinear
potentials. The presentation is arranged as follows. To provide the
necessary introduction to the general topic, Section II recapitulates basic
results which were firmly established in studies of integrable and
non-integrable versions of the one-dimensional NLS equations. Section III
leads the reader from 1D to the multidimensional world. In particular, it
introduces a concept of the Townes solitons (TSs), which are unstable, by
themselves, are closely related to various stabilization schemes, and
outlines the fundamental problem of the instability of NLS solitons in the
multidimensional space. Section IV provides a summary of basic stabilization
schemes, elaborated for the 2D and 3D solitons in systems of the NLS types.
Sections V and VI summarize essential predictions for the existence of
stable fundamental solitons, as well as topologically structured ones, such
as 2D and 3D vortex rings and 3D \textit{hopfions} (twisted vortex rings),
in models including, respectively, linear or nonlinear potentials as the
stabilizing factor. Section VII concludes the chapter.

\section{The one-dimensional NLS (nonlinear Schr\"{o}dinger) equation -- a
universal model of classical and semi-classical physics}


\subsection{The general setting}

The great amount of work performed on PDEs modeling the wave propagation in
1D\ dispersive nonlinear media had led, roughly 50 years ago, to the
discovery of several celebrated equations, which are fundamentally important
as universal models of the theory of nonlinear waves. These equations share
the unique property of \textit{integrability}, which was revealed with the
help of the mathematical technique known as the inverse-scattering transform
(IST). Three most important items in the list of classical integrable PDEs
are the Korteweg - de Vries (KdV), sine-Gordon (SG), and nonlinear Schr\"{o}%
dinger (NLS) equations. Methods for solving these equations and results
produced by those methods are summarized, in full detail, in several
well-known books (Zakharov \textit{et al}., 1980; Ablowitz and Segur, 1981;
Calogero and Degasperis, 1982; Newell, 1985; Takhtadjian and Faddeev, 1986;
Rogers and Schief, 2002; Yang. 2010).

The NLS equation plays the central role in the present chapter. In 1D, its
scaled form is commonly known:%
\begin{equation}
i\psi _{z}+\frac{1}{2}\psi _{xx}\pm |\psi |^{2}\psi -U(x)\psi =0.  \tag{2.1}
\end{equation}%
This equation is commonly used as the model for the propagation of light,
with local amplitude $\psi \left( x,z\right) $ of the electromagnetic field,
in planar optical waveguides with transverse coordinate $x$ and propagation
distance $z$ (Kivshar and Agrawal, 2003). In this case, term $\psi _{xx}$
represents the paraxial diffraction, the cubic term with the top or bottom
signs corresponds, respectively, to the self-focusing or defocusing Kerr
nonlinearity in the waveguiding material, and $-U(x)$ is proportional to the
local variation of the underlying refractive index, $\delta n\left( x\right)
$. On the other hand, Eq. (2.1) with $z$ replaced by scaled time $t$ is well
known as the semi-classical Gross-Pitaevskii (GP) equation for the
mean-field wave function, $\psi \left( x,t\right) $, of a Bose-Einstein
condensate (BEC) of ultracold bosonic atoms loaded into a tightly built
cigar-shaped trapping potential, which effectively eliminates the transverse
coordinates ($y$ and $z$), allowing BEC\ to evolve in time along the axial
direction, $x$ (Pitaevskii and Stringari, 2003). In this case, the top and
bottom signs in front of the cubic term in Eq. (2.1) imply, respectively,
attractive and repulsive interactions between atoms in the ultracold gas,
and a real axial potential, $U(x)$, is an essential ingredient of
experimental setups. The potential which is often used in the experiment
represents the harmonic oscillator (HO),%
\begin{equation}
U_{\mathrm{OH}}(x)=\left( \Omega ^{2}/2\right) x^{2}.  \tag{2.2}
\end{equation}

Equation (2.1) for the complex wave function $\psi (x)$ corresponds to the
Hamiltonian, which is considered as a functional of $\psi (x)$ and $\psi
^{\ast }(x)$, where $\ast $ stands for the complex conjugate function:%
\begin{equation}
H=\int_{-\infty }^{+\infty }\left[ \frac{1}{2}\left\vert \psi
_{x}\right\vert ^{2}+U(x)|\psi |^{2}\mp \frac{1}{2}|\psi |^{4}\right] dx.
\tag{2.3}
\end{equation}%
The NLS equation can be written in terms of the Hamiltonian in the standard
form (Takhtadjian and Faddeev, 1986),%
\begin{equation}
\frac{\partial \psi }{\partial z}=-i\frac{\delta H}{\delta \psi ^{\ast }},
\tag{2.4}
\end{equation}%
where $\delta /\delta \psi ^{\ast }$ stands for variational (Frech\'{e})
derivative, and the application of the derivative makes use of the
identities $\left\vert \psi (x)\right\vert ^{2}\equiv \psi (x)\psi ^{\ast
}(x)$ and $\left\vert \psi _{x}\right\vert ^{2}\equiv \psi _{x}\psi
_{x}^{\ast }$. In the presence of real potential $U(x)$, the non-integrable
equation (2.1) conserves two dynamical invariants: the total norm (alias the
integral power, in terms of the optical realization),%
\begin{equation}
N=\int_{-\infty }^{+\infty }\left\vert \psi (x)\right\vert ^{2}dx,  \tag{2.5}
\end{equation}%
and the Hamiltonian.

\subsection{The integrable NLS equation}

NLS equation (2.1) is integrable in the free space, with $U(x)=0$ (it is
also integrable in the case of a linear potential, $U(x)=Cx$, as this
potential may be eliminated from Eq. (2.1) by means of a gauge
transformation (Chen and Liu, 1976)). In the case of the top sign in front
of the cubic term (the self-focusing), the family of bright-soliton
solutions to this equation, with arbitrary amplitude $\eta $ and velocity $c$%
, is commonly known since it was obtained in the same classical work of
Zakharov and Shabat (1971), in which the integrability of Eq. (2.1) with $%
U(x)=0$ was discovered:%
\begin{equation}
\psi _{\mathrm{sol}}^{\mathrm{(1D)}}\left( x,t\right) =\eta ~\mathrm{sech}%
\left( \eta (x-cz)\right) \exp \left( icx+ik_{\mathrm{sol}}z\right) .
\tag{2.6}
\end{equation}%
Here, the soliton's propagation constant is%
\begin{equation}
k_{\mathrm{sol}}=\frac{1}{2}\left( \eta ^{2}-c^{2}\right) .  \tag{2.7}
\end{equation}%
In terms of the above-mentioned optics realization, this solution produces a
spatial (stripe) soliton, $c$ actually being not the velocity, but the slope
of soliton's stripe in the $\left( x,z\right) $ plane. The profile of
soliton (2.6) with $\eta =1$ and $c=0$ is displayed in Fig. 1(a).
\begin{figure}[tbp]
\begin{center}
\subfigure[]{\includegraphics[width=0.32\textwidth]{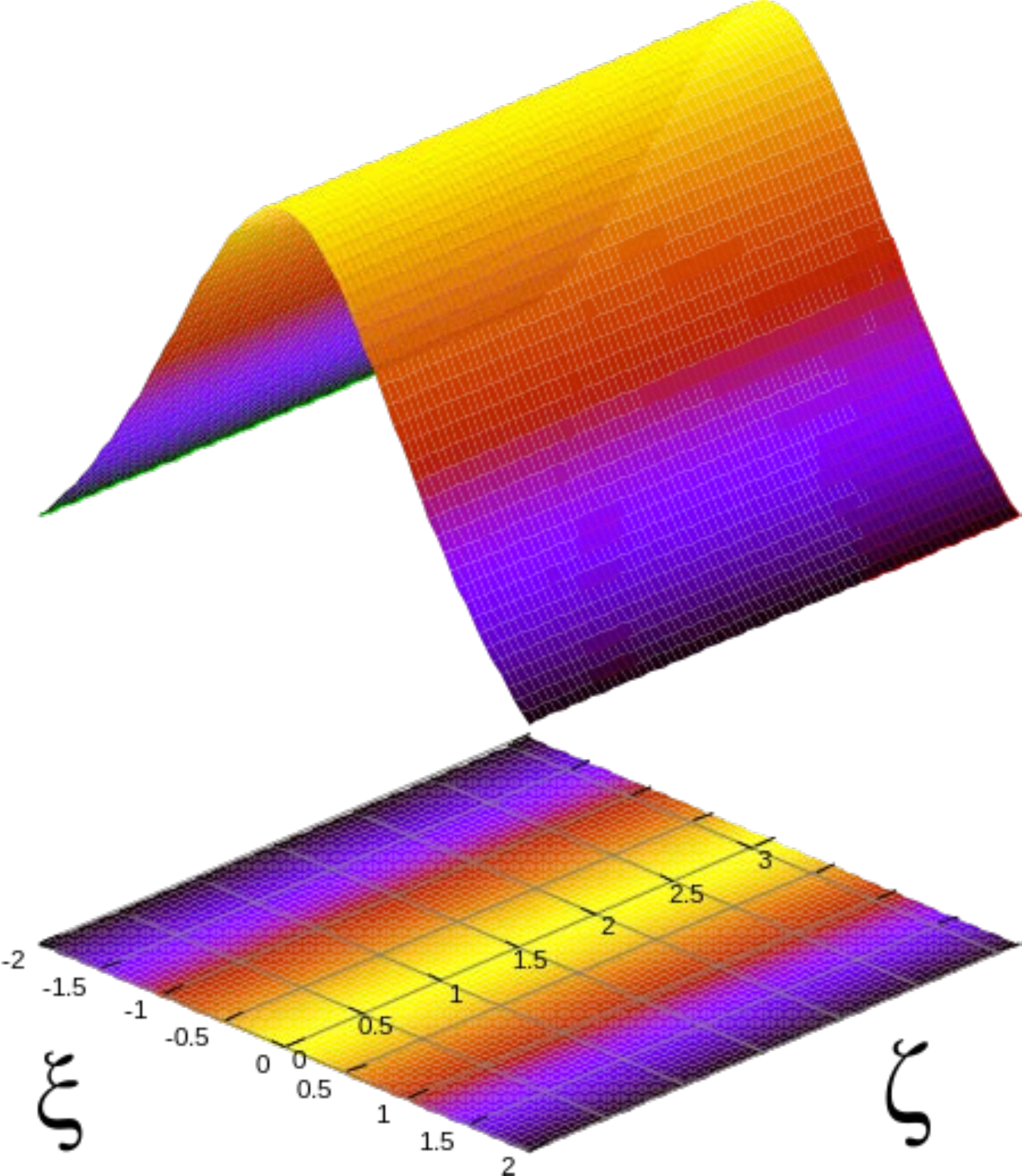}} %
\subfigure[]{\includegraphics[width=0.32\textwidth]{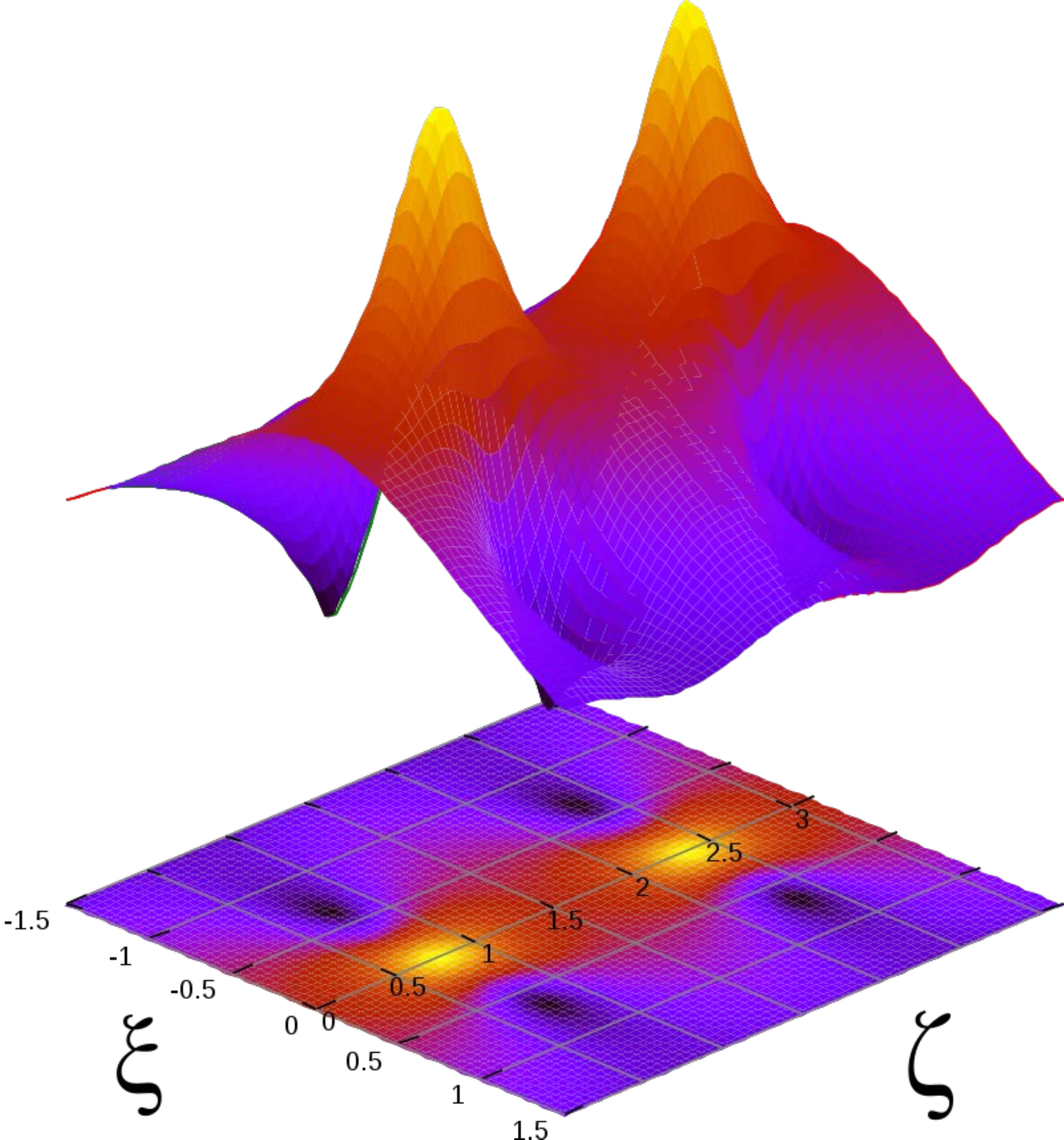}}
\end{center}
\caption{(a) The profile of the absolute value of the wave field, $%
\left\vert \protect\psi \left( x\equiv \protect\xi ,z\equiv \protect\zeta %
\right) \right\vert $, and its projection onto the $\left( \protect\xi ,%
\protect\zeta \right) $ plane, which represents the fundamental
bright-soliton solution (2.6) with $\protect\eta =1$ and $c=0$. (b) The same
for the second-order breather (two-soliton), which is produced by input
(2.10) with $\mathcal{N}=2$ (source: Wikipedia, creator: Alessio Damato,
https://commons.wikimedia.org/wiki/File:Soliton\_1st\_order.svg\#filelinks).
}
\label{fig1.3}
\end{figure}

The free-space NLS equation (not necessarily Eq. (2.1), but also ones with a
more general nonlinearity, which are not integrable) conserves the total
momentum,%
\begin{equation}
P=i\int_{-\infty }^{+\infty }\psi \psi _{x}^{\ast }dx.  \tag{2.8}
\end{equation}%
Integrable systems, such as Eq. (2.1) with $U(x)=0$, conserve an infinite
set of higher-order dynamical invariants, in addition to the three lowest
ones, $N$, $P$, and $H$, but they do not have a straightforward physical
meaning. For the fundamental bright-soliton solution (2.6), values of the
basic conserved quantities are%
\begin{equation}
N_{\mathrm{sol}}=2\eta ,P_{\mathrm{sol}}=2c\eta ,H_{\mathrm{sol}}=-\frac{1}{3%
}\eta ^{3}+\eta c^{2}.  \tag{2.9}
\end{equation}

The integrability of the NLS equation makes it possible to construct exact
solutions for collisions of solitons moving with different velocities (or
different spatial slopes, in terms of the spatial-domain light propagation
in the planar waveguide), $c_{1}$ and $c_{2}$. A well-known result is that
the collisions are fully elastic, i.e., the solitons reappear from the
collisions with precisely the same shapes, amplitudes, and velocities which
they had originally. The only effect produced by the collision is the shift
of both solitons along coordinate $x$, and a shift of their intrinsic
phases. In particular, solitons with equal amplitudes $\eta $, colliding
with velocities $\pm c$, shift in the direction of their motion by $\Delta
x=\pm \eta ^{-1}\ln \left( 1+4\eta ^{2}/c^{2}\right) $.

Another important manifestation of the integrability of the NLS equation
(2.1) in the free space ($U(x)=0$) with the top sign was discovered by
Satsuma and Yajima (1974): the IST technique gives rise to highly nontrivial
exact solutions in the form of $\mathcal{N}$-solitons, produced by the
initial condition
\begin{equation}
\psi (x,z=0)=\mathcal{N}\,\mathrm{sech}\,x  \tag{2.10}
\end{equation}%
with integer $\mathcal{N}$. These states may be considered as nonlinear
superpositions of $\mathcal{N}$ solitons with amplitudes $\eta =\left\{
1,3,...,2\mathcal{N}-1\right\} $. Although the binding energy of such $%
\mathcal{N}$-soliton complexes is exactly equal to zero, the solitons stay
together, forming an oscillatory state, which is often called a \textit{%
breather}. This solution can be written in a relatively simple form for $%
\mathcal{N}=2$:%
\begin{equation}
\psi _{\mathrm{two-sol}}\left( x,t\right) =4e^{iz/2}\frac{\cosh \left(
3x\right) +3e^{4iz}\cosh x}{\cosh \left( 4x\right) +4\cosh \left( 2x\right)
+3\cos \left( 4z\right) }.  \tag{2.11}
\end{equation}%
As shown in Fig. 1(b), the two-soliton breather, given by Eq. (2.11),
periodically oscillates between broad and narrow shapes, the latter one
featuring a single central peak and two small side peaks. For $\mathcal{N}=3$
in Eq. (2.10), the analytical form of the resulting three-soliton solution
is cumbersome, while its spatiotemporal evolution, displayed in Fig. 2(a),
along with the respective Fourier transform in Fig. 2(b), exhibit a new
feature, in comparison with the two-soliton breather: the central peak
periodically splits in two, which then recombine back. Only very recently
such a third-order soliton (breather) was observed experimentally in BEC
(Luo \textit{et al}., 2020).
\begin{figure}[tbp]
\begin{center}
\subfigure[]{\includegraphics[width=0.45\textwidth]{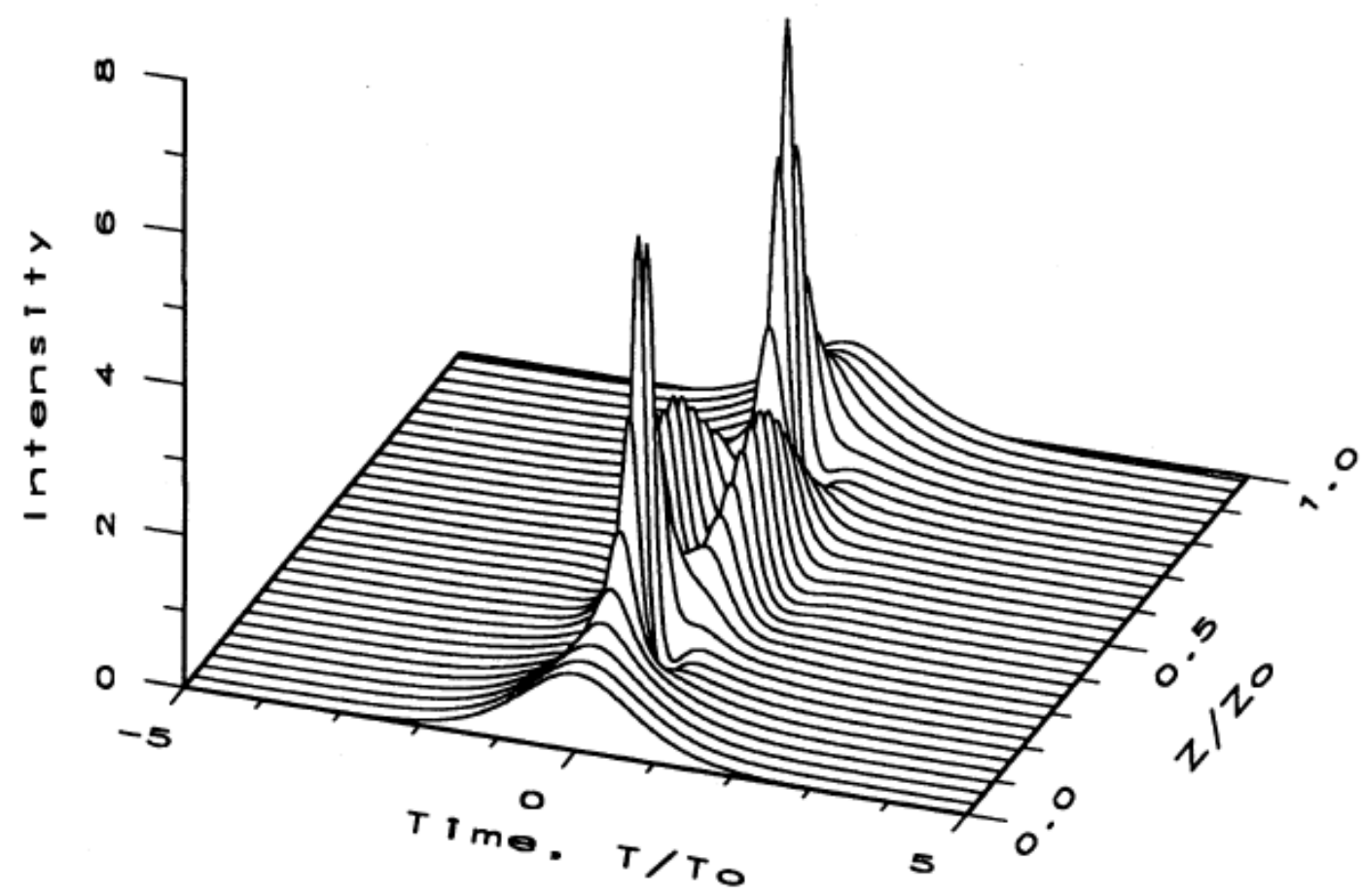}}%
\subfigure[]{\includegraphics[width=0.45\textwidth]{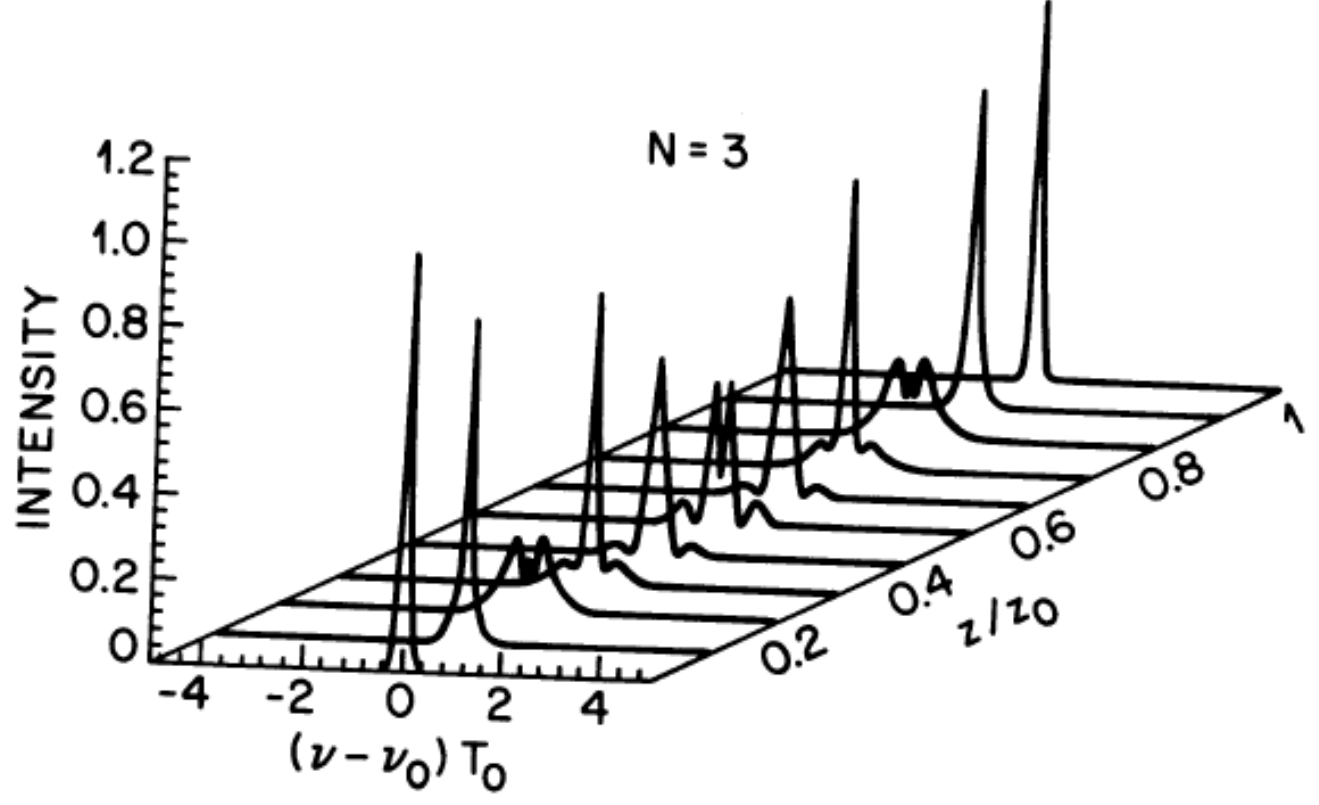}}
\end{center}
\caption{The profile of the local intensity of the wave field, $\left\vert
\protect\psi \right\vert ^{2}$, in the coordinate domain (a), and the
intensity of its Fourier transform (b), for the three-soliton (third-order
breather), generated by input (2.10) with $\mathcal{N}=3$. Note that the
central peak of the breather periodically splits in two secondary peaks,
which then recombine back into the single one. The evolution variable is
denoted $z$, corresponding to the propagation distance in fiber optics
(source: the book by Agrawal (2013). }
\label{fig1.4}
\end{figure}

The input in the form given by Eq. (2.10) is relevant in the general case
too, when $\mathcal{N}$ is not an integer. In this case, an explicit
solution for $\psi \left( x,t\right) $ is not available, but the respective
set of \textit{scattering data}, in terms of the IST method, was also found
in an exact form by Satsuma and Yajima (1974). The set contains a
higher-order soliton (breather) of order $\mathcal{N}_{\mathrm{sol}}=\left[
\mathcal{N}+1/2\right] $ (with $\left[ ...\right] $ standing for the integer
part), \textquotedblleft contaminated" by a dispersive radiation component.
In particular, input (2.10) with $1/2<\mathcal{N}<3/2$ creates the output
containing exactly one soliton (2.6) mixed with the radiation field.

The NLS equation (2.1) with $U(x)=0$ and the bottom sign in front of the
cubic term, which represents the self-defocusing nonlinearity, gives rise to
dark solitons, supported by the continuous-wave (CW) background with nonzero
intensity $\left\vert \psi \right\vert ^{2}\equiv n_{0}$ at $|x|\rightarrow
\infty $. In terms of optics, dark solitons represent a dark spot on top of
the uniformly lit backdrop:%
\begin{equation}
\psi _{\mathrm{dark}}\left( x,z\right) =\exp \left( in_{0}z\right) \left\{
\sqrt{n_{0}-c^{2}}\tanh \left( \sqrt{n_{0}-c^{2}}\left( x-cz\right) \right)
+ic\right\} .  \tag{2.6'}
\end{equation}%
In the case of $c=0$, the dark-soliton's field (2.6') vanishes at $x=0$. The
solution with positive or negative speed $c$, subject to constraint $|c|<%
\sqrt{n_{0}}$, represents the dark soliton moving across the background with
this speed (moving dark solitons are usually called gray solitons).

Both the bright and dark solitons are completely stable solutions of the
respective NLS equation (2.1). In particular, their stability agrees with
the necessary (but, generally speaking, not sufficient) condition for the
stability of solitons supported by the self-attractive nonlinearity, known
as the Vakhitov-Kolokolov (VK) stability criterion (Vakhitov and Kolokolov,
1973; this fundamentally significant criterion is considered in detail in
reviews of Berg\'{e} (1998) and Zakharov and Kuznetsov (2012), and in the
book of Fibich (2015)). The VK criterion is written for the norm of the
soliton family, considered as a function of the chemical potential:%
\begin{equation}
dN/dk>0.  \tag{2.12}
\end{equation}%
Indeed, it is obvious that the NLS-soliton's norm, given by Eq. (2.9), if
considered as a function of the chemical potential as per Eq. (2.7), i.e., $%
N_{\mathrm{sol}}=2\sqrt{2k_{\mathrm{sol}}}$ satisfies condition (2.12).
Here, this point is considered for the zero soliton's velocity, $c=0$,
because the velocity does not affect the stability of solutions of
Galilean-invariant equations, Eq. (2.1) with $U(x)=0$ being one of them.
Indeed, any quiescent solution of Eq. (2.1) generates a family of moving
ones, with velocity $c$, by means of the Galilean boost:%
\begin{equation}
\psi \left( x,z\right) _{c\neq 0}=\psi \left( x,z\right) _{c=0}\left(
x-cz,z\right) \exp \left( icx+\frac{i}{2}c^{2}z\right) .  \tag{2.13}
\end{equation}

Moreover, the bright soliton realizes the ground state of the NLS-based
setup, i.e., it is the state with the lowest value of Hamiltonian (2.3) for
fixed norm (2.5) (Zakharov and Kuznetsov, 2012). The simplest way to arrive
at this conclusion is to consider an ansatz for the stationary wave function,%
\begin{equation}
\left\vert \psi (x)\right\vert =A^{2}\,\mathrm{sech}\left( x/W\right) ,
\label{1.14}
\end{equation}%
with amplitude $A$ and width $W$. The norm (2.5) of the ansatz is%
\begin{equation}
N=2A^{2}W.  \tag{2.15}
\end{equation}%
Then, following the principle of the variational approximation (VA)
(Anderson and Bonnedal, 1979; Anderson, 1983; Malomed, 2002), one calculates
the value of Hamiltonian (2.3)\ for the ansatz (2.14), with $A^{2}$
expressed in terms of $N$ as per Eq. (2.15):
\begin{equation}
H_{\mathrm{ans}}=\frac{1}{6}\left( \frac{N}{W^{2}}-\frac{N^{2}}{W}\right) .
\end{equation}%
Finally, the minimization of this expression identifies the width of the
ground state, $W=2/N$, which exactly corresponds to the exact bright-soliton
solution (2.6).

\subsection{An example of exact solitons in a non-integrable NLS equation
with the delta-functional trapping potential}

To gain insight into families of NLS solitons in a non-integrable version of
the NLS equation, on can Eq. (2.1), with trapping potential $%
U(x)=-\varepsilon \delta (x)$, in which rescaling makes it possible to fix $%
\varepsilon =1$, both signs in front of the nonlinear term, and, moreover,
with a more general power of the nonlinearity, \textit{viz}., $2\sigma +1$:%
\begin{equation}
i\psi _{z}+\frac{1}{2}\psi _{xx}\pm |\psi |^{2\sigma +1}\psi +\delta (x)\psi
=0.  \tag{2.16}
\end{equation}%
It is easy to find exact solutions for solitons pinned to the trapping
center (Wang, Malomed, and Yan, 2019):%
\begin{equation}
\psi (x)=e^{ikz}\left[ \sqrt{k(\sigma +1)}\,\mathrm{sech}\left( \sigma \sqrt{%
2k}\,(|x|+\xi )\right) \right] ^{1/\sigma },  \tag{2.17}
\end{equation}%
with a real positive constant%
\begin{equation}
\xi =\left( 2\sigma \sqrt{2k}\right) ^{-1}\ln \left( \frac{\sqrt{2k}+1}{%
\sqrt{2k}-1}\right) .  \tag{2.18}
\end{equation}%
The squared amplitude of the pinned soliton is%
\begin{equation}
A^{2}(\sigma ,k)\equiv \left\vert \psi (x=0)\right\vert ^{2}=\left[ \left(
1+\sigma \right) \left( k-1/2\right) \right] ^{1/\sigma }.  \tag{2.19}
\end{equation}%
It follows from Eqs. (2.18) and (2.19) that the solutions exist with the
propagation constant exceeding a cutoff value,
\begin{equation}
k>k_{\mathrm{cutoff}}\equiv 1/2,  \tag{2.20}
\end{equation}%
the amplitude vanishing at $k\rightarrow 1/2$.

The norm of solution (2.19) can be explicitly calculated for the cubic
nonlinearity, with $\sigma =1$,%
\begin{equation}
N_{\sigma =1}(k)=2\left( \sqrt{2k}-1\right) ,  \tag{2.21}
\end{equation}%
and for the quintic nonlinearity, with $\sigma =2$:
\begin{equation}
N_{\sigma =2}(k)=\sqrt{\frac{3}{2}}\left[ \pi -2\tan ^{-1}\left( \sqrt{\frac{%
\sqrt{2k}+1}{\sqrt{2k}-1}}\right) \right] .  \tag{2.22}
\end{equation}%
Note that the norm given by Eq. (2.21) for $\sigma =1$ diverges at $%
k\rightarrow \infty $, while in the case of $\sigma =2$, which is the
critical one in the 1D setting, the value given by Eq. (2.22) at $%
k\rightarrow \infty $ is finite:%
\begin{equation}
N_{\sigma =2}(k\rightarrow \infty )=\sqrt{3}\pi \left( 2\sqrt{2}\right)
^{-1}\approx 1.92.  \tag{2.23}
\end{equation}%
Accordingly, there is a discontinuity in the dependence of $N_{\sigma
}(k\rightarrow \infty )$ on $\sigma $. First, it diverges in the \textit{%
subcritical} case, $\sigma <2$ -- in particular, as%
\begin{equation}
N_{\sigma <2}(k)\approx \sqrt{3}\pi \left( 2\sqrt{2}\right) ^{-1}k^{\left(
2-\sigma \right) /4}  \tag{2.24}
\end{equation}%
for $0<2-\sigma \ll 2$. Next, it takes the final value (2.23) in the \textit{%
critical} case, $\sigma =2$. Finally, $N(k\rightarrow \infty )$ decays $\sim
k^{-(\sigma -2)/4}$ in the \textit{supercritical} case, $\sigma >2$.

Dependences $N(k)$ for $\sigma \leq 2$, including the explicitly found ones,
given by Eqs. (2.21) and (2.22), satisfy the VK criterion (2.12) at all
values of $k>1/2$, at which the solitons exist. In full agreement with the
prediction of the criterion, all soliton families (2.17) are fully stable at
$\sigma \leq 2$ (Wang, Malomed, and Yan, 2019). In the supercritical case, $%
\sigma >2$, the family (2.17) satisfies the VK criterion in a finite
interval of the propagation constant, which can be explicitly found for $%
0<\sigma -2\ll 1$:%
\begin{equation}
\frac{1}{2}<k<k_{\mathrm{cr}}(\sigma )\approx \frac{8}{\left( \pi (\sigma
-2)\right) ^{2}}.  \tag{2.25}
\end{equation}%
A narrow stability interval exists even in the limit of $\sigma \gg 2$, with
$k_{\mathrm{cr}}(\sigma )-1/2\approx 1/\sigma $. In terms of the norm, the
stability interval has a finite size too,%
\begin{equation}
0<N<N_{\max }(\sigma )\equiv N_{\sigma }\left( k=k_{\mathrm{cr}}(\sigma
)\right) .  \tag{2.26}
\end{equation}%
In particular, $N_{\max }\left( \sigma =2\right) $ is given by Eq. (2.23),
and with the growth of $\sigma $, $N_{\max }(\sigma $) monotonously
decreases towards $N_{\max }\left( \sigma \rightarrow \infty \right) =1$.

\begin{figure}[t]
\begin{center}
\vspace{0.05in} {\includegraphics[width=0.75\textwidth]{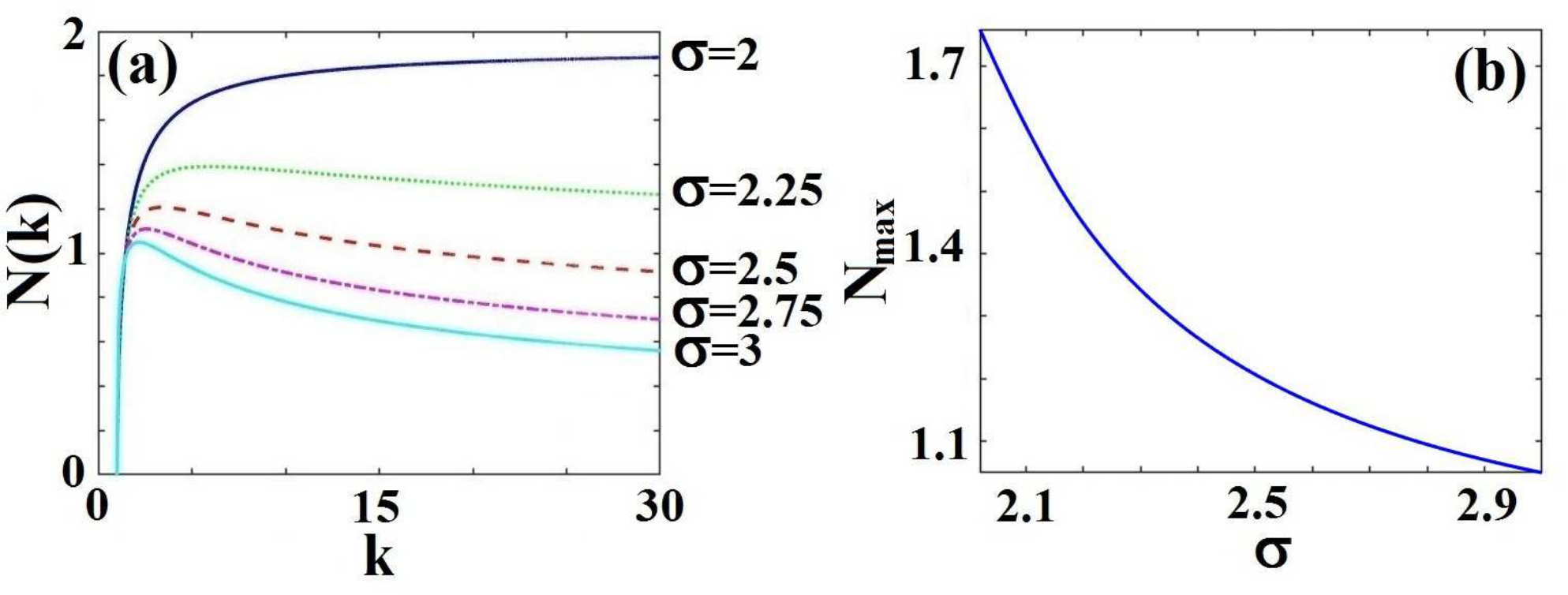}}
\end{center}
\par
\vspace{-0.2in}
\caption{(a) The dependence of norm $N_{\protect\sigma }(k)$ on propagation
constant $k$, numerically computed from Eq.~(2.16) with the top sign in
front o the nonlinear term, for $\protect\sigma =2$, and supercritical
values, $\protect\sigma =\ 2.25,\,2.5,\,2.75,\,3$. According to the VK
criterion, the stability interval is one with $dN/dk>0$. (b) The maximum
norm of the pinned solitons, defined as per Eq. (2.26), versus $\protect%
\sigma $\thinspace\ in the supercritical interval, $2<\protect\sigma \leq 3$%
. The figure is borrowed from the paper by Wang, Malomed, and Yan (2019).}
\label{fig1.5}
\end{figure}

At $\sigma >2$ and $k>k_{\mathrm{cr}}(\sigma )$, the pinned solitons are
unstable. These results, produced by means of the numerical solution of the
stationary version of Eq. (2.16) with the upper sign, are summarized in Fig. %
\ref{fig1.5}).

In addition to that, it is possible to produce a family of exact solutions
for solitons pinned to the attractive delta-functional potential in the case
of the self-repulsive nonlinearity, corresponding to the top sign in Eq.
(2.16) (Wang, Malomed, and Yan, 2019):%
\begin{equation}
\psi _{\mathrm{defoc}}(x)=e^{-kz}\left\{ \sqrt{k(\sigma +1)}/\sinh \left[
\sigma \sqrt{2k}\,(|x|+\xi )\right] \right\} ^{1/\sigma },  \tag{2.27}
\end{equation}%
with%
\begin{equation}
\xi =\frac{1}{2\sigma \sqrt{2k}}\ln \left( \frac{1+\sqrt{2k}}{1-\sqrt{2k}}%
\right) ,  \tag{2.28}
\end{equation}%
and the squared amplitude%
\begin{equation}
A^{2}(\sigma ,k)\equiv \left\vert \psi _{\mathrm{defoc}}(x)\right\vert ^{2}=%
\left[ \left( 1+\sigma \right) \left( 1/2-k\right) \right] ^{1/\sigma },
\tag{2.29}
\end{equation}%
cf. the solution given by Eqs. (2.17) - (2.19). As it follows from Eqs.
(2.28), the existence region for the localized modes pinned by the
attractive delta-functional potential embedded in the defocusing medium is $%
k<k_{\mathrm{cutoff}}\equiv 1/2$, which is exactly opposite to that in the
case of the self-focusing, cf. Eq. (2.20). As for the $N(k)$ dependence for
these solutions, it takes a simple form in the case of the cubic
self-repulsion, $\sigma =1$:%
\begin{equation}
N_{\mathrm{defoc}}(\sigma =1,k)=2\left( 1-\sqrt{2k}\right) ,  \tag{2.30}
\end{equation}%
cf. dependence (2.21) for $\sigma =1$ in the case of the cubic
self-attraction.

For localized states supported by self-repulsive nonlinearity, the VK
criterion, as the necessary stability condition, is replaced by the \textit{%
anti-VK} criterion, with the opposite sign (Sakaguchi and Malomed, 2010):%
\begin{equation}
dN/dk<0,  \tag{2.30}
\end{equation}%
cf. Eq. (2.12). The consideration demonstrates that the solutions given by
Eqs. (2.27) and (2.28) satisfy the anti-VK criterion at all values of $%
\sigma $ and $k$ (see, e.g., Eq. (2.30)). Accordingly, the full stability
analysis has demonstrated that all these solutions are indeed stable (Wang,
Malomed, and Yan, 2019).

\section{The exit to the multidimensional world}


\subsection{Two-dimensional Townes solitons (TSs)}

While there are important setups which make it possible to introduce
physically relevant 1D systems, as briefly mentioned above, the real world
is three-dimensional, and, in some cases, quasi-two-dimensional. This
obvious fact strongly suggests to consider 3D and 2D solitons in nonlinear
optics, BEC, plasmas, and other nonlinear physical media.

The simplest relevant model which admits direct extension from 1D to 3D and
2D is the cubic NLS/GP equation (in fact, as mentioned above, its 1D version
(2.1) was derived by the inverse reduction, 3D $\rightarrow $ 1D):%
\begin{equation}
i\psi _{t}+\frac{1}{2}\nabla ^{2}\psi \pm |\psi |^{2}\psi -U(r)\psi =0,
\tag{3.1}
\end{equation}%
where $\nabla ^{2}$ is the 3D or 2D Laplacian, the top and bottom signs in
front of the cubic term again correspond to the attractive and repulsive
nonlinearity, respectively, and the trapping potential, if any, is assumed
to be isotropic, depending on the radial coordinate,
\begin{equation}
r=\sqrt{x^{2}+y^{2}+z^{2}}  \tag{3.2}
\end{equation}%
(or $r=\sqrt{x^{2}+y^{2}}$ in 2D). The Hamiltonian corresponding to Eq.
(3.1) is%
\begin{equation}
H=\frac{1}{2}\int \left( |\nabla \psi |^{2}-|\psi |^{4}\right) d^{D}x,
\tag{3.3}
\end{equation}%
where $\int d^{D}x$ stands for the integration in the 3D or 2D space.

Equation (3.1) is written as the GP equation for the three-dimensional BEC.
In the realization of the NLS equation as the spatiotemporal propagation
equation in optics, the evolutional variable $t$ in Eq. (3.1) is replaced
(as in Eq. (2.1)) by the propagation distance, $z$, while one of transverse
coordinates is replaced by the \textit{local time}, $\tau \equiv t-x/V_{%
\mathrm{gr}}$ (Kivshar and Agrawal, 2003), while two other spatial
variables, $\left( x,y\right) $, keep the meaning of the transverse
coordinates in the bulk optical waveguide.

In the multidimensional form, the GP/NLS equation (3.1) is always
non-integrable. Powerful numerical methods have been developed for the
solution of equations of this type, see reviews by Bao, Jaksch, and
Markowich, 2003; Muruganandam and Adhikari, 2009; Vudragovi\'{c} \textit{et
al}., 2012; Bao and Cai, 2013. In particular, the ground state of many
settings modeled by the GP equation can be looked for by means of the
imaginary-time-integration (alias gradient-flow) method (Bao and Du, 2004).
In spite of the complexity of the multidimensional nonlinear GP/NLS
equations, in many cases analytical methods, such as the variational and
Thomas-Fermi (TF) approximations, can be applied efficiently to these
models, as shown, in particular, in sections V and VI of this chapter. In
exceptional cases, exact analytical solutions can be found too (see, e.g.,
Eq. (6.15) below).

Fundamental (isotropic) localized solutions of Eq. (3.1) are looked for in
the usual form,%
\begin{equation}
\psi \left( r,t\right) =e^{-i\mu t}\phi (r),  \tag{3.4}
\end{equation}%
where $\mu <0$ is a real chemical potential in the case of BEC ($-\mu $ is
the propagation constant in the optics model), $r$ is the radial coordinate,
and real function $\phi (r)$ obeys the equation%
\begin{equation}
\mu \phi +\frac{1}{2}\left( \frac{d^{2}\phi }{dr^{2}}+\frac{D-1}{r}\frac{%
d\phi }{dr}\right) +\phi ^{3}=0,  \tag{3.5}
\end{equation}%
where $D=2$ or $3$ is the spatial dimension. Obviously, localized solutions
of Eq. (3.5) have the asymptotic form%
\begin{equation}
\phi (r)\sim r^{-(D-1)/2}\exp \left( -\sqrt{-2\mu }r\right) .  \tag{3.6}
\end{equation}%
Furthermore, in the 2D case the ansatz%
\begin{equation}
\psi \left( r,\theta ,t\right) =e^{-i\mu t+iS\theta }\phi _{S}(r),  \tag{3.7}
\end{equation}%
where $\theta $ is the azimuthal coordinate, gives rise to 2D solitons with
embedded integer vorticity (alias the winding number) $S=\pm 1,\pm 2$, ...
(Kruglov and Vlasov, 1985; Kruglov \textit{et al}., 1988). In this case, Eq.
(3.5) for real function $\phi _{S}(r)$ is replaced by
\begin{equation}
\mu \phi _{S}+\frac{1}{2}\left( \frac{d^{2}\phi _{S}}{dr^{2}}+\frac{1}{r}%
\frac{d\phi _{S}}{dr}-\frac{S^{2}}{r^{2}}\phi _{S}\right) +\phi _{S}^{3}=0.
\tag{3.8}
\end{equation}%
The asymptotic form of relevant solutions to Eq. (3.8) at $r\rightarrow 0$
is obvious too:%
\begin{equation}
\phi _{S}(r)\sim r^{|S|}.  \tag{3.9}
\end{equation}%
The presence of the inner hole in the middle of the soliton, represented by
Eq. (3.9), lends the vortex solitons an annular (ring-like) shape (see Fig.
13 below, which shows the hole in a 3D soliton with embedded vorticity).

The stationary equations (3.5) and (3.8) are invariant with respect to the
conformal transformation,%
\begin{equation}
\tilde{\mu}=l^{-2}\mu ,\tilde{r}=lr,\tilde{\phi}=l^{-1}\phi ,  \tag{3.10}
\end{equation}%
which entails rescaling of the integral norm,%
\begin{equation}
\tilde{N}=l^{D-2}N\equiv l^{D-2}\int |\psi |^{2}d^{D}x,  \tag{3.11}
\end{equation}%
where $l$ is an arbitrary scaling factor. Thus, the norm of the 2D solitons
with any vorticity $S$ is invariant with respect to the conformal
transformation -- in other words, for the soliton family with given $|S|$,
the norm takes a single value, which does not depend on $\mu $.

\begin{figure}[tbp]
\begin{center}
\subfigure[]{\includegraphics[width=0.45\textwidth]{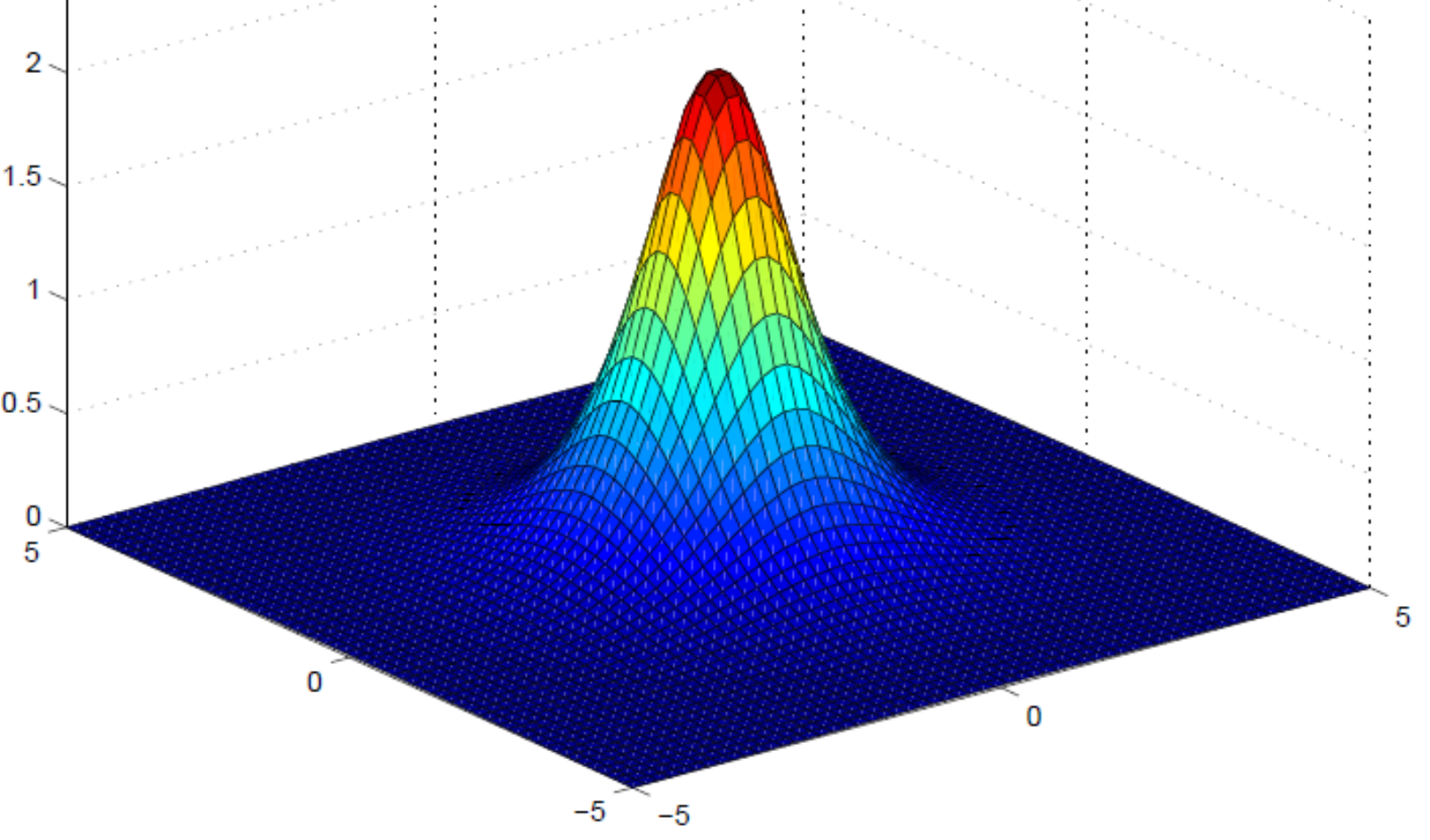}} %
\subfigure[]{\includegraphics[width=0.45\textwidth]{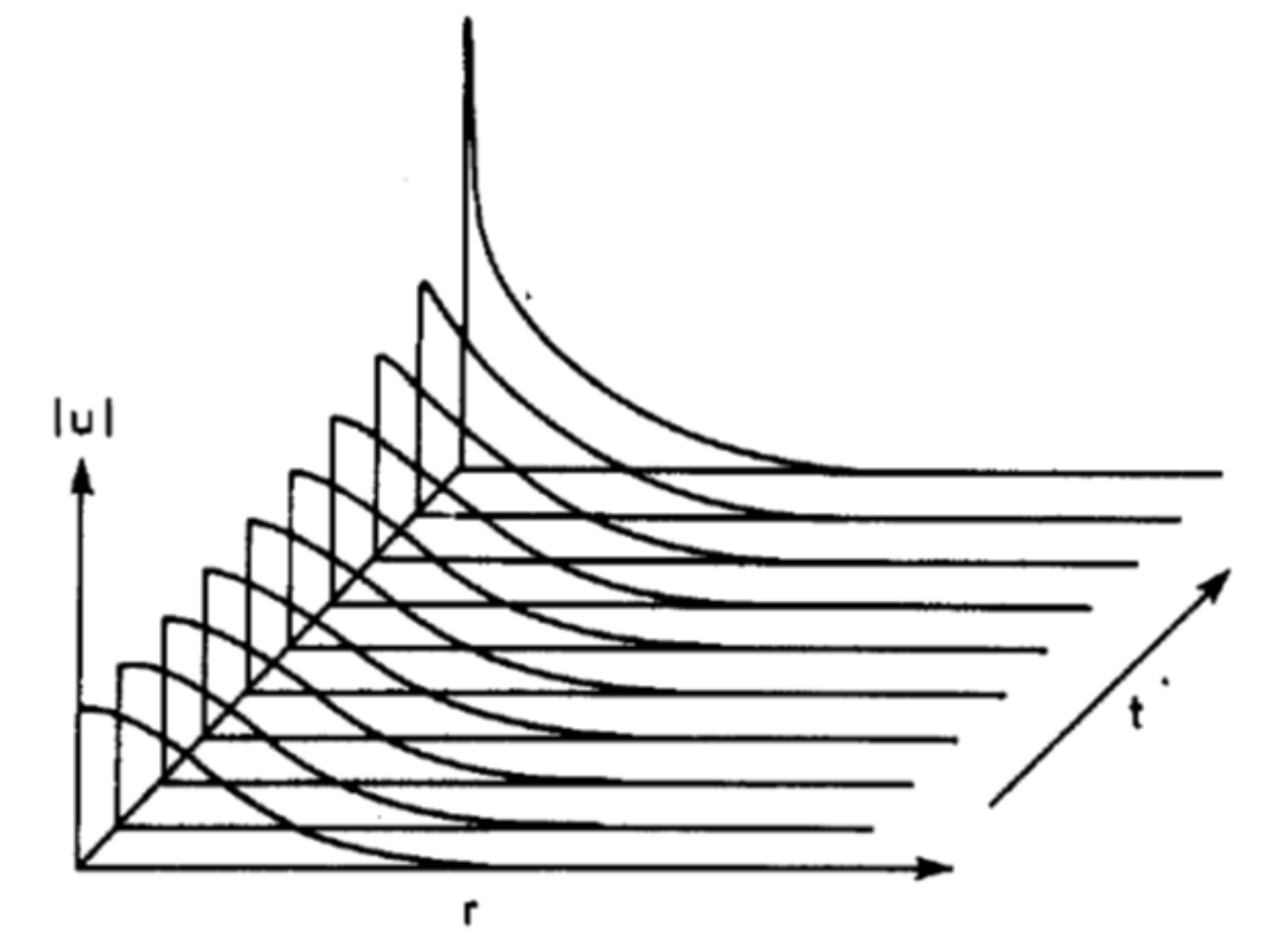}}
\end{center}
\caption{(a) A numerically found spatial profile of the two-dimensional TS
(Townes soliton). (b) Development of the TS's collapse shown in a radial
cross-section (source:
https://www2.mathematik.uni-halle.de/dohnal/SOLIT\_WAVES/NLS\_blowup.pdf).}
\label{fig1.15}
\end{figure}

In particular, for the 2D fundamental ($S=0$) NLS solitons, which are often
called \textit{Townes solitons} (TSs), which were first theoretically
considered by Chiao, Garmire, and Townes (1964), this universal value is%
\begin{equation}
N_{\mathrm{TS}}\approx 5.85.  \tag{3.12}
\end{equation}%
An analytical approximation for the same value was elaborated by Desaix,
Anderson, and Lisak (1991), who used the VA based on the Gaussian ansatz:%
\begin{equation}
\left( N_{\mathrm{TS}}\right) _{\mathrm{VA}}=2\pi ,  \tag{3.13}
\end{equation}%
thus the relative inaccuracy of the VA is $\approx 7\%$. A numerically found
plot of the TS is shown (in the Cartesian coordinates) in Fig. \ref{fig1.15}%
(a).

As a matter of fact, the TSs were introduced by Chiao, Garmire, and Townes
(1964) as the first example of solitons ever considered in optics (under the
name of the \textquotedblleft self-trapped optical beam", as this happened
before the term \textquotedblleft soliton" was coined by Zabusky and Kruskal
(1965)). However, the TSs are problematic objects, because they are subject
to intrinsic instability, as shown below (nevertheless, weakly unstable TSs
were recently created and observed in BEC, see Fig. \ref{fig1.16} below).

The same argument which leads to Eqs. (3.11) and (3.12), i.e., the
conclusion that the norm of the fundamental TSs does not depend on their
size (or chemical potential), applies to the TSs with embedded vorticity,
which were introduced, as mentioned above by Kruglov and Vlasov (1985), and
Kruglov \textit{et al}. (1988). At $S\geq 1$, the constant value of the norm
grows nearly linearly with $S$, see an approximate analytical result for
that, given below by Eq. (5.11).

\subsection{Similarities: one- and three-dimensional Townes solitons (TSs)}

It is relevant to mention that TSs can be defined in 1D as well as
(Abdullaev and Salerno, 2005), as solutions to the 1D NLS equation with the
self-attractive quintic term:%
\begin{equation}
i\psi _{t}+\frac{1}{2}\psi _{xx}+|\psi |^{4}\psi =0,  \tag{3.14}
\end{equation}%
with Hamiltonian%
\begin{equation}
H=\int_{-\infty }^{+\infty }\left( \frac{1}{2}\left\vert \psi
_{x}\right\vert ^{2}-\frac{1}{3}|\psi |^{6}\right) dx,  \tag{3.15}
\end{equation}%
cf. Eqs. (2.3) and (3.3). It admits a family of exact solutions in the form
of%
\begin{equation}
\psi _{\mathrm{TS}}^{\mathrm{(1D)}}=e^{-i\mu t}\frac{\left( -3/2\mu \right)
^{1/4}}{\sqrt{\cosh \left( 2\sqrt{-\mu }x\right) }},  \tag{3.16}
\end{equation}%
which exist for all $\mu <0$. Actually, this solution coincides with the one
given by Eq. (2.17) with $\xi =0$ (i.e., in the absence of the $\delta (x)$
term in Eq. (2.16)), in the case of $\sigma =2$, which was determined as the
critical one in the above consideration.

As well as their 2D counterparts, the entire family of these TSs has the
single value of the norm, which is identical to the one given above by Eq.
(2.23). Recall that the solitons produced by the same equation including the
$\delta $-functional potential term, i.e., solutions of Eq. (2.16) with the
top sign in front of the nonlinear term with $\sigma =2$, which are given by
expressions (2.16) and (2.17), are completely stable. On the contrary to
that, all solitons (3.16) are unstable. It is relevant to mention that the
substitution of solution (3.16) in expression (3.15) shows that the value of
the Hamiltonian (energy) for the entire family of the one-dimensional TSs is
exactly equal to zero:%
\begin{equation}
H_{\mathrm{TS}}^{\mathrm{(1D)}}\equiv \int_{-\infty }^{+\infty }\left( \frac{%
1}{2}\left\vert \frac{\partial \psi _{\mathrm{TS}}^{\mathrm{(1D)}}}{\partial
x}\right\vert ^{2}-\frac{1}{3}\left\vert \psi _{\mathrm{TS}}^{\mathrm{(1D)}%
}\right\vert ^{6}\right) dx=0.  \tag{3.17}
\end{equation}%
The same property is true for the two-dimensional TSs: the value of the 2D
Hamiltonian (3.3) for is zero for all the TSs belonging to the 2D family.

\subsection{The basic difficulty: instability of 2D and 3D solitons}

One-dimensional solitons appear, basically, as stable solutions of the
corresponding PDEs -- in particular, all fundamental soliton solutions of
the integrable KdV, NLS, and SG\ equations are stable. The above-mentioned $%
\mathcal{N}$-soliton compound solutions of the NLS equation (breathers),
such as the one given for $\mathcal{N}=2$ by Eq. (2.11), are subject to
slowly growing instability against splitting in free solitons, which is
possible because, as mentioned above, the binding energy of the compound is
zero.

The situation is dramatically different in the multidimensional settings,
because the 2D and 3D NLS/GP equations give rise to the above-mentioned
collapse, i.e., the appearance of a singularity in the solution (infinite
amplitude) after a finite evolution time. For the 2D equation (3.1) with the
cubic self-focusing in the free space ($U=0$), the occurrence of the
collapse is a consequence of the \textit{virial theorem} established by
Vlasov, Petrishchev, and Talanov (1971), in the form of a corollary
following from Eq. (3.1):%
\begin{equation}
\frac{d^{2}}{dt^{2}}\left( N\left\langle r^{2}\right\rangle \right) \equiv
\frac{d^{2}}{dt^{2}}\int \int r^{2}|\psi |^{2}dxdy=2H,  \tag{3.18}
\end{equation}%
where $H$ is Hamiltonian (3.3),
\begin{equation}
N=\int \int \left\vert \psi \left( x,y\right) \right\vert ^{2}dxdy
\tag{3.19}
\end{equation}
is the 2D norm, and $\left\langle r^{2}\right\rangle $ is the mean value of
the squared radius of the localized configuration of the wave function.
Because $N$ and $H$ are dynamical invariants (constants), a solution to Eq.
(3.18) gives%
\begin{equation}
\left\langle r^{2}\right\rangle (t)=\left\langle r^{2}\right\rangle
(t=0)+Ct+\left( H/N\right) t^{2},  \tag{3.20}
\end{equation}%
with a constant $C$. Thus, $\left\langle r^{2}\right\rangle $ vanishes $\sim
\left( t_{\mathrm{cr}}-t\right) $ at some critical moment of time, $t_{%
\mathrm{cr}}$, for $H<0$. The conservation of the norm suggests that,
simultaneously, the amplitude of the field diverges as $\left\vert \psi
\right\vert _{\max }\sim \left( t_{\mathrm{cr}}-t\right) ^{-1/2}$. The
vanishing of the mean squared radius at this moment implies the emergence of
the singularity. The shape of the collapsing state is asymptotically close
to that of the TS with $\mu \sim -\left( t_{\mathrm{cr}}-t\right) ^{-1}$
(Fibich, 2015).

On the other hand, for $H>0$ the same solution (3.20) implies that $%
\left\langle r^{2}\right\rangle $ diverges at $t\rightarrow \infty $, i.e.,
the localized configuration spreads out (decays). Actually, the TS solution
corresponds, as mentioned above, to $H=0$ (cf. Eq. (3.17) for the
one-dimensional TSs), thus the TS is a \textit{separatrix} between the
collapsing and decaying solutions. In any dynamical system, the separatrix
is obviously unstable against small perturbations. A typical example of the
instability of a TS in direct simulations, leading to the onset of the
collapse, is shown in Fig. \ref{fig1.15}(b).

The asymptotic stage of the supercritical collapse, governed by the 3D
version of Eq. (3.1) with the top sign in front of the cubic term and $U=0$,
is somewhat different, featuring $\left\langle r^{2}\right\rangle \sim
\left( t_{\mathrm{cr}}-t\right) ^{4/5}$ and $\left\vert \psi \right\vert
_{\max }\sim \left( t_{\mathrm{cr}}-t\right) ^{-3/5}$. These conclusion can
be obtained in an analytical form by means of the VA (Zakharov and
Kuznetsov, 2012).

The norm of the TS, given by Eq. (3.12), is a critical (threshold) value
necessary for the onset of the collapse in the framework of the 2D version
of Eq. (3.1). For this reason, it is named, as mention above, the critical
collapse (Zakharov and Kuznetsov, 2012). On the contrary, in the 3D version
of Eq. (3.1) the threshold value is zero (which was also mention above),
i.e., the collapse may be initiated by an input with an arbitrarily small
norm, for which reason it is called the supercritical collapse. Another
aspect of this issue is that the collapse in the 2D equation (3.1) always
includes a finite norm, $N>N_{\mathrm{TS}}$, therefore it is also called
\textit{strong collapse} (Zakharov and Kuznetsov, 2012). On the other hand,
the supercritical collapse governed by the 3D equation (3.1) is called
\textit{weak collapse} because, having no finite threshold in terms of the
norm, it may involve a small share of the total norm, while the rest will be
thrown away in the course of the blowup.

The analysis similar to that based on Eqs. (3.18) and (3.20), which aims to
predict the possibility of the collapse, can be developed in a less rigorous
but more general form, which applies to the NLS equation in the space of
dimension $D$ with the self-attraction term of an arbitrary power (not
necessarily cubic):%
\begin{equation}
i\psi _{t}+\frac{1}{2}\nabla ^{2}\psi +|\psi |^{2\sigma +1}\psi =0,
\tag{3.21}
\end{equation}%
the cubic term in Eq. (3.1) corresponding to $\sigma =1$, cf. Eq. (2.16).
This equation conserves the norm,
\begin{equation}
N=\int \left\vert \psi \right\vert d^{D}x,  \tag{3.22}
\end{equation}
and the Hamiltonian,%
\begin{equation}
H_{\sigma }=\int \left( \frac{1}{2}|\nabla \psi |^{2}-\frac{1}{\sigma +1}%
|\psi |^{2\sigma +2}\right) d^{D}x\equiv H_{\mathrm{grad}}+H_{\mathrm{%
self-focusing}},  \tag{3.23}
\end{equation}%
cf. Eq. (2.101). Then, following Zakharov and Kuznetsov (2012), one
considers a localized isotropic configuration of field $\psi $, with
amplitude $A$ and size (radius) $R$. An obvious estimate for the norm is%
\begin{equation}
N\sim A^{2}R^{D}.  \tag{3.24}
\end{equation}%
Similarly, the gradient and self-focusing terms in Hamiltonian (3.23) are
estimated as follows, eliminating $A^{2}$ in favor of $N$ by means of Eq.
(2.116):%
\begin{equation}
H_{\mathrm{grad}}\sim NR^{-2},H_{\mathrm{self-focusing}}\sim -N^{(\sigma
+1)/2}R^{-(\sigma -1)D/2}.~  \tag{3.24}
\end{equation}%
The collapse, i.e., catastrophic shrinkage of the state towards $%
R\rightarrow 0$, takes place if the consideration of $R\rightarrow 0$ for
fixed $N$ reveals that $H(R\rightarrow 0)\rightarrow -\infty $ (in other
words, the system's ground state formally corresponds to $H=-\infty $). The
comparison of the two terms in Eq. (3.24) readily demonstrates that the
unconditional (i.e., supercritical) collapse occurs if $\left\vert H_{%
\mathrm{self-focusing}}\right\vert $ diverges at $R\rightarrow 0$ faster
than $H_{\mathrm{grad}}$, which means%
\begin{equation}
\sigma D>2.  \tag{3.25}
\end{equation}%
In the critical case, which corresponds to
\begin{equation}
\sigma D=2  \tag{3.26}
\end{equation}%
(in particular, Eq. (3.26) holds for the 2D cubic ($\sigma =2$) NLS
equation), both terms in Eq. (3.24) feature the same scaling at $%
R\rightarrow 0$, the critical collapse taking place if $N$ exceeds a certain
threshold value, as shown above for the cases of $\sigma =1$ and $D=2$, as
well as $\sigma =2$ and $D=1$.

The same scaling arguments make it possible to establish a relation between
the norm and chemical potential of multidimensional solitons generated by
Eq. (3.21):%
\begin{equation}
N\sim (-\mu )^{\left( 2-\sigma D\right) /\left( 2\sigma \right) }.
\tag{3.27}
\end{equation}%
This dependence makes it possible to apply the above-mentioned VK \textit{%
criterion}. In the present notation, it takes the form of%
\begin{equation}
dN/d\mu <0,  \tag{3.28}
\end{equation}%
cf. Eq. (2.12). Obviously, the VK criterion, if applied to relation (3.27),
predicts instability precisely in the case when condition (3.25) holds.

Still more vulnerable to the instability are ring-shaped vortex solitons,
such as 2D ones constructed as per Eqs. (3.7) and (3.8). For them, in the
presence of the critical collapse, the \textquotedblleft most dangerous"
(fastest growing) instability mode is not the self-shrinkage driven by the
collapse, but spontaneous fission of the ring into a set of fragments.

The general approach to the study of stability is based on the consideration
of small perturbations added to the underlying stationary solution. As an
appropriate example, one can take 2D vortex solitons of Eq. (3.1), in the
form of expression (3.7). The perturbed solution is looked for, in the polar
coordinates, as%
\begin{equation}
\psi \left( r,\theta ,t\right) =e^{-i\mu t+iS\theta }\left\{ \phi
_{S}(r)+\varepsilon \left[ e^{iL\theta +\Gamma t}u(r)+e^{-iL\theta +i\Gamma
^{\ast }t}v^{\ast }(r)\right] \right\} ,  \tag{3.29}
\end{equation}%
where $\varepsilon $ is an infinitesimal amplitude of the perturbation,
integer $L$ is its azimuthal index, complex functions $u(r)$ and $v(r)$
represent the perturbation eigenmode, and $\Gamma $, which may be complex
too, is the eigenvalue. The substitution of ansatz (3.29) in Eq. (3.1) (with
$U=0$ and the top sign in front of the cubic term) and linearization with
respect to $\varepsilon $ leads to a system of the \textit{Bogoliubov - de
Gennes} (BdG)\ \textit{equations},%
\begin{equation}
\left( i\Gamma +\mu \right) u+\frac{1}{2}\left( \frac{d^{2}u}{dr^{2}}+\frac{1%
}{r}\frac{du}{dr}-\frac{\left( S+L\right) ^{2}}{r^{2}}u\right) +\phi
_{S}^{2}(r)\left( 2u+v\right) =0,  \tag{3.30}
\end{equation}%
\begin{equation}
\left( -i\Gamma +\mu \right) v+\frac{1}{2}\left( \frac{d^{2}v}{dr^{2}}+\frac{%
1}{r}\frac{dv}{dr}-\frac{\left( S-L\right) ^{2}}{r^{2}}v\right) +\phi
_{S}^{2}(r)\left( 2v+u\right) =0,  \tag{3.31}
\end{equation}%
which should be solved numerically, with boundary conditions (b.c.) $\left\{
u(r),v(r)\right\} \rightarrow 0$ at $r\rightarrow \infty $ and
\begin{equation}
u\sim r^{\left\vert S+L\right\vert },v\sim r^{\left\vert S-L\right\vert }~%
\mathrm{at}~r\rightarrow 0.  \tag{3.32}
\end{equation}%
The underlying solution is stable if all eigenvalues $\Gamma $ have zero
real parts. In the particular example corresponding to Eqs. (3.30)-(3.32),
all the vortex solitons are unstable, but the derivation of the BdG equation
outlined here sets a pattern for the derivation in other models, which may
produce stable solutions, as shown below.

As concerns the 2D TSs with $S=0$, produced by the cubic NLS equation (3.1),
the system of BdG equations (3.30) and (3.31) does not give rise to unstable
eigenvalues for them. In fact, the collapse-driven instability of the TS is
\textit{subexponential}, being accounted for by a pair of zero eigenvalues
(this fact explains the initially very slow growth of the instability
observed in Fig. \ref{fig1.15}(b)), while the splitting instability of
vortex-ring solitons carrying $S\neq 0$ is accounted for by a finite
exponential instability growth rate, with $\mathrm{Re}\left( \Gamma \right)
\neq 0$.

The VK criterion, given by Eq. (3.28), is related to the stability
eigenvalues: if the criterion holds, the spectrum does not contain purely
real eigenvalues $\Gamma >0$; however, the criterion ignores the possibility
of the existence of complex eigenvalues with $\mathrm{Re}\left( \Gamma
\right) >0$ and $\mathrm{Im}\left( \Gamma \right) \neq 0$. In particular,
the instability of the solitons in region (3.25), driven by the
supercritical collapse, is accounted for by a real positive $\Gamma $,
therefore it is detected by the VK criterion. On the other hand, as it is
shown below, the splitting instability of vortex rings is dominated by
complex eigenvalues, hence it is ignored by the criterion.

\subsection{The experimental situation: observation of weakly unstable
Townes solitons (TSs) in Bose-Einstein condensates (BECs)}

The fact that the instability of the TSs is weak, as outlined above, makes
it possible to create them in experiments. Very recently, such results were
reported in BEC. First, TSs composed of $\simeq 15\times 10^{3}$ atoms of
cesium were successfully made and observed by Chen and Hung (2020) in an
effectively two-dimensional setup, under the action of strong confinement
applied in the third direction. This result was achieved by means of the
\textit{quenching} method, i.e., switch of the nonlinearity sign from
repulsion to attraction by means of the Feshbach resonance (FR), i.e., the
possibility to alter the strength and sign of the effective interaction
between atoms in BEC, with the help of uniform dc magnetic field (Pollack
\textit{et al}., 2009; Bauer and Lettner, 2009; Chin \textit{et al}., 2010;
Tojo \textit{et al}., 2010). The FR allows one to switch the sign of the
interaction from repulsion to attraction, thus initiating the creation of
solitons. The observed profile of the TS was close to the one produced by
the numerical solution of Eq. (3.8) (see Fig. \ref{fig1.15}(a)). In fact,
the experiment reported in that work produced not a single TS, but a set of
them with different sizes, produced by the modulational instability of a
self-attractive condensate with a smooth distribution of the density.

In a subsequent experiment, Chen and Hung (2021) have demonstrated that the
profiles and sizes of individual TSs indeed obey the above-mentioned scaling
invariance, see Eqs. (3.10) and (3.11). Furthermore, it was observed that
non-negligible long-range dipole-dipole interactions between atoms in the
same BEC of cesium does not break the scaling invariance. The latter finding
may be explained by the analysis performed by Sakaguchi and Malomed (2011),
who had demonstrated that, in the \textquotedblleft additional" mean-field
(MF) approximation, which considers the interaction of the magnetic dipole
momentum of an individual atom with magnetostatic field created by the
distribution of the momentum density of all other atoms in the condensates,
amounts to a renormalization of the effective strength of the contact
interaction, represented by the cubic term in Eq. (3.1). Namely, the
scattering length of the contact interactions, $a_{s}$, to which the cubic
coefficient is proportional in the unscaled form of the GP equation
(Pitaevskii and Stringari, 2003), is replaced by%
\begin{equation}
a_{s}\rightarrow \left( a_{s}\right) _{\mathrm{eff}}\equiv
a_{s}+md^{2}/\hbar ^{2},  \tag{3.33}
\end{equation}

\noindent where $d$ is the dipole momentum, and $m$ is the atomic mass (in
fact, this relation was derived for a gas of particles (small molecules)
carrying an electric dipole moment, but the result for the magnetic moments
is essentially the same).

Other recent experimental results demonstrating the creation of observable
TSs were reported by Bakkali-Hassani \textit{et al}. (2021). They used a
mixture of two atomic states in $^{87}$Rb. It is known that, in this
species, the FR cannot switch the interaction sign from repulsion to
attraction. Instead, the effectively two-dimensional experimental setup used
a relatively small number of atoms in one state, $\sim 1000$, embedded into
a gas composed of a much larger number of atoms in the second state. The
respective system of scaled 2D GP equations for wave functions of the two
components, $\psi _{1,2}$, is (cf. Eq. (3.1))%
\begin{equation}
i\left( \psi _{1}\right) _{t}+\frac{1}{2}\nabla ^{2}\psi _{1}+\left(
g_{11}|\psi _{1}|^{2}+g_{12}|\psi _{2}|^{2}\right) \psi _{1}=0,  \tag{3.34}
\end{equation}%
\begin{equation}
i\left( \psi _{2}\right) _{t}+\frac{1}{2}\nabla ^{2}\psi _{2}+\left(
g_{22}|\psi _{2}|^{2}+g_{21}|\psi _{1}|^{2}\right) \psi _{2}=0,  \tag{3.35}
\end{equation}

\noindent where $g_{11,22}$ and $g_{12}\equiv g_{21}$ are, respectively,
positive coefficients accounting for the self-repulsion of each component
and cross-repulsion between them (the equality of $g_{12}$ and $g_{21}$ is
an obvious symmetry property of the system). While no direct attractive
interactions are possible in this case, the well-known condition,%
\begin{equation}
g_{11}g_{22}<g_{12}^{2}  \tag{3.36}
\end{equation}%
makes the binary BEC immiscible (Mineev, 1967; Timmermans, 1998). This
condition can be imposed in binary condensates of $^{87}$Rb atoms, as
demonstrated experimentally by Tojo \textit{et al}. (2010). Effectively,
this implies that the minority component features immiscibility-induced
self-attraction, which makes it possible to create a TS, on top of the
majority background with a nonzero density. The experimentally observed 2D
density plot demonstrating the creation of the TS, and its radial profile,
compared to the numerical solution of Eq. (2.103), are displayed in Fig. \ref%
{fig1.16}. The existence of the well-established TS was observed on the time
scale $\sim 50$ ms. Other experimental runs reported by Bakkali-Hassani
\textit{et al}. (2021) demonstrate, as well, slow decay of the TSs and the
start of their collapsing.
\begin{figure}[tbp]
\begin{center}
\includegraphics[width=0.55\textwidth]{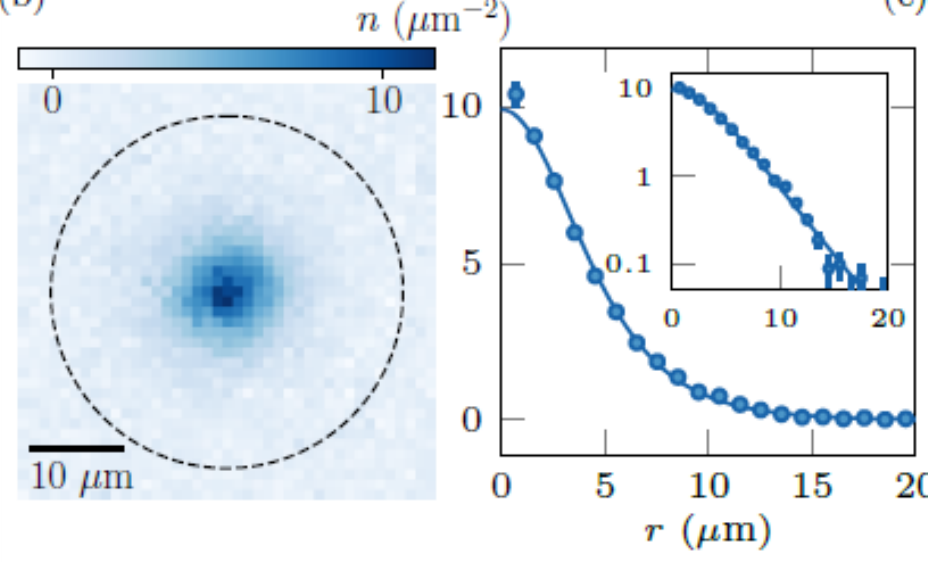}
\end{center}
\caption{Left: the experimentally observed density distribution in a
quasi-stable TS (Townes soliton) created in a minority component of a binary
condensate of $^{87}$Rb atoms, embedded in a majority background. The
effective self-attraction in the minority component, which is necessary to
build the TS, is induced by the immiscibility of the binary system. Right:
the experimentally observed radial profile of the same TS (the chain of
dots), fitted to the respective numerical solution of Eq. (3.5). The inset
shows the same data on the semilog scale, highlighting the exponentially
decaying structure of the TS's tails (source: Bakkali-Hassani \textit{et al}%
., 2021). }
\label{fig1.16}
\end{figure}

\section{The central issue: stabilization schemes for multidimensional
solitons}


\subsection{Scalar (single-component) models in free space (no external
potential)}

The simplest possibility to arrest the onset of the collapse, and thus
stabilize multidimensional fundamental solitons, is to introduce 2D and 3D
versions of the NLS equation with the cubic-quintic (CQ) nonlinearity, i.e.,%
\begin{equation}
i\psi _{t}+\frac{1}{2}\nabla ^{2}\psi +|\psi |^{2}\psi -|\psi |^{4}\psi =0,
\tag{4.1}
\end{equation}%
where all coefficients may be set equal to $\pm 1$ by means of obvious
rescaling. The 2D version of Eq. (4.1) can be implemented in optics,
considering the light propagation in a bulk waveguide filled by a dielectric
material whose intrinsic optical nonlinearity may be accurately approximated
by the combination of the cubic self-focusing and quintic defocusing, such
as liquid carbon disulfide (Tominaga and Yoshihara, 1995; Kong \textit{et al}%
., 2009; Edilson \textit{et al.}, 2013).

The same CQ nonlinearity in optics can be also realized in colloidal
suspensions of metallic nanoparticles (Reyna and de Ara\'{u}jo, 2017). In
the latter case, the size and concentration of the particles may be used as
parameters controlling actual values of coefficients in front of the cubic
and quintic terms in the counterpart of Eq. (4.1), written in the original
physical units (before the rescaling is applied to cast the NLS equation in
the form Eq. (4.1)). Furthermore, in a particular region of values of the
control parameters, the colloid may feature the effective optical
nonlinearity including a septimal (seventh-order) term too. In the
appropriately scaled form, the cubic-quintic-septimal generalization of Eq.
(4.1) takes the form of%
\begin{equation}
i\psi _{t}+\frac{1}{2}\nabla ^{2}\psi +\sigma _{3}|\psi |^{2}\psi -\sigma
_{5}|\psi |^{4}\psi -|\psi |^{6}\psi =0,  \tag{4.2}
\end{equation}%
where it is assumed that the highest septimal term is self-defocusing (to
prevent the onset of the collapse), while the coefficients in front of the
cubic and quintic ones, $\sigma _{3}$ and $\sigma _{5}$, may have any sign.
In fact, the rescaling makes it possible to set $\sigma _{5}=\pm 1$ (unless $%
\sigma _{5}=0$), keeping $\sigma _{3}$ as an irreducible parameter which is
allowed to take any positive or negative real value. Furthermore, Reyna and
de Ara\'{u}jo (2014) had experimentally demonstrated a regime in which the
competing nonlinearities are represented by the quintic and septimal terms,
while the cubic one nearly vanishes ($\sigma _{3}\approx 0$).

Similar nonlinearities of the CQ type occur in other optical media too,
including chalcogenide glasses, with a chemical composition different from
usual optical silica (Boudebs \textit{et al.}, 2003), organic materials
(stilbazolium derivatives, as shown experimentally by Zhan \textit{et al}.,
2002), and cold atomic gases (Greenberg and Gauthier, 2012). Furthermore, it
was demonstrated by Reshef \textit{et al}. (2017) that the CQ nonlinearity
is a generic feature of optical systems with a low linear refractive index
(they play an important role in modern photonics, known as \textquotedblleft
\textit{epsilon-near-zero}" media, see reviews by Liberal and Engheta
(2017), and Niu \textit{et al}. (2018)). It is also relevant to mention that
artificial CQ nonlinearity can be constructed by means of the \textit{%
cascading mechanism}, which induces a higher-order effective nonlinearity as
a chain of lower-order ones (Dolgaleva, Shin, and Boyd, 2009).

The addition of the defocusing cubic term represents an effect of \textit{%
saturation} of the underlying cubic self-focusing. In other cases, the
saturation is adequately modeled by the NLS equation with rational, rather
than polynomial, nonlinearity:%
\begin{equation}
i\psi _{t}+\frac{1}{2}\nabla ^{2}\psi +\frac{|\psi |^{2}}{1+|\psi |^{2}}\psi
=0.  \tag{4.3}
\end{equation}%
This equation was introduced by Anderson and Bonnedal (1979) for the
propagation of laser beams in plasmas. Essentially the same type of the
saturation finds important realizations in optics as a model for the
propagation of light beams in semiconductor-doped glasses (Rossignol \textit{%
et al}, 1987; Coutaz and Kull, 1991), as well as for beams with
extraordinary polarization in photorefractive (PhR) crystals (Segev \textit{%
et al}., 1992; Efremidis \textit{et al}. 2002). Note that, by means of
transformation $\psi =\exp \left( it\right) \tilde{\psi}$, Eq. (4.3) can be
written in another form, which is often used too:%
\begin{equation}
i\tilde{\psi}_{t}+\frac{1}{2}\nabla ^{2}\tilde{\psi}-\frac{1}{1+|\tilde{\psi}%
|^{2}}\tilde{\psi}=0.  \tag{4.4}
\end{equation}

A very essential difference of the saturable nonlinearity, as it is written
in Eqs. (4.3) or (4.4), from the CQ term in Eq. (4.1) is the fact that the
saturable nonlinearity may stabilize only fundamental 2D solitons, while
their counterparts with embedded vorticity remain completely unstable
against spontaneous splitting (Firth and Skryabin, 1997). On the contrary,
the combination of the cubic self-focusing and quintic defocusing can
stabilize 2D solitons with any integer value of embedded vorticity (winding
number) $S$ (Quiroga-Texeiro and Michinel, 1997; Malomed, Crasovan, and
Mihalache, 2002; Davydova and Yakimenko, 2004), although for $S\geq 3$ the
stability region becomes very narrow, see Eq. (4.14) below (Pego and
Warchall, 2002). Solutions for vortex solitons are looked for by means of
the same substitution as written above in Eq. (3.7), which leads to the
following radial equation for the stationary wave amplitude:
\begin{equation}
\mu \phi _{S}+\frac{1}{2}\left( \frac{d^{2}\phi _{S}}{dr^{2}}+\frac{1}{r}%
\frac{d\phi _{S}}{dr}-\frac{S^{2}}{r^{2}}\phi _{S}\right) +\phi
_{S}^{3}-\phi _{S}^{5}=0,  \tag{4.5}
\end{equation}%
cf. Eq. (3.8). The stability of these states is investigated by taking
perturbed solutions in the same form (3.29) as introduced above. Then, the
linearization leads to the system of equations for small perturbations, of
the BdG type, which is a straightforward extension of Eqs. (3.30) and (3.31):%
\begin{gather}
\left( i\Gamma +\mu \right) u+\frac{1}{2}\left( \frac{d^{2}u}{dr^{2}}+\frac{1%
}{r}\frac{du}{dr}-\frac{\left( S+L\right) ^{2}}{r^{2}}u\right)  \notag \\
+\phi _{S}^{2}(r)\left( 2u+v\right) -\phi _{S}^{4}\left( 3u+2v\right) =0,
\tag{4.6}
\end{gather}%
\begin{gather}
\left( -i\Gamma +\mu \right) v+\frac{1}{2}\left( \frac{d^{2}v}{dr^{2}}+\frac{%
1}{r}\frac{dv}{dr}-\frac{\left( S-L\right) ^{2}}{r^{2}}v\right)  \notag \\
+\phi _{S}^{2}(r)\left( 2v+u\right) -\phi _{S}^{4}\left( 2u+3v\right) =0.
\tag{4.7}
\end{gather}%
Equations (4.6) and (4.7) should be supplemented, as above, by b.c. (3.32).

The NLS equation with the saturable quintic self-focusing (without cubic
terms),%
\begin{equation}
i\psi _{t}+\frac{1}{2}\nabla ^{2}\psi +\frac{|\psi |^{4}}{1+|\psi |^{4}}\psi
=0,  \tag{4.8}
\end{equation}%
cf. Eq. (4.3), may be a relevant model as well, applying, in a certain
parameter region, to the light propagation in a waveguide filled with liquid
carbon disulfide (Reyna \textit{et al}., 2016).

In all these realizations of the NLS equation with the polynomial (in
particular, CQ) or saturable nonlinearity in optics, $t$ is actually the
propagation distance (usually denoted $z$, in that context), and the
relevant realization is two-dimensional, with transverse coordinates $\left(
x,y\right) $ in the bulk waveguides modeled by these equations. The full 3D
realization would make it necessary to consider the spatiotemporal
propagation in the same waveguides, which is far from implementation in
experiments.

\subsection{The wave field in the trapping potential}

Another generic possibility to stabilize 2D and 3D solitons against the
collapse driven by the cubic self-focusing, and to partly stabilize vortex
solitons against the splitting instability, is to add a trapping potential
to the multidimensional NLS/GP equation. The usually considered potential is
a straightforward multidimensional extension of the one-dimensional HO
potential (2.2). Thus, in the 3D case, Eq. (3.1) is replaced by%
\begin{equation}
i\psi _{t}+\frac{1}{2}\nabla ^{2}\psi +|\psi |^{2}\psi -\frac{1}{2}\left[
\Omega _{x,y}^{2}\left( x^{2}+y^{2}\right) +\Omega _{z}^{2}z^{2}\right] \psi
=0.  \tag{4.9}
\end{equation}%
This form of the HO potential implies a possibility of its anisotropy, with $%
\Omega _{z}^{2}\neq $ $\Omega _{x,y}^{2}$, which keeps the axial symmetry in
the $\left( x,y\right) $ plane (this is the usual situation which takes
place in experiments (Pitaevskii and Stringari, 2003)), but breaks the
spherical symmetry. In principle, one may also consider a fully anisotropic
potential, with $\Omega _{x}^{2}\neq \Omega _{y}^{2}$. In optics, the NLS
equation may be three-dimensional, modeling the spatiotemporal propagation
of light in the bulk waveguide, but the trapping potential may only be
two-dimensional, which corresponds to $\Omega _{z}=0$ in Eq. (4.10), as the
potential cannot be a quadratic function of time.

Stationary solutions of Eq. (4.10), with vorticity $S$, are naturally looked
for, in cylindrical coordinates,
\begin{equation}
\left\{ \rho \equiv \sqrt{x^{2}+y^{2}},\theta ,z\right\}  \tag{4.10}
\end{equation}
as a 3D version of ansatz (3.7):%
\begin{equation}
\psi =e^{-i\mu t+iS\theta }\phi _{S}\left( \rho ,z\right) ,  \tag{4.11}
\end{equation}%
with real function $\phi _{S}$ satisfying equation
\begin{gather}
\mu \phi _{S}+\frac{1}{2}\left( \frac{d^{2}\phi _{S}}{d\rho ^{2}}+\frac{1}{%
\rho }\frac{d\phi _{S}}{d\rho }-\frac{S^{2}}{\rho ^{2}}\phi _{S}+\frac{%
d^{2}\phi _{S}}{dz^{2}}\right) +\phi _{S}^{3}  \notag \\
-\frac{1}{2}\left( \Omega _{x,y}^{2}\rho ^{2}+\Omega _{z}^{2}z^{2}\right)
\phi _{S}=0.  \tag{4.12}
\end{gather}

The NLS/GP equation (4.9) with the axially symmetric but spherically
asymmetric trapping potential conserves three dynamical invariants, \textit{%
viz}., the norm (see also Eq. (3.22))%
\begin{equation}
N=\int \int \int dxdydz\left\vert \psi \left( x,y,z,t\right) \right\vert
^{2},  \tag{4.13}
\end{equation}%
$z$-component of the angular momentum,%
\begin{equation}
M_{z}=-i\int \int \int dxdydz\psi ^{\ast }\left( x\frac{\partial \psi }{%
\partial y}-y\frac{\partial \psi }{\partial x}\right) \equiv
-i\int_{0}^{\infty }\rho d\rho \int_{0}^{2\pi }d\theta \int_{-\infty
}^{+\infty }dz\psi ^{\ast }\frac{\partial \psi }{\partial \theta }
\tag{4.14}
\end{equation}%
(it is written here in terms of both the Cartesian coordinates and
cylindrical ones, defined as per Eq. (4.10)), and the Hamiltonian (cf. Eq.
(3.3)):%
\begin{equation}
H=\frac{1}{2}\int \int \int dxdydz\left\{ \left\vert \nabla \psi \right\vert
^{2}-|\psi |^{4}+\left[ \Omega _{x,y}^{2}\left( x^{2}+y^{2}\right) +\Omega
_{z}^{2}z^{2}\right] |\psi |^{2}\right\} .  \tag{4.15}
\end{equation}%
In the case when the HO potential is isotropic, with $\Omega
_{x,y}^{2}=\Omega _{z}^{2}$, Eq. (4.9) conserves not only the $z$-component
(4.14), but the full vectorial angular momentum,%
\begin{equation}
\mathbf{M}=-i\int \int \int dxdydz\psi ^{\ast }\left( \mathbf{r}\times
\nabla \right) \psi .  \tag{4.16}
\end{equation}

In the general form, the stability of the stationary states is addressed by
looking for a perturbed solution of Eq. (4.10), in the cylindrical
coordinates, as

\begin{equation}
\psi \left( r,\theta ,z,t\right) =e^{-i\mu t+iS\theta }\left\{ \phi
_{S}(\rho ,z)+\varepsilon \left[ e^{iL\theta +\Gamma t}u(\rho
,z)+e^{-iL\theta +i\Gamma ^{\ast }t}v^{\ast }(\rho ,z)\right] \right\} ,
\tag{4.17}
\end{equation}%
which is a straightforward extension of the 2D ansatz (3.29). The
substitution of this in Eq. (4.9) and linearization with respect to the
small amplitude $\varepsilon $ of the perturbation leads to the
corresponding system of the BdG equations,%
\begin{gather}
\left( i\Gamma +\mu \right) u+\frac{1}{2}\left[ \frac{d^{2}u}{d\rho ^{2}}+%
\frac{1}{\rho }\frac{du}{d\rho }-\frac{\left( S+L\right) ^{2}}{\rho ^{2}}u+%
\frac{d^{2}u}{dz^{2}}\right]  \notag \\
+\phi _{S}^{2}(\rho ,z)\left( 2u+v\right) -\frac{1}{2}\left( \Omega
_{x,y}^{2}\rho ^{2}+\Omega _{z}^{2}z^{2}\right) u=0,  \tag{4.18}
\end{gather}%
\begin{gather}
\left( -i\Gamma +\mu \right) v+\frac{1}{2}\left[ \frac{d^{2}v}{dr^{2}}+\frac{%
1}{r}\frac{dv}{dr}-\frac{\left( S-L\right) ^{2}}{r^{2}}v+\frac{d^{2}v}{dz^{2}%
}\right]  \notag \\
+\phi _{S}^{2}(\rho ,z)\left( 2v+u\right) -\frac{1}{2}\left( \Omega
_{x,y}^{2}\rho ^{2}+\Omega _{z}^{2}z^{2}\right) v=0,  \tag{4.19}
\end{gather}%
cf. Eqs. (3.30), (3.31) and (4.6), (4.7). Equations (4.18) and (4.19) should
be solved numerically, subject to b.c.
\begin{equation}
u\sim \rho ^{\left\vert S+L\right\vert },v\sim \rho ^{\left\vert
S-L\right\vert }~\mathrm{at}~\rho \rightarrow 0,  \tag{4.20}
\end{equation}%
cf. Eq. (3.32).

\subsection{Spatially periodic potentials}

Optical-lattice (OL) potentials, which may be induced, as interference
patterns, by pairs of counter-propagating laser beams illuminating the BEC,
provide a versatile toolbox for experimental and theoretical studies of
atomic condensates (Brazhnyi and Konotop, 2004; Morsch and Oberthaler, 2006;
Lewenstein, Sanpera, and Ahufinger, 2012; Dutta \textit{et al}., 2015). The
general 3D form of the OL potential is
\begin{equation}
U\left( x,y,z\right) =-\left[ \left( V_{0}\right) _{x}\cos \left( \frac{2\pi
}{L_{x}}x\right) +\left( V_{0}\right) _{y}\cos \left( \frac{2\pi }{L_{y}}%
y\right) +\left( V_{0}\right) _{z}\cos \left( \frac{2\pi }{L_{z}}z\right) %
\right] ,  \tag{4.21}
\end{equation}%
where different amplitudes $\left( V_{0}\right) _{x,y,z}$ and periods $%
L_{x,y,z}$ imply a possibility to consider anisotropic lattice potentials,
and the respective NLS/GP equation is written as%
\begin{equation}
i\psi _{t}+\frac{1}{2}\nabla ^{2}\psi +|\psi |^{2}\psi +U\left( x,y,z\right)
\psi =0.  \tag{4.22}
\end{equation}%
The 3D lattice potential is sufficient to stabilize both fundamental and
vortex solitons. In particular, although the periodic potential breaks the
spatial isotropy, the soliton's vorticity (alias the winding number) may be
unambiguously defined in this case too (Baizakov, Malomed, and Salerno,
2003; Yang and Musslimani, 2003).

The 2D reduction of the three-dimensional OL potential (4.21) is obvious. An
essential finding is that the quasi-2D lattice potential in the 3D space,
i.e., potential (4.21) with $\left( V_{0}\right) _{z}=0$, is sufficient for
the stabilization of fully three-dimensional solitons; similarly, the
quasi-1D potential, with only $\left( V_{0}\right) _{x}\neq 0$, is
sufficient for the stabilization of 2D solitons, although the quasi-1D
lattice cannot stabilize full 3D solitons (Baizakov, Malomed, and Salerno,
2004; Mihalache \textit{et al}., 2004a; Leblond, Malomed, and Mihalache,
2007).

Another relevant form of the OL\ potential in the 2D space (as well as a
quasi-2D potential in 3D) is a radial lattice. The corresponding 2D NLS/GP
equation takes the form of%
\begin{equation}
i\psi _{t}+\frac{1}{2}\nabla ^{2}\psi +|\psi |^{2}\psi -V\left( \rho \equiv
\sqrt{x^{2}+y^{2}}\right) \psi =0,  \tag{4.23}
\end{equation}%
where the radial potential was considered in various forms -- in particular,
as%
\begin{equation}
V_{\mathrm{Bessel}}(\rho )=-V_{0}J_{0}(\rho /a),  \tag{4.24}
\end{equation}%
with Bessel function $J_{0}(\rho /a)$, by Kartashov, Vysloukh, and Torner
(2004), and as the radially periodic potential,%
\begin{equation}
V_{\cos }(\rho )=-V_{0}\cos \left( 2k\rho \right) ,  \tag{4.25}
\end{equation}%
(cf. Eq. (2.62)) by Baizakov, Malomed, and Salerno (2006).

In optics, the underlying NLS equation may be three- or two-dimensional, for
the spatiotemporal or spatial-domain propagation, respectively, The
effective periodic potential, representing the photonic-crystal structure in
this equation (including its 3D version) may be solely two-dimensional,
corresponding to the spatially-periodic modulation of the local refractive
index in the transverse plane (see books by Joannopoulos \textit{et al}.
(2008), Skorobogatiy and Yang (2009), and Yang (2010)).

A powerful method, which has made it possible to create many species of 2D
optical solitons in PhR materials (see the review by Lederer \textit{et al}.
2008), was proposed and implemented by Efremidis \textit{et al}. (2002,
2003). It makes use of the fact that\ the light with ordinary polarization
propagates linearly in PhR crystals, hence interference of laser beams with
such polarization can be used to induce an effective (virtual) lattice
pattern in the bulk material, while light with the extraordinary
polarization, illuminating the same material, is subject to the action of
strong saturable nonlinearity (Segev \textit{et al}., 1992). The result
includes self-focusing of the extraordinarily-polarized probe beam (as in
Eq. (4.4)) and the XPM-mediated action of the virtual lattice potential,
which is optically induced by the interference pattern in the ordinary
polarization. The effective 2D equation for the spatial-domain propagation
of the probe field $u$ is (Efremidis \textit{et al}. (2002, 2003))%
\begin{equation}
iu_{z}+\frac{1}{2}\left( u_{xx}+u_{zz}\right) -\frac{E_{0}u}{1+|u|^{2}+U_{0}%
\left[ \cos \left( 2\pi x/L_{x}\right) +\cos \left( 2\pi y/L_{y}\right) %
\right] },  \tag{4.26}
\end{equation}%
where $z$ is the propagation constant, $\left( x,y\right) $ are the
transverse coordinates, $U_{0}$ is the strength of the virtual lattice
potential, and $E_{0}$ is the dc electric field which induces the saturable
nonlinearity (self-focusing and defocusing for $E_{0}>0$ and $E_{0}<0$,
respectively). Thus, Eq. (4.26) introduces a 2D model similar to one based
on Eqs. (4.22) and (4.21), but with a more complex structure. Note that the
lowest-order expansion of Eq. (4.26) for small $|u|^{2}$ and $U_{0}$ reduces
to the 2D version of Eq. (4.22).

The above models with lattice potentials, based on Eqs. (4.22) and (4.23),
are written with the self-focusing (self-attractive) sign in front of the
cubic term. In realization provided by optical media, this is virtually
always the case. In BEC, the opposite, self-repulsive nonlinearity, is more
typical, corresponding to natural repulsive interactions between atoms in
the BEC, although the sign may be switched to the attraction by means of FR.
In the case of the self-repulsive nonlinearity, 2D and 3D self-trapped
states can be created, with the help of the OL potential, as \textit{gap
solitons}. Further, stable 2D gap solitons with intrinsic vorticity were
constructed by Sakaguchi and Malomed (2004), see examples in Fig. \ref%
{fig2.1}.
\begin{figure}[tbp]
\begin{center}
\includegraphics[width=0.68\textwidth]{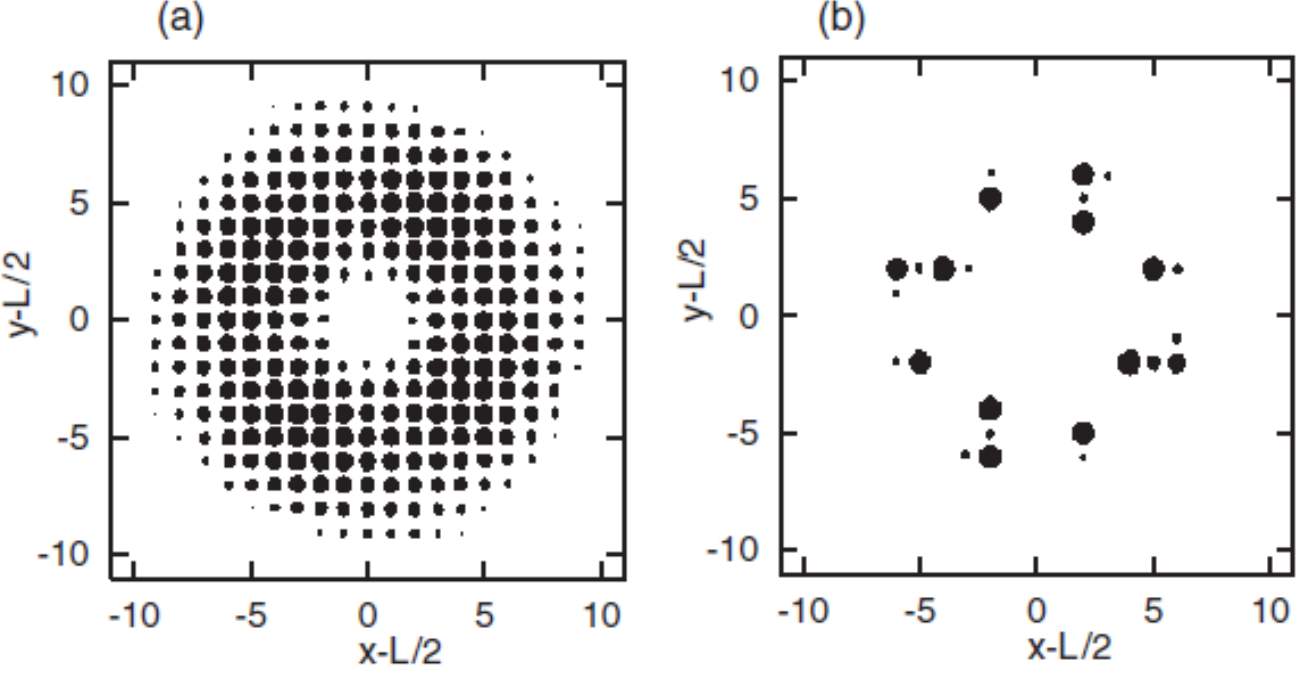}
\end{center}
\caption{Typical examples of stable 2D gap solitons with embedded vorticity $%
S=2$, produced by the two-dimensional NLS/GP equation with a lattice
potential (the 2D version of potential (4.21)) and self-repulsive cubic
nonlinearity. Panels (a) and (b) display density distributions in loosely-
and tightly bound vortex gap solitons, respectively (source: Sakaguchi and
Malomed, 2004b).}
\label{fig2.1}
\end{figure}

\subsection{Nonlinear trapping potentials}

\paragraph{Potentials based on attractive nonlinearities}

Quite interesting predictions for the creation of stable multidimensional
soliton-like (self-trapped) states were reported using the NLS/GP equation
with effective nonlinear potentials induced by local modulation of the
coefficient in front of the cubic term in the equation. The general form of
the equation is
\begin{equation}
i\psi _{t}+\frac{1}{2}\nabla ^{2}\psi +g(x,y,z)|\psi |^{2}\psi =0.
\tag{4.27}
\end{equation}%
Following the terminology used since long ago in solid-state physics
(Harrison, 1966), such settings are sometimes referred to as \textit{%
pseudopotentials}. In optics, 1D and 2D versions of this model can be
realized in media with spatially inhomogeneous distributions of density of
resonant dopants, which affect the local strength of the self-focusing
(Hukride, Runde, and Kip, 2003). In BEC, Eq. (4.27) can be implemented in
the space of any dimension by means of FR with the local strength controlled
by a spatially inhomogeneous dc magnetic field (Ghanbari \textit{et al}.,
2006; Bauer \textit{et al.}, 2009) or by an appropriately patterned optical
field, which also can also control the FR (Yamazaki \textit{et al}., 2010;
Clark \textit{et al}., 2015).

Many theoretical and experimental results obtained for solitons supported by
spatially periodic nonlinear potentials, as well as by combinations of
nonlinear and linear lattices, were summarized in a review by Kartashov,
Malomed, and Torner (2011). A majority of these results pertain to 1D
setups. 2D nonlinear lattices can support stability of solitons against the
collapse under special conditions, such as a necessity to use nonlinear
potentials with sharp edges. Examples of 2D settings of the latter type were
elaborated, in particular, for the 2D version of Eq. (4.27) with radial
modulation profiles in the form of an annulus (Sakaguchi and Malomed, 2006)
or step-wise structure (Sakaguchi and Malomed, 2012), i.e.,%
\begin{equation}
i\psi _{t}+\frac{1}{2}\nabla ^{2}\psi +g(r)|\psi |^{2}\psi =0,  \tag{4.28}
\end{equation}%
with, respectively,%
\begin{equation}
g_{\mathrm{annulus}}(r)=%
\begin{array}{c}
1,~\mathrm{at~}r_{\min }<r<r_{\max }, \\
0,~\mathrm{at~}r<r_{\min }~\mathrm{and}~r>r_{\max },%
\end{array}
\tag{4.29}
\end{equation}%
or%
\begin{equation}
g_{\mathrm{step}}(r)=%
\begin{array}{c}
1,~\mathrm{at~}r<r_{\max }, \\
1-\Delta g,~\mathrm{at~}r>r_{\max },%
\end{array}
\tag{4.30}
\end{equation}%
where constraint $0<\Delta g<1$ is imposed. The modulation profile (4.29)
maintains stable 2D solitons, provided that the annulus is not too narrow,
\textit{viz}., for $r_{\min }/r_{\max }<0.47$, while profile (4.30) provides
stability of 2D solitons in an interval $\Delta N$ of the norm around its
TS\ value (3.12) which scales $\sim \Delta g$ for $\Delta g\ll 1$.

\paragraph{Potentials based on the repulsive nonlinearity}

A different option for the creation of highly stable self-trapped states was
proposed by Borovkova \textit{et al}. (2011a,b). It is based on Eq. (4.28)
with the self-defocusing (repulsive) nonlinearity, $g(r)<0$, whose local
strength, $\left\vert g(r)\right\vert $, diverges at $r\rightarrow \infty $
as%
\begin{equation}
g(r\rightarrow \infty )\approx -g_{0}r^{\alpha },  \tag{4.31}
\end{equation}%
with positive constant $g_{0}$ (Borovkova \textit{et al}. 2011a), or faster
-- in particular, as an \textit{anti-Gaussian}, i.e.,%
\begin{equation}
g(r\rightarrow \infty )\approx -g_{0}\exp \left( r^{2}/r_{0}^{2}\right)
\tag{4.32}
\end{equation}%
(Borovkova \textit{et al}. 2011b). The corresponding stationary 3D and 2D
solutions with chemical potential $\mu >0$ are looked for as%
\begin{equation}
\psi _{\mathrm{3D}}=e^{-i\mu t}\phi (r),\psi _{\mathrm{2D}}=e^{-i\mu
t+iS\theta }\phi (r),  \tag{4.33}
\end{equation}%
where $S=0,1,2,...$ is vorticity, $S=0$ corresponding to the ground state
(fundamental stationary solution). The substitution of ansatz (4.33), with
dimension $D=3$ or $2$, in Eq. (4.28) leads to the respective stationary
equation,%
\begin{equation}
\mu \phi +\frac{1}{2}\left( \frac{d^{2}\phi }{dr^{2}}+\frac{D-1}{r}\frac{%
d\phi }{dr}-\frac{S^{2}}{r^{2}}\phi \right) +g(r)\phi ^{3}  \tag{4.34}
\end{equation}%
(term $S^{2}/r^{2}$ appears here only in the 2D case).

Although the nonlinearity in Eq. (4.28) is self-repulsive at all values of $r
$, the fact that the repulsion is stronger at large $r$ suggest a
possibility to produce self-trapped states. This possibility is clearly
confirmed by the TF approximation, which neglects derivatives in Eq. (4.34)
and thus yields the following approximate solution of this equation:%
\begin{equation}
\phi _{\mathrm{TF}}^{2}(r)=\left\{
\begin{array}{c}
\left( \mu -S^{2}r^{-2}\right) /\left\vert g(r)\right\vert ,~\mathrm{at}~r>S/%
\sqrt{\mu }, \\
0,~\mathrm{at}~r<S/\sqrt{\mu }%
\end{array}%
\right.   \tag{4.35}
\end{equation}%
(obviously, this approximation makes sense only for $\mu >0$). The
\textquotedblleft inner hole" in the 2D version of this expression at $r<S/%
\sqrt{\mu }$ (with $S\geq 1$) is a known artifact of the TF approximation
(see a review by Fetter (2009)), while the actual solution has the usual
asymptotic form at $r\rightarrow 0$, namely, $\phi (r)\approx \mathrm{const}%
\cdot r^{|S|}$. The applicability condition for the TF approximation is
\begin{equation}
\left( dg/dr\right) ^{2}\ll \mu g^{2},  \tag{4.36}
\end{equation}%
which, in particular, always holds at large $r$ for $g(r)$ given by Eq.
(4.31).

A necessary condition for the self-trapping is the convergence of the total
norm corresponding to the TF approximation,
\begin{equation}
N_{\mathrm{TF}}=2\pi (D-1)\int_{0}^{\infty }\phi _{\mathrm{TF}%
}^{2}(r)r^{D-1}dr.  \tag{4.37}
\end{equation}%
Thus, for $g(r)$ with the power-law asymptotic form (4.31), the integral in
Eq. (4.37) converges for $\alpha >D$, as well as for $g(r)$ with the
anti-Gaussian asymptotic form (4.32).

For $g(r)$ with $\alpha <D$ in Eq. (4.31), it was demonstrated by Zeng and
Malomed (2017) that 2D solutions of Eq. (4.34), as well as their 1D
counterparts (they are produced by Eq. (4.34) with $D=1$, $S=0$, and $r $
replaced by $x$, while $r$ is replaced by $|x|$ in Eq. (4.31)), are
meaningful too. They include \textit{localized continuous waves} (CWs),
globally approximated by the TF expression (4.35) with $S=0$, as well as
\textit{localized dark vortices} in 2D, and \textit{localized dark solitons}
in 1D (including bound states of dark solitons). These are modes of such
types created on top of the weakly localized CWs. They feature nontrivial
stability boundaries in the respective parameter spaces (in particular,
bound states of localized dark solitons are weakly unstable, while localized
dark vortices with $S=2$ may be stable).

\subsection{The Gross-Pitaevskii (GP) equation with Lee-Huang-Yang (LHY)
corrections for quantum droplets (QDs)}

\paragraph{The 3D model}

The recent progress in both theoretical and experimental work with
multidimensional soliton-like (self-trapped) objects has been greatly
advanced by the advent of the concept of QDs, proposed by Petrov (2015). It
offers a practically feasible possibility to suppress the collapse in 3D and
2D binary (two-component) BEC with attractive interactions, thus making it
possible to create stable self-trapped \textquotedblleft droplets", which
are usually not called solitons, but are quite similar states. Actually, the
term \textquotedblleft droplet" is adopted because the condensate density in
them cannot exceed a certain maximum value, making this quantum state of
matter effectively incompressible, similar to fluids.

In terms of the theory, the analysis of Petrov (2015) led to the 3D\ GP
equation with an extra quartic term suppressing the onset of the collapse
driven by the usual cubic self-attraction. The extra term is induced by the
\textit{Lee-Hung-Yang effect} (Lee, Huang, and Yang, 1957), which represents
an effective correction induced by quantum fluctuations around the MF
(mean-filed) states described by the GP equation.

To derive the LHY correction, one starts with the energy density of the
binary BEC in the MF\ approximation:
\begin{equation}
\mathcal{E}_{\mathrm{MF}}=\frac{1}{2}g_{11}n_{1}^{2}+\frac{1}{2}%
g_{22}n_{2}^{2}+g_{12}n_{1}n_{2},  \tag{4.38}
\end{equation}%
where $g_{11,22}>0$, and $g_{12}<0$ are coupling constants which determine,
respectively, the strengths of the self-repulsive interaction between atoms
in the same species (component), and cross-attraction between atoms
belonging to different species, while $n_{1,2}$ are atomic densities of the
two species. The combination of the intra-component repulsion and
cross-component attraction makes the binary condensate obviously miscible.
In the MF approximation, the collapse occurs if the cross-attraction is
effectively stronger than the self-repulsion, which means%
\begin{equation}
-g_{12}>\sqrt{g_{11}g_{22}}.  \tag{4.39}
\end{equation}

In the experiments of Cabrera \textit{et al}. (2018), Cheiney \textit{et al}%
. (2018), and Semeghini \textit{et al}. (2018), the two species were
realized as atomic states of $^{39}$K, \textit{viz}., $\left\vert
F,m_{F}\right\rangle =\left\vert 1,-1\right\rangle $ and $\left\vert
1,0\right\rangle $, where $F$ and $m_{F}$ stand, respectively, for the total
angular momentum of the potassium atom and its $z$-component. In this case,
scattering lengths $a_{s}$ of repulsive collisions are the same for both
species, suggesting to define%
\begin{equation}
g_{11}=g_{22}\equiv g=4\pi \hbar ^{2}m/a_{s},  \tag{4.40}
\end{equation}%
while the inter-species attraction is controlled by means of FR.

The LHY correction to the energy density (4.38) was derived by Petrov (2015)
as a contribution from the zero-point energy of the Bogoliubov excitations
around the MF state with equal densities of the mixed components, $%
n_{1}=n_{2}\equiv n$:
\begin{equation}
\mathcal{E}_{\mathrm{LHY}}=\frac{128}{30\sqrt{\pi }}gn^{2}\sqrt{na_{s}^{3}},
\tag{4.41}
\end{equation}%
where $m$ is the atomic mass. In a dilute condensate, the LHY term (4.41), $%
\sim n^{5/2}$ is, generally, much smaller than the MF ones, $\sim n^{2}$ ,
in Eq. (4.38), hence the LHY correction is negligible, as one might expect.
However, when FR is used to tune $g_{12}$ so that it is close to the
equilibrium point, $g_{12}=-g$ (see Eq. (4.39)), at which the MF
self-repulsion is nearly balanced by a slightly stronger cross-attraction
between the components, the LHY term becomes essential. Thus, one can define
the parameter of the weak residual MF attraction,
\begin{equation}
0<-\delta g\equiv -\left( g_{12}+g\right) \ll g.  \tag{4.42}
\end{equation}%
The final result is the LHY-corrected GP equation in 3D for equal wave
functions $\psi $ of both components, derived by Petrov (2015) in the
following scaled form:
\begin{equation}
i\psi _{t}+\frac{1}{2}\nabla ^{2}\psi +3|\psi |^{2}\psi -\frac{5}{2}|\psi
|^{3}\psi ,  \tag{4.43}
\end{equation}%
cf. the usual GP equation (4.27). Here, the quartic self-repulsive term, $%
\left( 5/2\right) |\psi |^{3}\psi $, represents the LHY correction which
prevents the onset of the collapse.

The derivation of the LHY-amended GP was revised by Hu and Liu (2020), who
have taken into regard the pairing field. Finite-temperature effects, which
can essentially change the structure and stability of QDs were considered in
detail by Wang, Liu, and Hu (2021).

It is relevant to mention that, in addition to theoretically predicted
(Petrov, 2015) and experimentally demonstrated (Cabrera \textit{et al}.,
2018, Cheiney \textit{et al}., 2018, and Semeghini \textit{et al}., 2018)
fundamental QD states, their stable counterparts with embedded vorticity $%
S=1 $ and $2$ were predicted too (Kartashov \textit{et al}., 2018).

\paragraph{Lower dimensions: 2D and 1D}

Under the action of extremely tight confinement applied in direction ($z$),
reduction of the 3D equation (4.43) to a 2D model was carried out by Petrov
and Astrakharchik (2016). In this case, the energy density of the binary BEC
with equal densities $n$ of the two components becomes
\begin{equation}
\mathcal{E}_{\mathrm{2D}}=\frac{8\pi n^{2}}{\ln ^{2}(\left\vert
a_{12}\right\vert /a_{s})}\left[ \ln \left( \frac{n}{\left( n_{0}\right) _{%
\mathrm{2D}}}\right) -1\right] ,  \tag{4.44}
\end{equation}%
where $a_{s}>0$ is the same as in Eq. (4.40), $a_{12}<0$ is the scattering
length corresponding of the the attractive inter-species interaction, and
the reference density is

\begin{equation}
\left( n_{0}\right) _{\mathrm{2D}}=\left( 2\pi \right) ^{-1}\exp \left(
-2\gamma -3/2\right) \left( a_{s}\left\vert a_{12}\right\vert \right)
^{-1}\ln (\left\vert a_{12}\right\vert /a_{s})  \notag
\end{equation}
($\gamma \approx 0.5772$ is the Euler's constant), cf. Eqs. (4.38) and
(4.41) in the 3D situation. The corresponding 2D LHY-amended GPE for equal
wave functions of both components is

\begin{equation}
i\partial _{t}\psi =-\frac{1}{2}\nabla ^{2}\psi +\frac{8\pi }{\text{ln}^{%
{\large 2}}(\left\vert a_{{\large 12}}\right\vert /a_{{\large s}})}\ln
\left( \frac{|\psi |^{{\Large 2}}}{\sqrt{e}\left( n_{{\Large 0}}\right) _{%
\mathrm{2D}}}\right) |\psi |^{2}\psi .  \notag
\end{equation}

\noindent For the theoretical analysis, it is convenient to cast this
equation in the following scaled form:

\begin{equation}
i\partial _{t}\psi =-\frac{1}{2}\nabla ^{2}\psi +\ln \left( |\psi
|^{2}\right) |\psi |^{2}\psi .  \tag{4.45}
\end{equation}%
The increase of density from $|\psi |^{2}<1$ to $|\psi |^{2}>1$ leads to the
change of the sign of the logarithmic factor in Eq. (4.45). As a result, the
cubic term in this equation is self-focusing at $|\psi |^{2}<1$, maintaining
the formation of QDs, and defocusing at $|\psi |^{2}>1$, thus arresting the
transition to the collapse, and securing the stability of 2D QDs.
Furthermore, Eq. (4.45) gives rise to stable QDs with embedded vorticity $S$
-- at least, up to $S=5$ (Li \textit{et al}., 2018).

For the 1D setting with extremely tight confinement in the two transverse
directions, the analysis performed by Petrov and Astrakharchik (2016) yields
the following effective energy density:
\begin{equation}
\mathcal{E}_{\mathrm{1D}}=\delta g\cdot n^{2}-\frac{4\sqrt{2}}{3\pi }%
(gn)^{3/2},  \tag{4.46}
\end{equation}%
cf. Eqs. (4.38), (4.41), and (4.44), where $\delta g$ and $g$ are the same
coefficients as in Eqs. (4.40) and (4.42) The respective LHY-amended GP
equation features a combination of the usual MF cubic nonlinearity and a
quadratic term, representing the LHY correction in the 1D setting:
\begin{equation}
i\partial _{t}\psi =-(1/2)\partial _{xx}\psi +\delta g\cdot |\psi |^{2}\psi
-(\sqrt{2}/\pi )g^{3/2}|\psi |\psi .  \tag{4.47}
\end{equation}%
Note that the LHY-induced quadratic term is \emph{self-focusing} in Eq.
(4.47), on the contrary to the defocusing sign of the quartic term in the 3D
equation (4.43). Because the most interesting results for QDs are obtained
in the case of the competition between the residual MF term and the
LHY-induced correction, in the 1D case the relevant situation is one with $%
\delta g>0$, when the residual MF self-interaction is \emph{repulsive}, in
contrast with the residual self-attraction adopted in the 3D setting, as
mentioned above.

The effectively 2D and 1D description outlined above is valid for extremely
strong transverse confinement, with the characteristic size $l_{\mathrm{%
confinement}}\ll l_{\mathrm{healing}}\sim 30$ nm, where $l_{\mathrm{healing}%
} $ is the healing length in the BEC for experimentally relevant settings,
such as those realized in the experimental works by Cabrera \textit{et al}.
(2018), Cheiney \textit{et al}. (2018), and Semeghini \textit{et al} (2018).
In reality, the values of $l_{\mathrm{confinement}}$ in the experiment is $%
\gtrsim 0.5$ $\mathrm{\mu }$m. For this reason, the dimension crossover $%
\mathrm{3D\rightarrow 2D}$ requires a more careful consideration. In
particular, for a relatively loosely confined (\textquotedblleft thick")
quasi-2D layer of BEC it may be relevant to consider the 2D version of Eq.
(4.43), keeping the quartic LHY term (Shamriz, Chen, and Malomed, 2020a).
Detailed consideration of the dimension reductions, $\mathrm{3D\rightarrow 2D%
}$ and $\mathrm{3D\rightarrow 1D}$, beyond the first approximation presented
by Petrov and Astrakharchik, was elaborated by Zin \textit{et al}. (2018),
Ilg \textit{et al}. (2018), and Lavoine and Bourdel (2021).

\subsection{Spinor (two-component) BEC models}

\subsubsection{Spin-orbit-coupled BEC in two dimensions}

Ultracold atomic gases in the BEC state are often used as testbeds for
emulating, in a simple clean form, various effects known in complex settings
of condensed matter physics (Lewenstein, Sanpera, and Ahufinger, 2012; Hauke
\textit{et al.}, 2012). One of important effects emulated by binary BECs is
the spin-orbit coupling (SOC), originally discovered in semiconductors, as
the weakly-relativistic interaction of the spin of moving electrons with the
electrostatic field of the ionic lattice (Dresselhaus, 1955; Bychkov and
Rashba, 1984). Although the true spin of bosonic atoms, used for this
purpose, is zero, the spinor wave function of electrons may be mapped into
the two-component wave function of the condensate, thus realizing \textit{%
pseudospin} $1/2$ (Lin, Jim\'{e}nez-Garc\'{\i}a, and Spielman, 2011;
Galitski and Spielman, 2013). While most experimental work on this topic
addressed effectively 1D settings, the realization of the SOC in the 2D
binary BEC was reported too, by Wu \textit{et al}. (2016). The 2D and 3D
realizations are obviously necessary for the creation of multidimensional
solitons.

In the MF approximation, the system of effectively two-dimensional GP
equations for the two-component wave function, $\left( \psi _{+},\psi
_{-}\right) $, can be written as follows (Zhang, Mao, and Zhang, 2012;
Sakaguchi \textit{et al}., 2018):%
\begin{equation}
\hspace{-11mm}i\frac{\partial \psi _{+}}{\partial t}=-\frac{1}{2}\nabla
^{2}\psi _{+}+\left( \lambda _{R}\widehat{D}^{[-]}\psi _{-}-i\lambda _{D}%
\widehat{D}^{[+]}\psi _{-}\right) -(|\psi _{+}|^{2}+\gamma |\psi
_{-}|^{2})\psi _{+}-\Omega \psi _{+},  \tag{4.48}
\end{equation}%
\begin{equation}
\hspace{-11mm}i\frac{\partial \psi _{-}}{\partial t}=-\frac{1}{2}\nabla
^{2}\psi _{-}-\left( \lambda \widehat{_{R}D}^{[+]}\psi _{+}+i\lambda _{D}%
\widehat{D}^{[-]}\psi _{+}\right) -(|\psi _{-}|^{2}+\gamma |\psi
_{+}|^{2})\psi _{-}+\Omega \psi _{-},  \tag{4.49}
\end{equation}%
where linear operators%
\begin{equation}
\widehat{D}^{[\pm ]}\equiv \partial /\partial x\pm i\partial /\partial y
\tag{4.50}
\end{equation}%
represent SOC of the \textit{Rashba} and \textit{Dresselhaus} types, with
real strengths $\lambda _{R}$ and $\lambda _{D}$, respectively. Thus, the
SOC terms linearly couple two components of the pseudospinor wave function,
in Eqs. (4.48) and (4.49), by means of the first spatial derivatives.

The last terms in these equations, with real $\Omega $, represent the Zeeman
splitting between the components (if it is present), and the signs in front
of the cubic terms, including the cross-nonlinear ones, with $\gamma >0$,
correspond to attractive interactions between atoms in the condensate. In
this case, the system of Eqs. (4.48) and (4.49) gives rise to 2D solitons
which may be stable states realizing the system's ground state in the 2D
setting with cubic attraction (Sakaguchi, Li, and Malomed, 2014). Prior to
the publication of the latter result, it was commonly believed that NLS\
equations with cubic self-attraction can never create stable solitons in 2D
settings.

A relevant characteristic of the SOC system (4.48)-(4.50) is the linear
dispersion relation for its plane-wave solutions with 2D wave vector $%
\mathbf{k=}\left( k_{x},k_{y}\right) $,%
\begin{equation}
\psi _{\pm }\sim \exp \left( i\mathbf{k}\cdot \mathbf{r}-i\omega t\right) .
\tag{4.51}
\end{equation}%
The substitution of ansatz (4.51) in the linearized version of Eqs. (4.48)
and (4.49) yields two branches of the dispersion relation:%
\begin{equation}
\omega =\frac{1}{2}k^{2}\pm \sqrt{(\lambda _{R}^{2}+\lambda
_{D}^{2})k^{2}+4\lambda _{R}\lambda _{D}k_{x}k_{y}+\Omega ^{2}},  \tag{4.52}
\end{equation}%
which is anisotropic in the plane of $\left( k_{x},k_{y}\right) $. In the
special case of
\begin{equation}
\lambda _{D}=\pm \lambda _{R}\equiv \lambda ,  \tag{4.53}
\end{equation}%
which was actually realized by Lin, Jim\'{e}nez-Garcia and Spielman (2011)
in the first experimental demonstration of the SOC in binary BEC, the
pseudospin-dependent (second) term in right-hand side of Eq. (4.52) becomes
effectively one-dimensional, as it contains a single wave-vector component,
either $\left( k_{x}+k_{y}\right) $ or $\left( k_{x}-k_{y}\right) $:%
\begin{equation}
\omega =\frac{1}{2}k^{2}\pm \sqrt{2\lambda ^{2}(k_{x}\pm k_{y})^{2}+\Omega
^{2}}.  \tag{4.54}
\end{equation}

The 2D model with a quasi-1D form of SOC, which is actually tantamount to
the system of Eqs. (4.48) and (4.49), was recently considered by Kartashov
\textit{et al}. (2020). The respective system is written as%
\begin{equation}
\left[ i\partial _{t}+\frac{1}{2}\nabla ^{2}+i\lambda \sigma _{x}\partial
_{x}+\left(
\begin{array}{cc}
\left\vert \psi _{+}\right\vert ^{2}+\gamma |\psi _{-}|^{2} & 0 \\
0 & \left\vert \psi _{-}\right\vert ^{2}+\gamma |\psi _{+}|^{2}%
\end{array}%
\right) +\Omega \sigma _{z}\right] \Psi ,  \tag{4.55}
\end{equation}%
where the pseudospinor weave functions is define as $\Phi =\left\{ \psi
_{+},\psi _{-}\right\} $, $\sigma _{x}$ and $\sigma _{z}$ being the usual
Pauli matrices.

The branch of dispersion relation (4.52) with the bottom sign in front of
the pseudospin-dependent term splits the axis of $\omega $ into a \textit{%
semi-infinite band}, $\omega >\omega _{\min }$, where the branch takes it
values, and a \textit{semi-infinite spectral gap}, $\omega <\omega _{\min }$%
, where the plane-wave solutions (4.51) of the linearized system of Eqs.
(4.48) and (4.49) do not exist. In particular, in the case of $\lambda
_{D}=0 $ and $\lambda _{R}\equiv \lambda $, or vice versa, the boundary of
the semi-infinite gap is%
\begin{equation}
\omega _{\min }=\left\{
\begin{array}{c}
-(1/2)\left( \lambda ^{2}+\Omega ^{2}/\lambda ^{2}\right) ,~\mathrm{if}%
~|\Omega |<\lambda ^{2}, \\
-|\Omega |,~\mathrm{if}~|\Omega |>\lambda ^{2}.%
\end{array}%
\right.  \tag{4.56}
\end{equation}

The full nonlinear system of Eqs. (4.48) and (4.49) conserves the total
norm,
\begin{equation}
N=\iint (|\psi _{+}|^{2}+|\psi _{-}|^{2})dxdy\equiv N_{+}+N_{-},  \tag{4.57}
\end{equation}%
which is proportional to the number of atoms in the condensate, the
vectorial momentum (which is conserved in spite of the absence of the
Galilean invariance),
\begin{equation}
\mathbf{P}=i\iint (\psi _{+}^{\ast }\nabla \psi _{+}+\psi _{-}^{\ast }\nabla
\psi _{-})dxdy,  \tag{4.58}
\end{equation}%
cf. Eq. (2.18), and the energy,
\begin{equation}
E=E_{\mathrm{kin}}+E_{\mathrm{int}}+E_{\mathrm{pseudospin}}+E_{\mathrm{Zeeman%
}},  \tag{4.59}
\end{equation}%
which includes kinetic, interaction, pseudospin, and Zeeman terms:
\begin{equation}
\hspace{-11mm}E_{\mathrm{kin}}=\frac{1}{2}\iint \left( |\nabla \psi
_{+}|^{2}+|\nabla \psi _{-}|^{2}\right) dxdy,  \tag{4.60}
\end{equation}

\begin{equation}
\hspace{-11mm}E_{\mathrm{int}}=-\frac{1}{2}\iint \left[ \left( |\psi
_{+}|^{4}+|\psi _{-}|^{4}\right) +2\gamma |\psi _{+}|^{2}|\psi _{-}|^{2}%
\right] dxdy,  \tag{4.61}
\end{equation}%
\begin{equation}
E_{\mathrm{pseudospin}}=\iint \left[ \psi _{+}^{\ast }\left( \lambda _{R}%
\widehat{D}^{[-]}-i\lambda _{D}\widehat{D}^{[+]}\right) \psi _{-}-\psi
_{-}^{\ast }\left( \lambda _{R}\widehat{D}^{[+]}+i\lambda _{D}\widehat{D}%
^{[-]}\right) \psi _{+}\right] dxdy,  \tag{4.62}
\end{equation}%
\begin{equation}
\hspace{-11mm}E_{\mathrm{Zeeman}}=-\Omega \iint \left( |\psi _{+}|^{2}-|\psi
_{-}|^{2}\right) dxdy.  \tag{4.63}
\end{equation}

The comparison of Eqs. (4.48) and (4.49) with the underlying system of the
GP equations, written in physical units, shows that the unit length in Eqs.
(4.48) and (4.49) typically corresponds to the spatial scale $\sim 1$ $%
\mathrm{\mu }$m. Further, assuming typical values of the transverse
confinement length (in the third direction) $\simeq 3$ $\mathrm{\mu }$m and
the scattering length of the interatomic attraction $\sim -0.1$ nm, one
concludes that $N=1$ in the scaled notation is tantamount to $\simeq 3\times
10^{3}$ atoms in the condensate. In addition to that, a characteristic
strength $\Omega =1$ of the Zeeman splitting in Eqs. (4.48) and (4.49)
translates, in physical units, to strengths between $2\pi \times 100$ Hz and
$2\pi \times 1$ KHz (Sakaguchi \textit{et al}., 2018).

A system of three (rather than two) coupled GP equations in 2D, with the
interaction between the three components mediated by linear SOC terms, was
addressed by Adhikari (2021). These components form a spinor wave function
with spin $1$. A similar system of GP equations for the three-component
spinor wave function in 3D was studies by Gautam and Adhikari (2018).

\paragraph{The reduced two-dimensional\ system with strong spin-orbit
coupling and a finite spectral bandgap}

\begin{figure}[tbp]
\begin{center}
\includegraphics[width=0.50\textwidth]{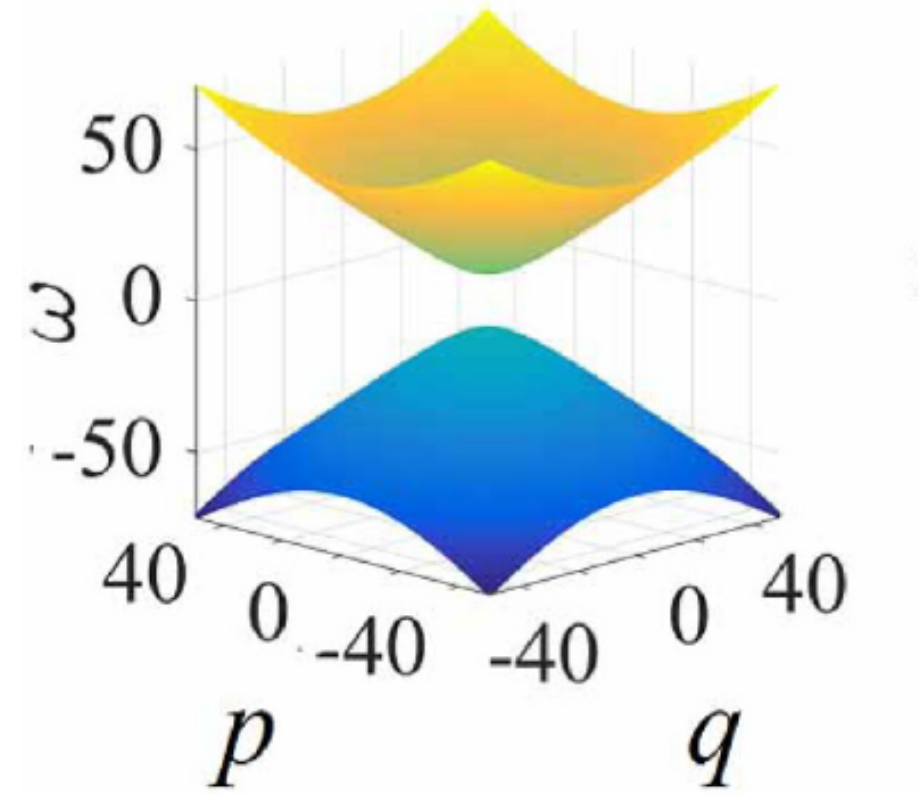}
\end{center}
\caption{The spectrum (4.66), produced by the linearization of the
simplified SOC equations (4.63) and (4.64), plotted for $\Omega =10$, The
spectrum clearly displays the presence of the finite bandgap (4.67) (source:
Li \textit{et al}. (2017)).}
\label{fig2.2}
\end{figure}

The system of Eqs. (4.48) and (4.49) can be essentially simplified, still
keeping the ability to produce stable 2D solitons, in the limit case when
the presence of strong SOC makes it possible to neglect the kinetic energy
in comparison to it, see Eqs. (4.59) and (4.61). In other words, in this
case one may drop terms $\sim \nabla ^{2}$ in Eqs. (4.48) and (4.49) (Li
\textit{et al}. 2017; Sakaguchi and Malomed, 2018). Following the latter
works, the respective 2D system is considered with $\lambda _{D}=0$ and,
eventually, setting $\lambda _{R}=1$ by means of rescaling:
\begin{equation}
i\frac{\partial \psi _{+}}{\partial t}=\left( \frac{\partial \psi _{-}}{%
\partial x}-i\frac{\partial \psi _{-}}{\partial y}\right) -(|\psi
_{+}|^{2}+\gamma |\psi _{-}|^{2})\psi _{+}-\Omega \psi _{+},  \tag{4.64}
\end{equation}%
\begin{equation}
i\frac{\partial \psi _{-}}{\partial t}=-\left( \frac{\partial \psi _{+}}{%
\partial x}+i\frac{\partial \psi _{+}}{\partial y}\right) -(\gamma |\psi
_{+}|^{2}+|\psi _{-}|^{2})\psi _{-}+\Omega \psi _{-}.  \tag{4.65}
\end{equation}%
Then, looking for plane-wave solutions to the linearized version of Eqs.
(4.64) and (4.65) as%
\begin{equation}
\psi _{\pm }\sim \exp \left( ipx+iqy-i\omega t\right)  \tag{4.66}
\end{equation}%
(cf. Eq. (4.51)) one obtains two branches of the dispersion relation,%
\begin{equation}
\omega =\pm \sqrt{\Omega ^{2}+p^{2}+q^{2}},  \tag{4.67}
\end{equation}%
as shown in Fig. \ref{fig2.2}. Unlike its counterpart (4.52) produced by the
full system of Eqs. (4.48) and (4.49), this spectrum features not a
semi-infinite bandgap, but, instead, an obvious \textit{finite bandgap},%
\begin{equation}
|\omega |<|\Omega |,  \tag{4.68}
\end{equation}%
in which stable 2D\ gap solitons can be found as solutions to the full
system of nonlinear equations\ (4.52) and (4.53) (Sakaguchi and Malomed,
2018).

Actually, branches of the full spectrum (4.52) cover (eliminate) the finite
bandgap (4.67), but this happens at very large values of $k$. This fact
implies that, in the framework of the full system, the gap solitons will
start decay through emission of radiation, but the emission rate will be
exponentially small.

\subsection{Nonlinear optical couplers emulating the spin-orbit coupling
(SOC)}

\subsubsection{ A spatiotemporal coupler emulating SOC in two dimensions}

In the spirit of possibilities to emulate matter-wave phenomenology by means
of optics and vice versa, a realization of the stabilization mechanism for
2D optical solitons, operating similar to that in Eqs. (4.48), (4.49) and
(4.55), was proposed by Kartashov \textit{et al}. (2015). The model is based
on the consideration of the spatiotemporal propagation of light in a
dual-core planar waveguide (coupler) with the Kerr nonlinearity acting in
each core. In this case, amplitudes of the electromagnetic waves in the two
cores of the coupler, $q_{1}$ and $q_{2}$, form a pseudospinor wave function
which obeys the following system of equations.%
\begin{equation}
i\frac{\partial q_{1}}{\partial \xi }=-\frac{1}{2}\left( \frac{\partial ^{2}%
}{\partial \eta ^{2}}+\frac{\partial ^{2}}{\partial \tau ^{2}}\right)
q_{1}-\left\vert q_{1}\right\vert ^{2}q_{1}-\left( C+i\delta \frac{\partial
}{\partial \tau }\right) q_{2}-\beta q_{1},  \tag{4.69}
\end{equation}%
\begin{equation}
i\frac{\partial q_{2}}{\partial \xi }=-\frac{1}{2}\left( \frac{\partial ^{2}%
}{\partial \eta ^{2}}+\frac{\partial ^{2}}{\partial \tau ^{2}}\right)
q_{2}-\left\vert q_{2}\right\vert ^{2}q_{2}-\left( C+i\delta \frac{\partial
}{\partial \tau }\right) q_{1}+\beta q_{2},  \tag{4.70}
\end{equation}%
Here $\xi $ is the scaled propagation distance, $\eta $ are $\tau $ are,
respectively, the transverse and temporal coordinates, the
group-velocity-dispersion (GVD) coefficient (assuming the anomalous sign of
the GVD), effective diffraction coefficient, and Kerr coefficients are
scaled to be $1$, $C$ is the real coupling coefficient, $\delta $ accounts
for the temporal dispersion of the coupling, and $\beta $ determines the
phase-velocity mismatch between the cores. This system is similar to Eq.
(4.55), with the SOC terms represented by the first derivatives acting on
the single coordinate, $\tau $, which is sufficient for the stabilization of
spatiotemporal solitons in this system (solitons which feature self-trapping
in both temporal and spatial directions in optical media are often called
\textquotedblleft light bullets", as was first proposed by Silberberg
(1990)). As concerns the dispersion relation for the linearized system of
Eqs. (4.69) and (4.70), the respective plane-wave solutions are sought for
as
\begin{equation}
q_{1,2}\sim \exp \left( ik\eta -i\omega \tau +ib\xi \right) ,  \tag{4.71}
\end{equation}%
cf. Eqs. (4.51) and (4.66). The result is the relation between the real
propagation constant $b$, transverse wavenumber $k$ and frequency $\omega $:%
\begin{equation}
b=-\frac{1}{2}\left( k^{2}+\omega ^{2}\right) \pm \sqrt{(C+\delta \cdot
\omega )^{2}+\beta ^{2}},  \tag{4.72}
\end{equation}%
which is similar to Eq. (4.54).

\subsubsection{The parity-time ($\mathcal{PT}$) symmetric SOC-emulating
optical-coupler model}

Another possibility for the emulation of 2D SOC by means of spatiotemporal
propagation of optical waves in planar dual-core couplers with the Kerr
nonlinearity was elaborated by Sakaguchi and Malomed (2016). That setting
makes it possible to combine SOC, nonlinearity, and the effect known as
parity-time ($\mathcal{PT}$) symmetry.

Settings featuring the $\mathcal{PT}$ symmetry may be considered as ones
designed at the border between conservative and dissipative systems. This
concept had appeared in quantum mechanics as a possibility to realize
non-Hermitian Hamiltonians which, nevertheless, produce purely real spectra
of energy eigenvalues (see a review of the topic by Bender (2007), and a
book by Moiseyev (2011)). The basic idea is to construct a Hamiltonian which
includes a complex potential $U(\mathbf{r})$, whose real and imaginary parts
are, respectively, even and odd functions of coordinates, $\mathbf{r=}%
\{x,y,z\}$:%
\begin{equation}
U(\mathbf{r})=U_{\mathrm{even}}(\mathbf{r})+iU_{\mathrm{odd}}(\mathbf{r}),
\tag{4.73}
\end{equation}%
hence the potential satisfies the symmetry condition
\begin{equation}
U(-\mathbf{r})=U^{\ast }(\mathbf{r})  \tag{4.74}
\end{equation}%
(recall $\ast $ stands for the complex conjugate). A well-known examples is
the 1D potential,%
\begin{equation}
U(x)=U_{0}x^{2}\left( ix\right) ^{\varepsilon },  \tag{4.75}
\end{equation}%
which, with $U_{0}>0$ and real $\varepsilon $, obviously satisfies condition
(4.74). It is known that each eigenvalue of the HO potential, to which
potential (4.75) reduces at $\varepsilon =0$, continues, at all values $%
\varepsilon >0$, as a real positive eigenvalue, monotonously increasing with
the growth of $\varepsilon $. On the other hand, at $\varepsilon <0$ the $%
\mathcal{PT}$ symmetry suffers breaking through a chain of bifurcations at
which pairs of adjacent eigenvalues collide and become complex (unphysical)
with the increase of $|\varepsilon |$. The last surviving eigenvalue is one
originating at $\varepsilon =0$ from the HO's ground state. The complex
potential (4.75) gives rise to no real eigenvalues at $\varepsilon <-1$.

While experimental realization of the $\mathcal{PT}$ symmetry in quantum
mechanics is a challenge, it can be readily emulated in optics, as reviewed
by Makris \textit{et al}. (2011). Indeed, the paraxial light propagation in
the spatial domain obeys the equation of the Schr\"{o}dinger type for local
amplitude $u(z,x)$ of the electromagnetic field. In the planar waveguide,
its scaled form is%
\begin{equation}
i\frac{\partial u}{\partial z}+\frac{1}{2}\frac{\partial ^{2}u}{\partial
x^{2}}-U_{\mathrm{even}}(x)u=iU_{\mathrm{odd}}(x)u,  \tag{4.76}
\end{equation}%
where an even real function $-U_{\mathrm{even}}(x)$ represents a spatially
even modulation of the local refractive index, while an odd real function $%
U_{\mathrm{odd}}(x)$ represents a globally balanced distribution of local
gain ($U_{\mathrm{odd}}(x)>0$) and loss ($U_{\mathrm{odd}}(x)<0$). This
setup was used by R\"{u}ter \textit{et al}. (2010) in the experimental
realization of the $\mathcal{PT}$ symmetry in optics. Although the model
represented by Eq. (4.76) is a dissipative one, it shares basic properties,
such as the possibility of generating a purely real spectrum of eigenvalues,
with conservative systems.

While the concept of the $\mathcal{PT}$ symmetry is a linear one, its
realization in optics suggests to combine it with the Kerr nonlinearity of
optical media. This possibility opens the way to creation of a large variety
of $\mathcal{PT}$-symmetric solitons (see reviews by Konotop, Yang, and
Zezyulin, 2016, and Suchkov \textit{et al}. 2016).

The blend of SOC and $\mathcal{PT}$ symmetry, proposed Sakaguchi and Malomed
(2016), is based on the use of a dual-core waveguide (coupler). This setup
is natural, as couplers provide optical platforms for the emulation of both
the SOC (see Eqs. (4.69), (4.70)) and $\mathcal{PT}$ symmetry, provided that
the two cores of the coupler carry mutually balanced gain and loss (Driben
and Malomed, 2011; Alexeeva \textit{et al}., 2012). The $\mathcal{PT}$%
-symmetric coupler may be designed in the 2D form as well (Burlak and
Malomed, 2013).

\begin{figure}[tbp]
\begin{center}
\includegraphics[width=0.50\textwidth]{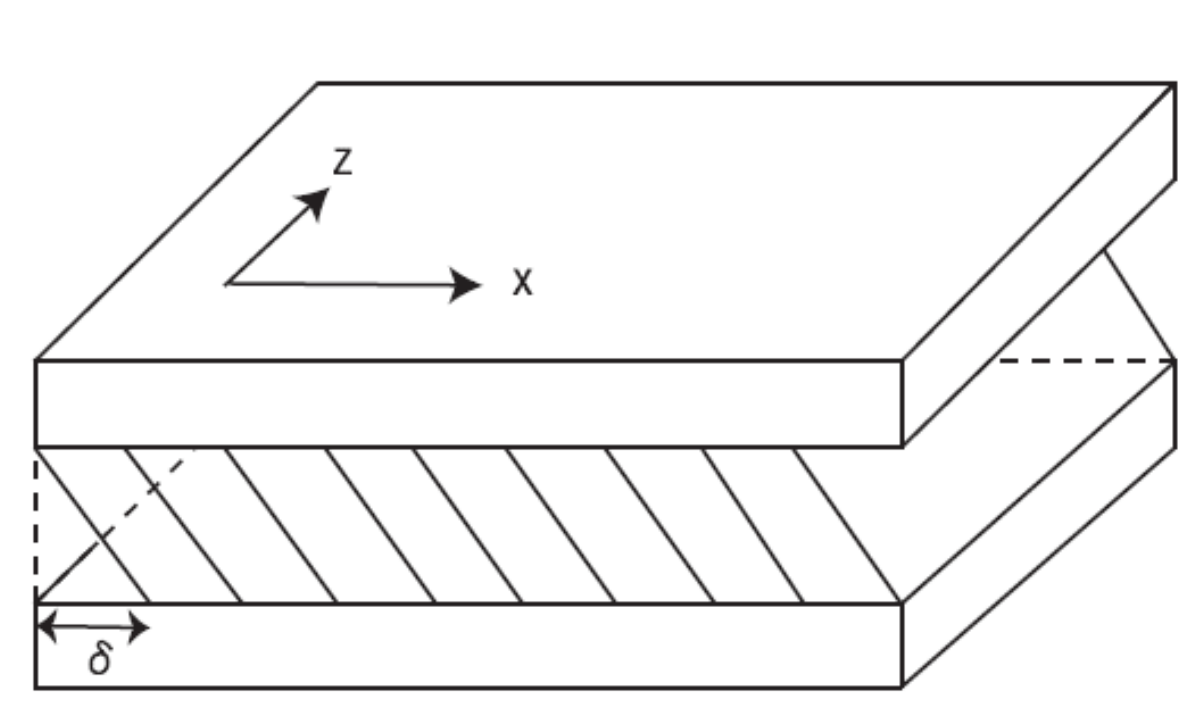}
\end{center}
\caption{A scheme of the planar coupler with an oblique layer between the
guiding cores, which realizes the model based on Eqs. (4.77) and (4.78).
Amplitudes of the electromagnetic waves in the top and bottom cores are $%
u_{2}$ and $u_{1}$, respectively. Slanted lines designate the skew of the
intermediate layer, measured by transverse shift $\protect\delta $ (source:
Sakaguchi and Malomed, 2016).}
\label{fig2.3}
\end{figure}

The \textquotedblleft blended" model is based on the following system of NLS
equations for amplitudes $u_{1}$ and $u_{2}$ of the electromagnetic waves in
the coupled cores:
\begin{equation}
i\left( u_{1}\right) _{z}+\frac{1}{2}\left[ \left( u_{1}\right) _{tt}+\left(
u_{1}\right) _{xx}\right] -\delta \cdot \left( u_{2}\right)
_{x}+u_{2}+\left\vert u_{1}\right\vert ^{2}u_{1}=i\Gamma u_{1},  \tag{4.77}
\end{equation}%
\begin{equation}
i\left( u_{2}\right) _{z}+\frac{1}{2}\left[ \left( u_{2}\right) _{tt}+\left(
u_{2}\right) _{xx}\right] +\delta \cdot \left( u_{1}\right)
_{x}+u_{1}+\left\vert u_{2}\right\vert ^{2}u_{2}=-i\Gamma u_{2}.  \tag{4.78}
\end{equation}%
Here, $z$ is the propagation distance, $x$ and $t$ are the transverse and
temporal coordinates, assuming the anomalous GVD, real $\Gamma >0$
represents real gain and loss coefficients in the two cores, while the
coefficients of the Kerr nonlinearity, paraxial diffraction, and inter-core
coupling are scaled to be $1$. Unlike Eqs. (4.69) and (4.70), this system
does not include dispersion of the coupling, while the pseudo-SOC terms $\mp
\delta \cdot \left( U_{2,1}\right) _{x}$ account for \textquotedblleft
skewness" of the coupling in the transverse direction, assuming that the
layer separating the guiding cores has an oblique structure, as
schematically shown in Fig. \ref{fig2.3}: roughly speaking, light couples
point $x$ in the top core to point $x+\delta $ in the bottom one.

The dispersion relation for the plane-wave solutions of the linearized
version of Eqs. (4.77) and (4.78), looked for as%
\begin{equation}
u_{1,2}\sim \exp \left( ikx-i\omega t+ibz\right)  \tag{4.79}
\end{equation}%
(cf. Eq. (4.71)), is%
\begin{equation}
\left( b+\frac{1}{2}\omega ^{2}+\frac{1}{2}k^{2}\right) ^{2}=1-\Gamma
^{2}+\left( \delta \cdot k\right) ^{2},  \tag{4.80}
\end{equation}%
cf. Eq. (4.72). This dispersion relation demonstrates that the $\mathcal{PT}$
symmetry holds at $\Gamma \leq 1$, i.e., under the condition that the
gain-loss coefficient is smaller than the inter-core coupling constant,
which is scaled to be $1$. Further consideration of Eq. (4.80) demonstrates
that it gives rise to solutions for propagation constant $b$ taking values
in the semi-infinite band:
\begin{equation}
b<b_{\max }\equiv \left\{
\begin{array}{c}
(1/2)\left[ \delta ^{2}+\left( 1-\Gamma ^{2}\right) \delta ^{-2}\right] ,~%
\mathrm{at}~~\delta ^{2}>\sqrt{1-\Gamma ^{2}}, \\
\sqrt{1-\Gamma ^{2}},~\mathrm{at}~~\delta ^{2}<\sqrt{1-\Gamma ^{2}}.%
\end{array}%
\right.  \tag{4.81}
\end{equation}%
Accordingly, solitons may exist in the remaining \textit{semi-infinite
bandgap}, i.e., at $b>b_{\max }$ (Sakaguchi and Malomed, 2016).

\subsubsection{Stabilization of 1D solitons in the SOC-emulating skewed
coupler with the quintic self-focusing}

A similar, although effectively one-dimensional, scheme, based on the
dual-core system with the skewed coupling between the cores and the \emph{%
quintic} self-focusing carried by the cores, was elaborated by Shamriz,
Chen, and Malomed (2020b). As mentioned above, the quintic self-focusing
plays the special role in 1D equations of the NLS type, leading to the
critical collapse in that case. Accordingly, the one-dimensional pseudo-SOC
makes it possible to suppress the critical collapse driven by the quintic
self-attraction, and thus stabilize solitons in the system.

The corresponding system of coupled one-dimensional NLS equations is

\begin{equation}
iu_{z}+\frac{1}{2}u_{xx}+|u|^{4}u+v-\delta \cdot v_{x}=0,  \tag{4.82}
\end{equation}%
\begin{equation}
iv_{z}+\frac{1}{2}v_{xx}+|v|^{4}v+u+\delta \cdot u_{x}=0,  \tag{4.83}
\end{equation}%
where $\delta $ has the same meaning as in Eqs. (4.77) and (4.78). Looking
for solutions to the linearized system of Eqs. (4.83) and (4.84) as
\begin{equation}
\left\{ u,v\right\} \sim \exp \left( ikz+iqx\right) ,  \tag{4.84}
\end{equation}
\ one derives the dispersion relation between the propagation constant ($k$)
and transverse wavenumber, $q$:%
\begin{equation}
k=-(1/2)q^{2}\pm \sqrt{1+\delta ^{2}q^{2}}.  \tag{4.85}
\end{equation}

The spectrum determined by Eq. (4.85) includes bandgaps, i.e., intervals of $%
k$ in which the linear plane waves do not exist, hence they may be populated
by solitons, in the framework of the full nonlinear system. In the case of $%
\delta ^{2}>1$ (when the pseudo-SOC is strong enough), it is the
semi-infinite \textit{main bandgap},
\begin{equation}
k>k_{\max }\equiv \frac{\delta ^{4}+1}{2\delta ^{2}}.  \tag{4.86}
\end{equation}%
Formal values of $q$ in the main bandgap are complex, hence solitons
populating it have oscillatory tails.

In the case of $\delta ^{2}<1$, there is an additional finite \textit{annex
bandgap}, adjacent to the main one:%
\begin{equation}
1<k<\frac{\delta ^{4}+1}{2\delta ^{2}}.  \tag{4.87}
\end{equation}%
In the annex bandgap, formal values of $q$ are purely imaginary, hence the
respective solitons have tails monotonously decaying at $|x|\rightarrow
\infty $. Further details can be found in the paper by Shamriz, Chen, and
Malomed (2020b)

\subsection{Three-dimensional SOC systems}

The most ambitious objective of the work with systems of GP equations
incorporating SOC terms is to use such terms to predict a possibility of the
stabilization of 3D solitons. This problem was addressed by Zhang \textit{et
al}. (2015) in the work which made use of the scaled GP system, including
SOC of the \textit{Weyl type} and the attractive cubic nonlinearity, written
for the two-component pseudospinor wave function $\Psi =\left\{ \psi
_{+},\psi _{-}\right\} $:
\begin{equation}
\left[ i\partial _{t}+\frac{1}{2}\nabla ^{2}+i\lambda \nabla {\small \cdot }{%
\boldsymbol{\sigma }}+\left(
\begin{array}{cc}
|\psi _{+}|^{2}+\eta |\psi _{-}|^{2} & 0 \\
0 & |\psi _{-}|^{2}+\eta |\psi _{+}|^{2}%
\end{array}%
\right) \right] \Psi =0.  \tag{4.88}
\end{equation}%
Here $\lambda $ is the SOC coefficient, $\mathbf{\sigma }=\left\{ \sigma
_{x},\sigma _{y},\sigma _{z}\right\} $ is the vector of the Pauli matrices,
and the coefficient of the cubic self-attraction is scaled to be $1$, while $%
\eta $ is the relative strength of the cross-attraction between the
components of the wave function. The energy corresponding Eq. (4.88) is
written as%
\begin{equation}
E=E_{\mathrm{kin}}+E_{\mathrm{int}}+E_{\mathrm{pseudospin}}~,\,  \tag{4.89}
\end{equation}%
\begin{equation}
E_{\mathrm{kin}}=\frac{1}{2}\int \int \int \left\vert \nabla \Psi
\right\vert ^{2}dxdydz,~  \tag{4.90}
\end{equation}%
\begin{equation}
E_{\mathrm{int}}=-\frac{1}{2}\int \int \int \left( |\psi _{+}|^{4}+|\psi
_{-}|^{4}+2\eta |\psi _{+}\psi _{-}|^{2}\right) dxdydz\,,  \tag{4.91}
\end{equation}%
\begin{equation}
E_{\mathrm{pseudospin}}=-i\lambda \int \int \int \Psi ^{\dag }\left( \nabla
{\small \cdot }{\boldsymbol{\sigma }}\right) \Psi dxdydz,  \tag{4.92}
\end{equation}%
cf. Eqs. (4.59) -- (4.63).

While the system represented by Eqs. (4.88) - (4.92) is a fully
three-dimensional one, it was demonstrated by Sherman \textit{et al}. (2020)
that 3D solitons may be stabilized by SOC terms which, by themselves, are
two-dimensional, including only derivatives $\partial _{x}$ and $\partial
_{y}$. The respective system of 3D GP equations for the two-component wave
function with the quasi-2D SOC terms is%
\begin{equation}
\left[ i\partial _{t}+\frac{1}{2}\nabla ^{2}+i\sum_{j=x,y}\lambda _{j}\sigma
_{j}\partial _{j}+\left(
\begin{array}{cc}
\left\vert \psi _{+}\right\vert ^{2}+\gamma |\psi _{-}|^{2} & 0 \\
0 & \left\vert \psi _{-}\right\vert ^{2}+\gamma |\psi _{+}|^{2}%
\end{array}%
\right) +\Omega \sigma _{z}\right] \Psi ,  \tag{4.93}
\end{equation}%
where real coefficients $\lambda _{x}$ and $\lambda _{y}$ represent the SOC
strengths. The 3D GP system (4.93) is a straightforward generalization of
its 2D counterpart written above in the form of Eq. (4.55), which provides
for the stabilization of 2D solitons by means of the quasi-1D SOC terms.

In the 2D setting, the addition of the SOC terms to the system of GP
equations with the cubic attraction makes it possible to create absolutely
stable solitons, which play the role of the otherwise missing ground state
of the system (Sakaguchi, Li, and Malomed, 2014). The difference of the 3D
systems with SOC terms, such as (4.88) and (4.93), is that they can maintain
metastable solitons, while the true ground state remains missing in the
system, as the cubic attraction is too strong (supercritical, as mentioned
above) in the 3D case.

\section{Stabilization of 3D and 2D fundamental solitons and vortices by
linear trapping potentials}


\subsection{The formulation of the problem}

In the appropriately normalized form, the three-dimensional GP equation for
the mean-field wave function $\psi $ of the self-attracting BEC\ trapped in
the axisymmetric (but not necessarily fully isotropic) HO potential, with
the scaled planar and axial strengths $1$ and $\Omega ^{2}$, is obtained
from Eq. (4.9):%
\begin{equation}
i\psi _{t}+\frac{1}{2}\nabla ^{2}\psi +|\psi |^{2}\psi -\frac{1}{2}\left[
\left( x^{2}+y^{2}\right) +\Omega ^{2}z^{2}\right] \psi =0.  \tag{5.1}
\end{equation}%
Coefficient $\Omega $ determines the \textit{aspect ratio} between the
trapping lengths imposed by the HO potential in the $\left( x,y\right) $
plane and long the $z$ axis:
\begin{equation}
\mathrm{a.r.}\equiv \sqrt{\Omega }  \tag{5.2}
\end{equation}%
The 2D version of Eq. (5.1) corresponds to dropping coordinate $z$ in Eq.
(5.1). Actually, the 2D limit corresponds to very strong confinement of the
BEC along $z$, with $\mathrm{a.r.}\rightarrow \infty $, imposed by $\Omega
^{2}\rightarrow \infty $. The opposite limit, with $\mathrm{a.r.}\rightarrow
0$, corresponds to $\Omega ^{2}\rightarrow 0$. Up to rescaling, it
represents \textquotedblleft tubular" quasi-one-dimensional solitons with
embedded vorticity, maintained by a cigar-shaped (strongly prolate) trap,
which are stable states (Salasnich, Malomed and Toigo, 2007).

It is relevant to mention that, while the trapping magnetic, optical, or
combined potentials indeed had the HO (parabolic) form in a majority of
experimental works with ultracold gases, the creation of a 3D confining
optical potential $U(r)$ in the form approximated by the infinitely deep
spherical box, with
\begin{equation}
U_{\mathrm{box}}\left( r\equiv \sqrt{x^{2}+y^{2}+z^{2}}\right) =\left\{
\begin{array}{c}
0,r<r_{0}, \\
+\infty ,r>r_{0},%
\end{array}%
\right.  \tag{5.3}
\end{equation}%
was reported by Gaunt \textit{et al}. (2013) and Navon \textit{et al}.
(2013). A 2D version of the box potential (5.3) was created and used by
Hueck \textit{et al}. (2018).

Equation (5.1) conserves the energy,
\begin{equation}
E=\frac{1}{2}\int \int \int \left[ \left( \left\vert \psi _{x}\right\vert
^{2}+\left\vert \psi _{y}\right\vert ^{2}+\left\vert \psi _{z}\right\vert
^{2}\right) +(x^{2}+y^{2}+\Omega ^{2}z^{2})|\psi |^{2}-|\psi |^{4}\right]
dxdydz,  \tag{5.4}
\end{equation}%
along with the norm and $z$-component of the angular momentum, which are
defined, respectively, as per Eqs. (4.13) and (4.14). In cylindrical
coordinates $\left( r,\theta ,z\right) $, defined as per Eq. (4.10),
stationary states of Eq. (5.1), with chemical potential $\mu $ and
integer vorticity $S$, are looked for as%
\begin{equation}
\psi =R(r,z)\exp \left( iS\theta -i\mu t\right) ,  \tag{5.5}
\end{equation}%
(cf. Eq. (4.11)), with real function $R(r,z)$ satisfying the stationary
equation,
\begin{equation}
\frac{\partial ^{2}R}{\partial r^{2}}+\frac{1}{r}\frac{\partial R}{\partial r%
}+\frac{\partial ^{2}R}{\partial z^{2}}+\left( 2\mu -\frac{S^{2}}{r^{2}}%
-r^{2}-\Omega ^{2}z^{2}\right) R-2R^{3}=0,  \tag{5.6}
\end{equation}%
cf. Eq. (4.12). The 2D version of Eq. (5.6) is obtained by dropping terms $%
\partial ^{2}R/\partial z^{2}$ and $-\Omega ^{2}z^{2}R$.

Stability of the numerically found stationary solutions is inspected through
computation of eigenvalues $\lambda $ of small perturbations, introduced by
looking for solutions to Eq. (5.1) in the form of
\begin{equation}
\psi (x,y,z,t)=[R(r,z)+u(r,z)\exp (\lambda t+iL\theta )+v^{\ast }(r,z)\exp
(\lambda ^{\ast }t-iL\theta )]\exp \left( iS\theta -i\mu t\right) ,
\tag{5.7}
\end{equation}%
where $\left( u,v\right) $ are eigenmodes of infinitesimal perturbations
with integer values of azimuthal index $L$, and $\ast $ stands, as above,
for the complex conjugate. The substitution of ansatz (5.7) in Eq. (5.1) and
linearization with respect to the perturbations leads to the system of BdG
equations,
\begin{equation}
\left( i\lambda +\mu \right) u+\frac{1}{2}\left[ \frac{\partial ^{2}}{%
\partial r^{2}}+\frac{1}{r}\frac{\partial }{\partial r}+\frac{\partial ^{2}}{%
\partial z^{2}}-\frac{(S+L)^{2}}{r^{2}}u-r^{2}-\Omega ^{2}z^{2}\right]
u+R^{2}(v+2u)=0,  \tag{5.8a}
\end{equation}%
\begin{equation}
\left( -i\lambda +\mu \right) v+\frac{1}{2}\left[ \frac{\partial ^{2}}{%
\partial r^{2}}+\frac{1}{r}\frac{\partial }{\partial r}+\frac{\partial ^{2}}{%
\partial z^{2}}-\frac{(S-L)^{2}}{r^{2}}v-r^{2}-\Omega ^{2}z^{2}\right]
v+R^{2}(u+2v)=0,  \tag{5.8b}
\end{equation}%
which have to be solved with b.c. demanding that $u(r)$ and $v(r)$ decay
exponentially at $r\rightarrow \infty $ and $\left\vert z\right\vert
\rightarrow \infty $, and decay as $r^{\left\vert S\pm L\right\vert }$ at $%
r\rightarrow 0$.

In physical units, assuming a weak isotropic trap with frequency $2\pi
\times 10$ Hz, into which a gas of $^{7}$Li atoms is loaded (for this atomic
species, the application of FR makes it possible to induce the attractive
interactions, with scattering length $a_{s}\simeq -0.1$ nm (Bradley \textit{%
et al}. 1995)), a typical radius of the vortex is $\sim 10$ $\mathrm{\mu }$%
m, a characteristic time scale is $\sim 0.1$ s, and a typical number of
atoms in a stable vortex, $\sim 10^{5}$, corresponds to values $\sim 10$ of
the dimensionless norm.

\subsection{Stability of 2D fundamental solitons and vortices}

\begin{figure}[tbp]
\begin{center}
\includegraphics[width=0.38\textwidth]{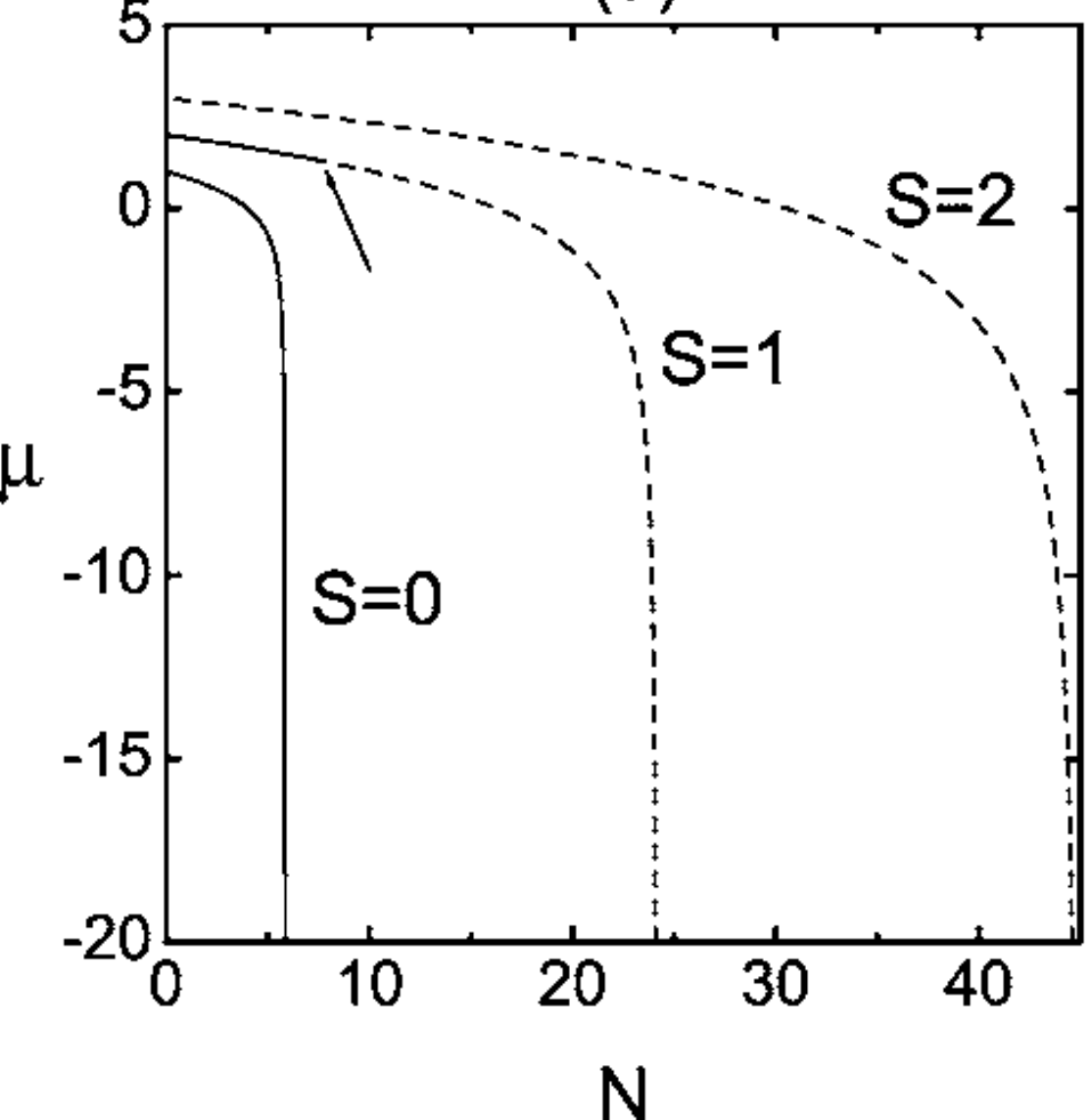} %
\includegraphics[width=0.38\textwidth]{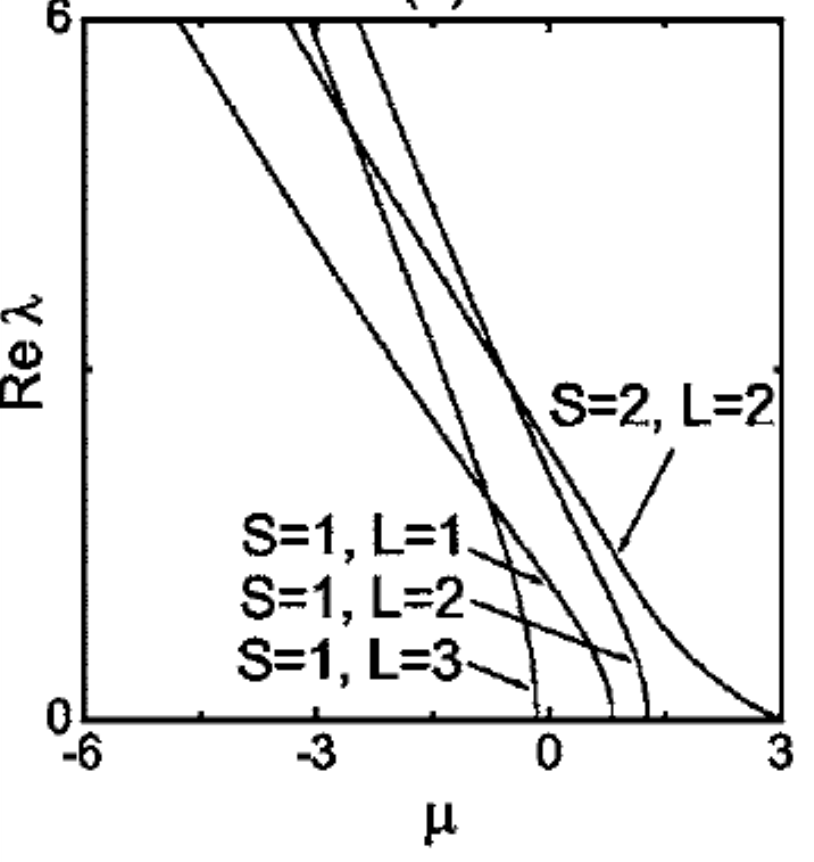}
\end{center}
\caption{The left panel: dependence $\protect\mu (N)$ of the chemical
potential of 2D solitons trapped in the HO potential (see Eqs. (5.1) and
(5.6)), with embedded vorticity $S=0,1,2$, on their norm, defined as per Eq.
(5.9) ($S=0$ pertains to the fundamental solitons). Continuous and dashed
lines represent stable and unstable (sub)families. The solitons with $S=0$
and $S=2$ are, severally, completely stable and unstable. For the family
with $S=1$, the tilted arrow indicates the boundary between stable ($N<7.79$%
) and unstable ($7.79<N<N_{\mathrm{TS}}^{(S=1)}\approx 24.1$) subfamilies.
The right panel displays dependences of the instability growth rates, $%
\mathrm{Re}\protect\lambda $, on $\protect\mu $ for the same families, as
obtained from numerical solution of the 2D version of Eqs. (5.8). The curves
are labeled by integer azimuthal index $L$ of the perturbation eigenmode,
defined as per Eq. (5.7) (source: Mihalache\textit{\ et al}., 2006). }
\label{fig5.1}
\end{figure}

Systematic results for the existence and stability of 2D solitons
(fundamental and vortical ones), produced by numerical solutions of the 2D
versions of Eqs. (5.1), (5.6), and (5.8), were reported by Mihalache \textit{%
et al}. (2006) and Carr and Clark (2006), following an earlier work by
Alexander and Berg\'{e} (2002). Families of solutions, with $S=0$ (the
fundamental solitons) and $S=1$ and $2$ (the solitons with embedded
vorticity), in the form of dependences of their chemical potentials on the
2D norm,%
\begin{equation}
N_{\mathrm{2D}}=\int \int \left\vert \psi \left( x,y\right) \right\vert
^{2}dxdy\equiv 2\pi \int_{0}^{\infty }R^{2}(r)rdr  \tag{5.9}
\end{equation}%
(see Eq. (5.5)), are displayed in the left panel of Fig. \ref{fig5.1}, which
also indicates their stability. In the limit of $\mu \rightarrow -\infty $,
the norm of the fundamental solitons is approaching the TS value, $N_{%
\mathrm{TS}}\approx 5.85$ (see Eq. (3.12)). For $S=1$ and $S=2$, the
respective limit values correspond to the TSs with embedded vorticity, which
were introduced (actually, as unstable states) by Kruglov and Vlasov (1985),
and Kruglov \textit{et al}. (1988). In particular, for the TS with $S=1$,
this value is%
\begin{equation}
N_{\mathrm{TS}}^{(S=1)}\approx 24.1.  \tag{5.10}
\end{equation}%
It was demonstrated by Qin, Dong, and Malomed (2016) that, for all $S\geq 1$%
, the TS norm is accurately predicted by formula%
\begin{equation}
N_{\mathrm{TS}}^{(S)}\approx 4\sqrt{3}\pi S  \tag{5.11}
\end{equation}%
(in particular, for $S=1$ Eq. (5.11) yields $N_{\mathrm{TS}}^{(S=1)}\approx
\allowbreak 21.8$, with a relative difference $\approx 9.5\%$ from numerical
value (5.10)). The fact that the norm of all solitons approaches the TS
value in the limit of $\mu \rightarrow -\infty $ is explained by the
circumstance that the width of these solitons shrinks $\sim 1/\sqrt{-\mu }$,
and for very narrow solitons the effect of the trapping potential becomes
negligible, i.e., they may be indeed considered as TSs in the free space.

Further, Fig. \ref{fig5.1} demonstrates that the family of the fundamental
solitons is \emph{entirely stable}, in agreement with the prediction of the
VK criterion, $d\mu /dN<0$, which obviously holds for all branches shown in
the left panel of Fig. \ref{fig5.1}. Thus, even an arbitrarily weak trapping
potential makes the family of the fundamental TSs, which is fully unstable
in the free space, a \emph{completely stable} one. This finding is explained
by the fact that, a seen in Fig. \ref{fig5.1}, the trap lifts the
above-mentioned norm degeneracy of the TSs in the free space. As a results,
all the solitons drop their norms to values \emph{below the collapse
threshold}, $N<N_{\mathrm{TS}}$, making the onset of the collapse impossible.

For the family of the trapped solitons with $S=1$, the VK criterion is not
sufficient for securing their stability. The point is that this criterion
indeed protects the vortex states against collapsing, but not against
spontaneous splitting into fragments. The instability spectrum, i.e.,
dependences of the instability growth rate on the chemical potential $\mu $
of the vortex solitons with $S=1$ and $2$, is displayed in the right panel
of Fig. \ref{fig5.1}. The plot is produced by numerical solution of the
two-dimensional variant of BdG equations (5.8). As a result, it is found
that the solitons with $S=1$ are stable in the interval of norms%
\begin{equation}
0<N<N_{\mathrm{crit}}^{(S=1)}\approx 7.79\approx 0.32N_{\mathrm{TS}}^{(S=1)}
\tag{5.12}
\end{equation}%
(cf. Eq. (5.10)). In the remaining part of their existence region, \textit{%
viz}., at%
\begin{equation}
7.79<N<24.1,  \tag{5.13}
\end{equation}%
the solitons with $S=1$ are unstable against splitting, in accordance with
the fact that the dominant instability mode in Fig. \ref{fig5.1} corresponds
to $L=2$ in Eq. (5.7). Instability growth rates which account for the
splitting instability are complex ones. Accordingly, they form a quartet of
the form
\begin{equation}
\lambda _{1,2,3,4}=\pm \mathrm{Re}(\lambda )\pm i\mathrm{Im}(\lambda ),
\tag{5.14}
\end{equation}%
with two independent signs $\pm $, which represent the so-called \textit{%
Hamiltonian-Hopf bifurcation} (van der Meer, 1990). In terms of the chemical
potential, the stability region (5.12) corresponds to%
\begin{equation}
\mu _{\mathrm{crit}}\approx 1.276<\mu <\mu _{\max }\equiv 2.  \tag{5.15}
\end{equation}%
Here, $\mu _{\max }=1+S$ is commonly known quantum-mechanical energy
eigenvalue of the 2D isotropic HO potential, which correspond to the limit
of $N\rightarrow 0$, i.e., to Eq. (5.1) replaced by its linearized version.

\begin{figure}[tbp]
\begin{center}
\includegraphics[width=0.98\textwidth]{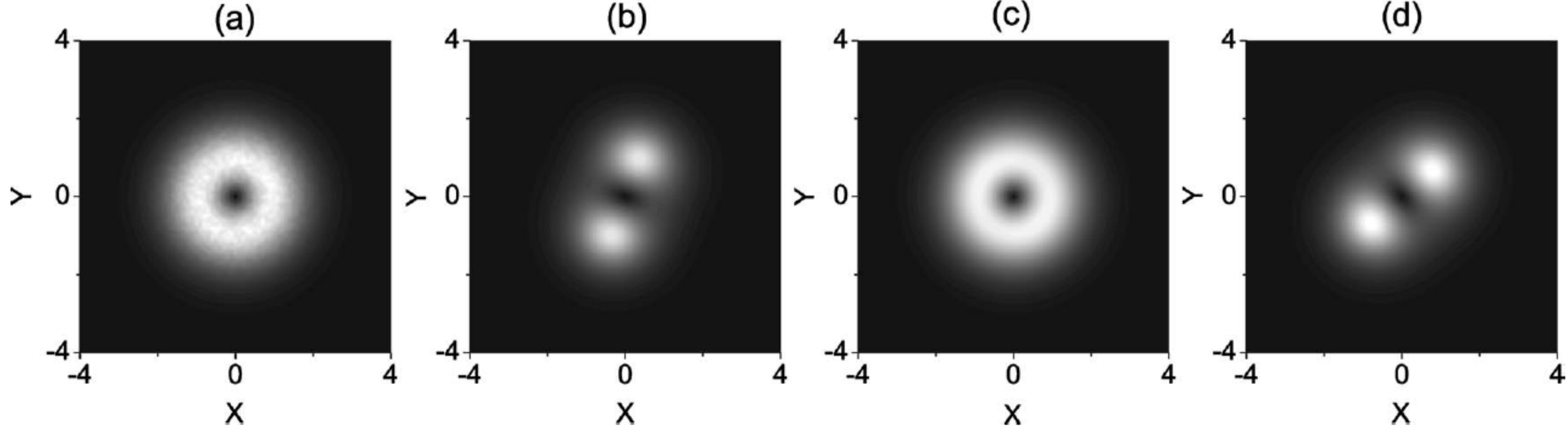}
\end{center}
\caption{Density profiles of the 2D solitons with vorticity $S=1$ trapped in
the HO potential, with initial norm $N=8.476$, which belongs to interval
(5.16). In this case, simulations of Eq. (5.1), starting from a numerically
exact solution of stationary equation (5.6) with $\protect\mu =1.2$, give
rise to a periodic sequence of splittings and recombinations. Shown here are
the density prfiles as $t=0$ (a), $t=100$ (b), $t=140$ (c), and $t=180$ (d)
(source: Mihalache\textit{\ et al}., 2006).}
\label{fig5.2}
\end{figure}

The numerical analysis demonstrates that irreversible splitting of the
unstable vortex solitons into fragments actually takes place in a part of
region (5.12), namely,
\begin{equation}
10.30<N<N_{\mathrm{TS}}^{(S=1)}\approx 24.1.  \tag{5.16}
\end{equation}%
Because the norm of the fragments produced by the splitting of the vortex in
interval (5.16) exceeds the above-mentioned TS norm for the fundamental
soliton, $N_{\mathrm{TS}}^{(S=0)}\approx 5.85$ (see Eq. (3.12)), each
fragment eventually develops the intrinsic collapse. On the other hand, in
the narrower interval,
\begin{equation}
7.79<N<10.30,  \tag{5.17}
\end{equation}%
which occupies $\approx 10\%$ of the existence region, $0<N<$ $N_{\mathrm{TS}%
}^{(S=1)}\approx 24.1$, the norm of the fragments falls below the collapse
threshold. For this reason, unstable vortex solitons with the norm falling
in interval (5.17) periodically splits in two fragments and recombines back,
as shown in Fig. \ref{fig5.2}. The corresponding dynamical pattern keeps its
vorticity, while the relative loss of its norm in the course of the
evolution from $t=0$ to $t=240$ is $\approx 1.9\times 10^{-3}$, which
corroborates the robustness of the pattern.

All the vortex solitons with $S\geq 2$ are completely unstable solutions, in
the framework of Eq. (5.1) with the self-attractive cubic nonlinearity.
Indeed, the entire existence region for these states, $\mu \leq 3$, is
covered, in the right panel of Fig. 46, by the splitting-instability branch
corresponding to $L=2$.

\subsection{Variational and numerical results for 3D fundamental and
vortical solitons trapped in the HO\ potential}

\begin{figure}[tbp]
\begin{center}
\includegraphics[width=0.45\textwidth]{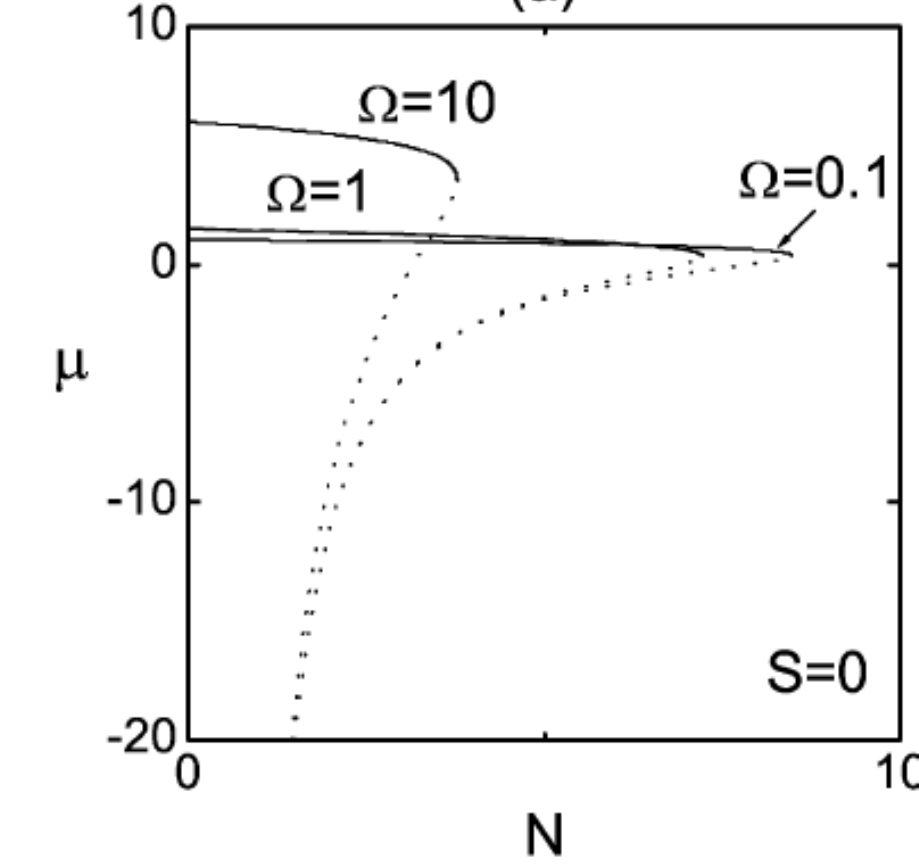}
\end{center}
\caption{Dependences of the chemical potential on the norm for families of
3D fundamental (zero-vorticity) solitons, produced by the numerical solution
of Eq. (5.6) for three different values of the HO-trap parameter $\Omega $.
Solid and dashed segments of the curves represent, respectively, stable and
unstable soliton subfamilies, in exact agreement with the VK stability
criterion, $d\protect\mu /dN<0$ (source: Malomed\textit{\ et al}., 2007).}
\label{fig5.3}
\end{figure}

Results for 3D fundamental (zero-vorticity) solitons, produced by numerical
solution of Eqs. (5.6) and (5.8), are quite simple. They are presented in
Fig. \ref{fig5.3} by means of $\mu (N)$ curves (similar to the 2D situation,
shown in Fig. \ref{fig5.1}) for three values of the anisotropy parameter, $%
\Omega =10$, $1$, and $0.1$, which correspond, respectively, to the oblate,
spherically isotropic, and prolate trapping potential in Eq. (5.1). As well
as in the 2D case, the stability of the fundamental solitons exactly obeys
the VK criterion (the negative slope), $d\mu /dN<0$; however, unlike the 2D
situation, the $\mu (N)$ curves in Fig. \ref{fig5.3} contain segments with
negative and positive slopes, only the former ones being stable. The norm of
the soliton family attains it maximum at the point where the two segments
connect.

Families of 3D vortex solutions to Eq. (5.6) were found in a numerical form
by Adhikari (2001), and stability eigenvalues for some vortices with $S=1$
were computed by Saito and Ueda (2002). A rather accurate analytical
approximation for these families is provided by the variational method,
which was developed by Malomed \textit{et al}. (2007). To apply the VA to
Eq. (5.6), note that this equation can be derived from the Lagrangian,
\begin{equation}
L=\int_{0}^{\infty }rdr\int_{0}^{+\infty }dz\left[ R_{z}^{2}+R_{r}^{2}+%
\left( \frac{S}{r}\right) ^{2}R^{2}-2\mu R^{2}+\left( r^{2}+\Omega
^{2}z^{2}\right) R^{2}-R^{4}\right] .  \tag{5.18}
\end{equation}%
A natural 3D ansatz for vortex states with integer winding number $S\geq 0$
is (cf. Eq. (4.32))%
\begin{equation}
R(r,z)=Ar^{S}\exp \left[ -r^{2}/\left( 2\rho ^{2}\right) -z^{2}/\left(
2h^{2}\right) \right] ,  \tag{5.19}
\end{equation}%
where $A,\rho $, and $h$ are free parameters. The 3D norm (4.13) of this
ansatz is%
\begin{equation}
N=\left( \pi ^{3/2}S!\right) M,~M\equiv A^{2}\rho ^{4}h,  \tag{5.20}
\end{equation}%
and the substitution of the ansatz in Lagrangian (5.18) yields the
corresponding effective Lagrangian (written here for $S=1$, as all the
states with $S\geq 2$ are completely unstable):

\begin{equation}
L_{\mathrm{eff}}=\frac{\sqrt{\pi }}{8}M\left( -4\mu +\frac{1}{h^{2}}+\frac{4%
}{\rho ^{2}}+4\rho ^{2}+\Omega ^{2}h^{2}-\frac{M}{2\sqrt{2}\rho ^{2}h}%
\right) .  \tag{5.21}
\end{equation}%
Then, values of variational parameters $A$, $h$, and $\rho $ in ansatz
(5.19) are predicted by the Euler-Lagrange equations,%
\begin{equation}
\partial L_{\mathrm{eff}}/\partial \left( A,h,\rho \right) =0.  \tag{5.22}
\end{equation}%
In particular, the last equation in system (5.22) takes a simple form,
\begin{equation}
M=8\sqrt{2}h\left( 1-\rho ^{4}\right) ,  \tag{5.23}
\end{equation}%
which predicts that the vortex state may only exist if it is narrow enough
in the radial direction, with $\rho <1$, which is a natural consequence of
the presence of the trapping potential.

The results for the vortex solitons with $S=1$ are displayed by means of $%
\mu (N)$ curves, along with their counterparts produced by the numerical
solution of Eq. (5.6), in the left panel of Fig. 49. It is seen that the VA,
based on ansatz (5.19), provides sufficiently accurate predictions, in
comparison to the numerical findings, at all values of the anisotropy
parameter $\Omega $.

\begin{figure}[tbp]
\begin{center}
\includegraphics[width=0.38\textwidth]{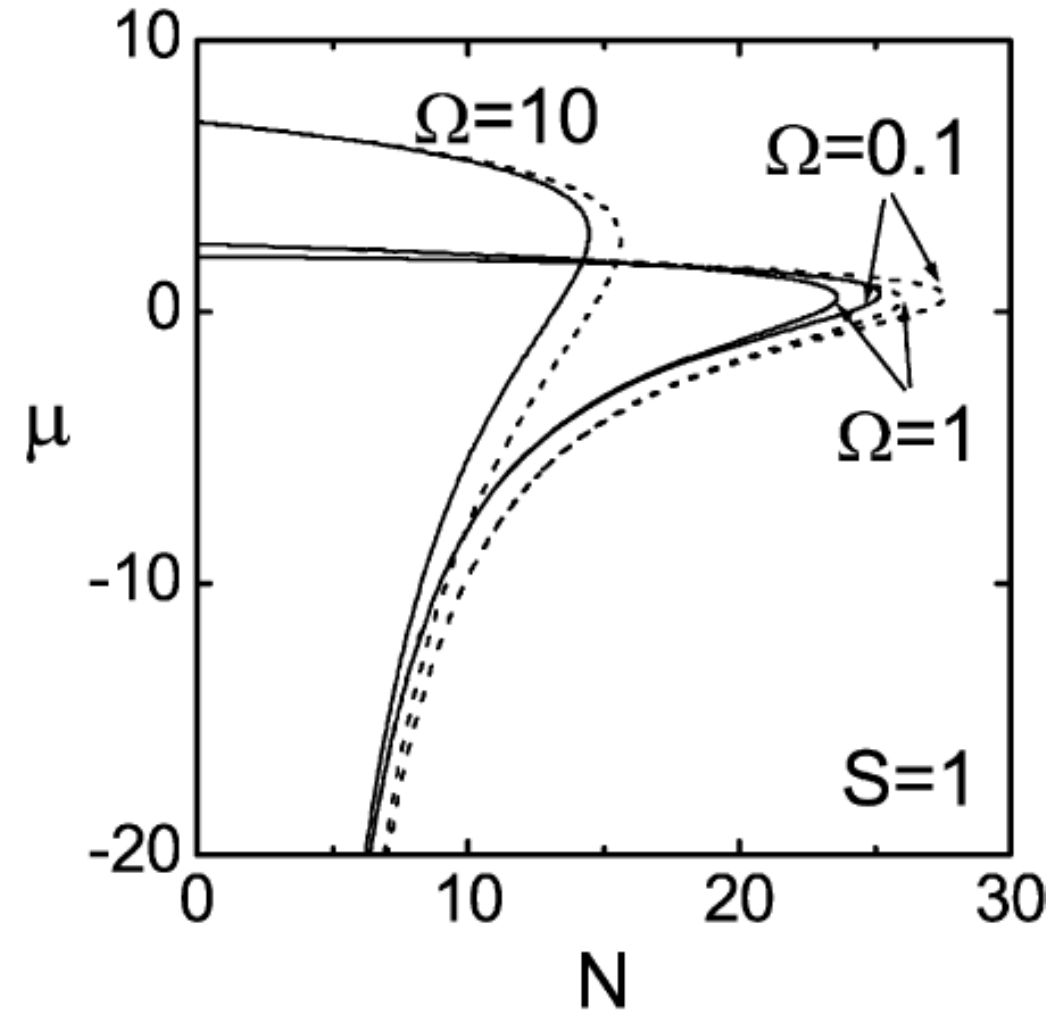} %
\includegraphics[width=0.38\textwidth]{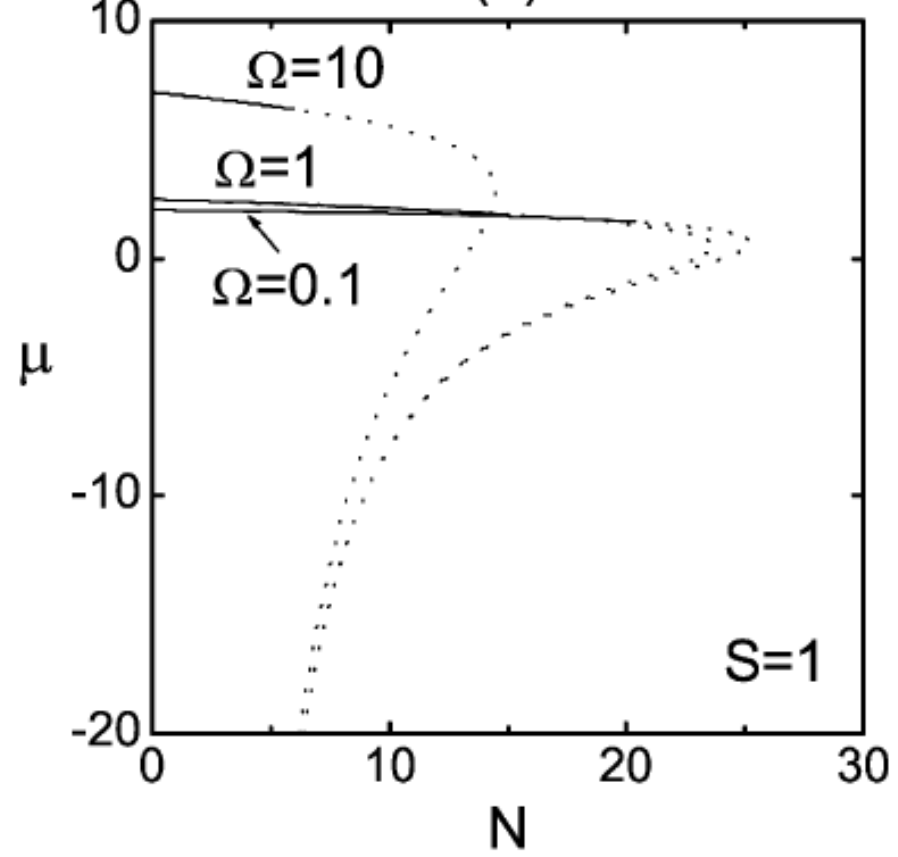}
\end{center}
\caption{The left panel: dependences $\protect\mu (N)$ for 3D solitons with
embedded vorticity $S=1$, as predicted by the VA based on ansatz (5.19) and
Euler-Lagrange equations (5.21)-(5.23) (dashed lines), and as obtained from
the numerical solution of Eq. (5.6) (solid lines). The dependences are
presented for three values of the anisotropy parameter $\Omega $. The right
panel: the same numerically produced $\protect\mu (N)$ curves, separated
into stable (solid) and unstable (dashed) segments, as predicted by the
numerical solution of the BdG equations (5.8) (source: Malomed\textit{\ et al%
}., 2007). }
\label{fig5.4}
\end{figure}

The stability of the vortex-soliton families with $S=1$ was identified from
the numerical solution of the BdG equations (5.8). The results show in the
right panel of Fig. \ref{fig5.4} that, similar to the 2D case (cf. Fig. 46),
the VK criterion is only necessary but not sufficient for the stability of
the vortex modes, as it determines the stability only against the collapse,
which is represented by real eigenvalues $\lambda $, but not against the
splitting, which is accounted for by a quartet of complex eigenvalues, in
the general form given by Eq. (5.14). It is worthy to note that the
stability interval of the norm is much wider for the prolate trap (with $%
\Omega =0.1$), $0<N<N_{\max }\approx 20.347$, and for the isotropic one
(with $\Omega =1$), $0<N<N_{\max }\approx 15.023$, than for the oblate trap
with $\Omega =10$, for which the stability interval is $0<N<N_{\max }\approx
5.947$.

\begin{figure}[tbp]
\begin{center}
\includegraphics[width=0.67\textwidth]{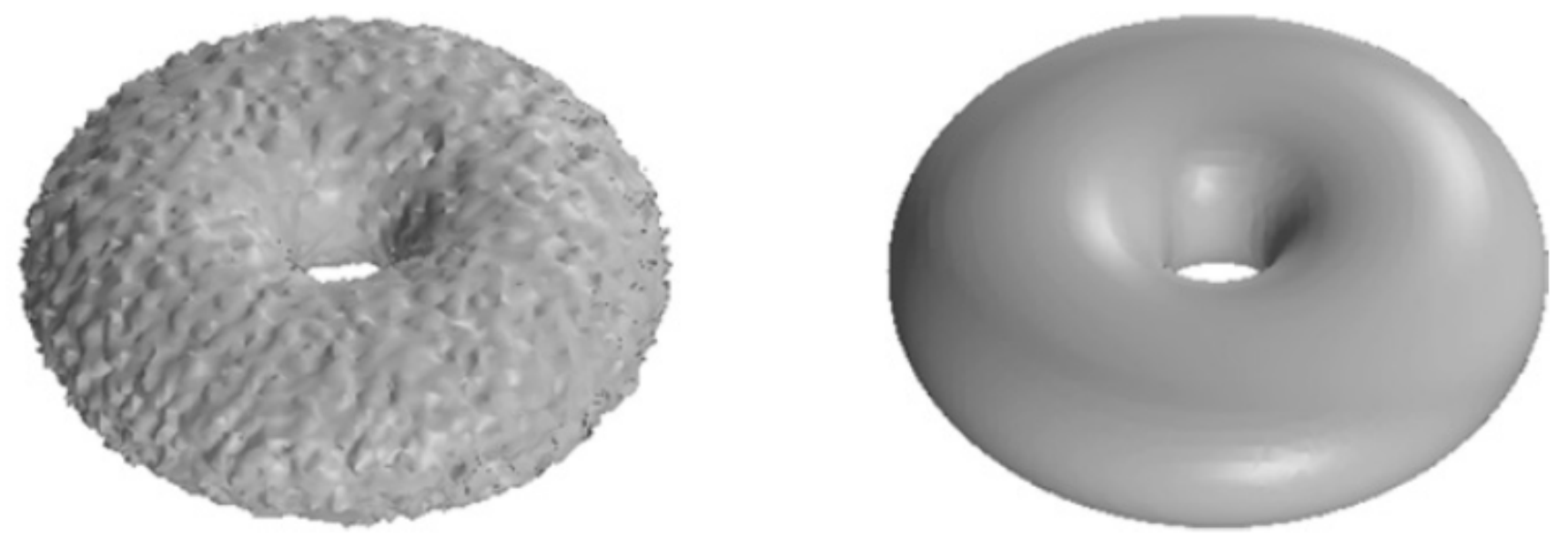}
\end{center}
\caption{Self-cleaning of a stable 3D vortex soliton (\textquotedblleft
donut") with $S=1$ in the isotropic model ($\Omega =1$ in Eq. (5.1)) after
the application of a random initial perturbation at the amplitude level of $%
10\%$. The left and right panels display, respectively, the shape of the
perturbed vortex at the initial moment, $t=0$, and at $t=120$. The
unperturbed vortex soliton has chemical potential $\protect\mu =2$ and norm $%
N=12.55$ (source: Malomed\textit{\ et al}., 2007).}
\label{fig5.5}
\end{figure}

The predictions for the stability of the 3D solitons produced by the
solution of the BdG equations (5.8) were completely confirmed by direct
simulations of Eq. (5.1) for perturbed evolution (Malomed \textit{et al}.,
2007). In particular, the robustness of stable vortices with $S=1$ is
illustrated by Fig. \ref{fig5.5}, which demonstrates that the stable
donut-shaped vortex soliton is able to absorb strong disturbance and restore
its unperturbed shape.

\subsection{Stable 2D two-component solitons with \textit{hidden vorticity}}

Proceeding to the consideration of two-component systems under the action of
the self-attractive nonlinearity and trapping potential, it is
straightforward to investigate stability of vortex solitons with equal wave
functions of both components (Yakimenko, Zaliznyak, and Lashkin, 2009). A
more intriguing possibility is to seek for stable two-component solitons
with HV\ (hidden vorticity), which are built of components carrying opposite
winding numbers, $S_{1,2}=\pm S$ (Brtka, Gammal, and Malomed, 2010).

In this case, the system of nonlinearly coupled GP equations for wave
functions $\psi _{1,2}$ of the two components is
\begin{subequations}
\begin{equation}
i\frac{\partial \psi _{1}}{\partial t}=\left[ -\frac{1}{2}\left( \frac{%
\partial ^{2}}{\partial x^{2}}+\frac{\partial ^{2}}{\partial y^{2}}\right)
-\left( |\psi _{1}|^{2}+\beta \left\vert \psi _{2}\right\vert ^{2}\right) +%
\frac{1}{2}\left( x^{2}+y^{2}\right) \right] \psi _{1},  \tag{5.24a}
\end{equation}%
\end{subequations}
\begin{equation}
i\frac{\partial \psi _{2}}{\partial t}=\left[ -\frac{1}{2}\left( \frac{%
\partial ^{2}}{\partial x^{2}}+\frac{\partial ^{2}}{\partial y^{2}}\right)
-\left( |\psi _{2}|^{2}+\beta \left\vert \psi _{1}\right\vert ^{2}\right) +%
\frac{1}{2}\left( x^{2}+y^{2}\right) \right] \psi _{2},  \tag{5.24b}
\end{equation}%
The system is assumed to be fully symmetric, with strengths of the
self-attraction in each component and HO potential scaled to be $1$, like in
the single-component GP equation (5.1), while $\beta $, that may be both
positive and negative, is the relative strength of the interaction between
the components (attraction for $\beta >0$ and repulsion for $\beta <0$). In
this notation, the collapse threshold for the norm of each component of the
symmetric ($\psi _{1}=\psi _{2}$) soliton with $S_{1,2}=\pm 1$ follows from
Eq. (5.10):
\begin{equation}
N_{\mathrm{TS}}^{(S=\pm 1)}\approx 24.1\left( 1+\beta \right) ^{-1}.
\tag{5.25}
\end{equation}

\begin{figure}[tbp]
\begin{center}
\includegraphics[width=0.50\textwidth]{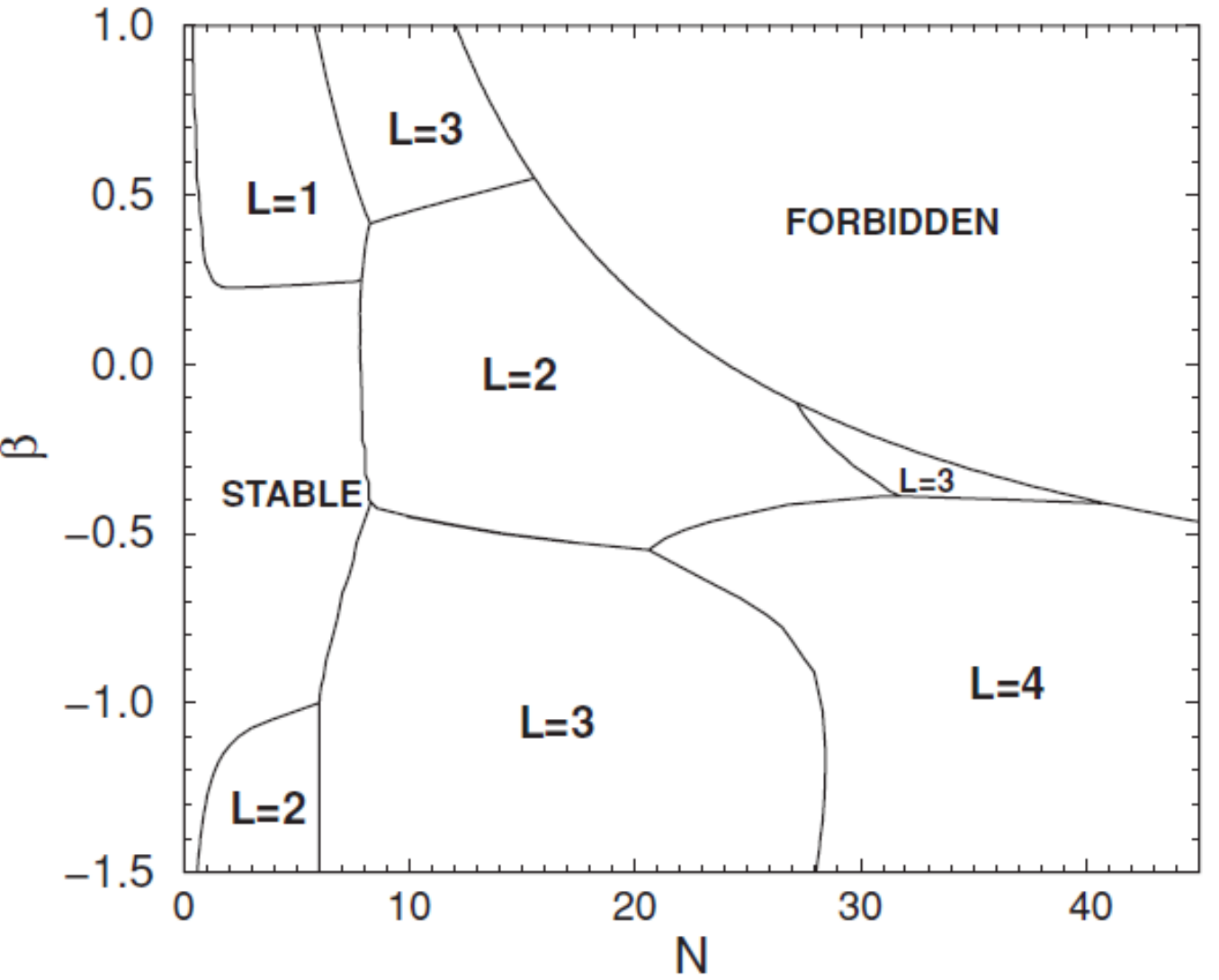}
\end{center}
\caption{The stability chart for the HV (hidden-vorticity) 2D solitons with $%
S_{1,2}=\pm 1$ (see Eq. (5.26)) in the plane of $\left( N,\protect\beta %
\right) $, produced by the numerical solution of Eqs. (5.24) and the
respective BdG equations for small perturbations. Here $N$ is the norm of
one component of the unperturbed state (the total norm is $2N$). In domains
labeled by different values of $L$ the HV states are unstable against
perturbation modes with the respective value of azimuthal index $L$, see Eq.
(5.28). In the \textquotedblleft forbidden" area, bounded by curve $N\approx
24.1/\left( 1+\protect\beta \right) $ (see Eq. (5.25)), vortex solitons with
$S=1$ do not exist (source: Brtka, Gammal, and Malomed\textit{\ }2010). }
\label{fig5.6}
\end{figure}

HV states are stationary solutions of Eqs. (5.24), written in the polar
coordinates as%
\begin{equation}
\psi _{1}=\exp \left( -i\mu t+iS\theta \right) \phi (r),\psi _{2}=\exp
\left( -i\mu t-iS\theta \right) \phi (r),  \tag{5.26}
\end{equation}%
with $S\geq 1$ and real function $\phi (r)$ determined by the same radial
equation as in the case of the state with the explicit vorticity (i.e., with
identical phase terms $S\theta $ in both components, rather than $\pm
S\theta $ in Eq. (5.26)):%
\begin{equation}
\mu \phi =-\frac{1}{2}\left( \frac{d^{2}\phi }{dr^{2}}+\frac{1}{r}\frac{%
d\phi }{dr}-\frac{S^{2}}{r^{2}}-r^{2}\right) -\left( 1+\beta \right) \phi
^{3}.  \tag{5.27}
\end{equation}%
While the density profile of the HV modes is the same as of the usual 2D
vortex solitons, up to obvious rescaling $\phi \rightarrow \left( 1+\beta
\right) ^{-1/2}\phi $ (provided that $\beta >-1$), the stability is
different. To address this point, a perturbed HV solution is looked for as%
\begin{equation}
\psi _{1,2}(r,t)=[R(r)+u_{1,2}(r)\exp \left( \lambda t+iL\theta \right)
+v_{1,2}^{\ast }(r)\exp \left( \lambda ^{\ast }t-iL\theta \right) ]\exp
\left( -i\mu t\pm iS\theta \right) ,  \tag{5.28}
\end{equation}%
where $+S$ and $-S$ correspond to $\psi _{1}$ and $\psi _{2}$, respectively,
and $L$ is an integer azimuthal index, cf. Eq. (5.7). The substitution of
this expression in Eqs. (5.24) and the linearization with respect to small
perturbations $\left( u_{1,2},v_{1,2}\right) $ leads to the respective BdG
equations. Then, the numerical solution of the BdG equations produces the
stability chart for $S=1$, displayed in Fig. \ref{fig5.6}, in which the
stability area for the HV states is shown in the plane of $\left( N,\beta
\right) $ ($N$ is defined as the norm of one component). For $S\geq 2$, all
the HV states are completely unstable.

In other parts of the plane displayed in Fig. \ref{fig5.6}, instability
domains dominated by perturbation eigenmodes with different values of
azimuthal index $L$ are identified, as per the results of Brtka, Gammal, and
Malomed (2010). It is worthy to note an essential difference from the
instability for the usual vortex solitons which, as shown above in Fig. \ref%
{fig5.1}, is always dominated by the perturbation eigenmode with $L=2$,
hence the unstable vortex spontaneously splits in two fragments (Fig. \ref%
{fig5.2}). Figure \ref{fig5.6} demonstrates that the eigenmode with $L=2$
remains dominant for the HV solitons at, roughly speaking, $|\beta |<0.5$;
on the other hand, $L=1$, $3$ and $4$ may be the azimuthal indices of the
leading perturbation eigenmode at $|\beta |>0.5$. When, in particular, the
leading unstable perturbation has $L=1$, it does not split the HV soliton.
Instead, the weak instability slowly deforms it into a rotating
crescent-shaped mode, as shown in Fig. 52 (eventually, this mode implodes
under the action of the intrinsic collapse).
\begin{figure}[tbp]
\begin{center}
\includegraphics[width=0.80\textwidth]{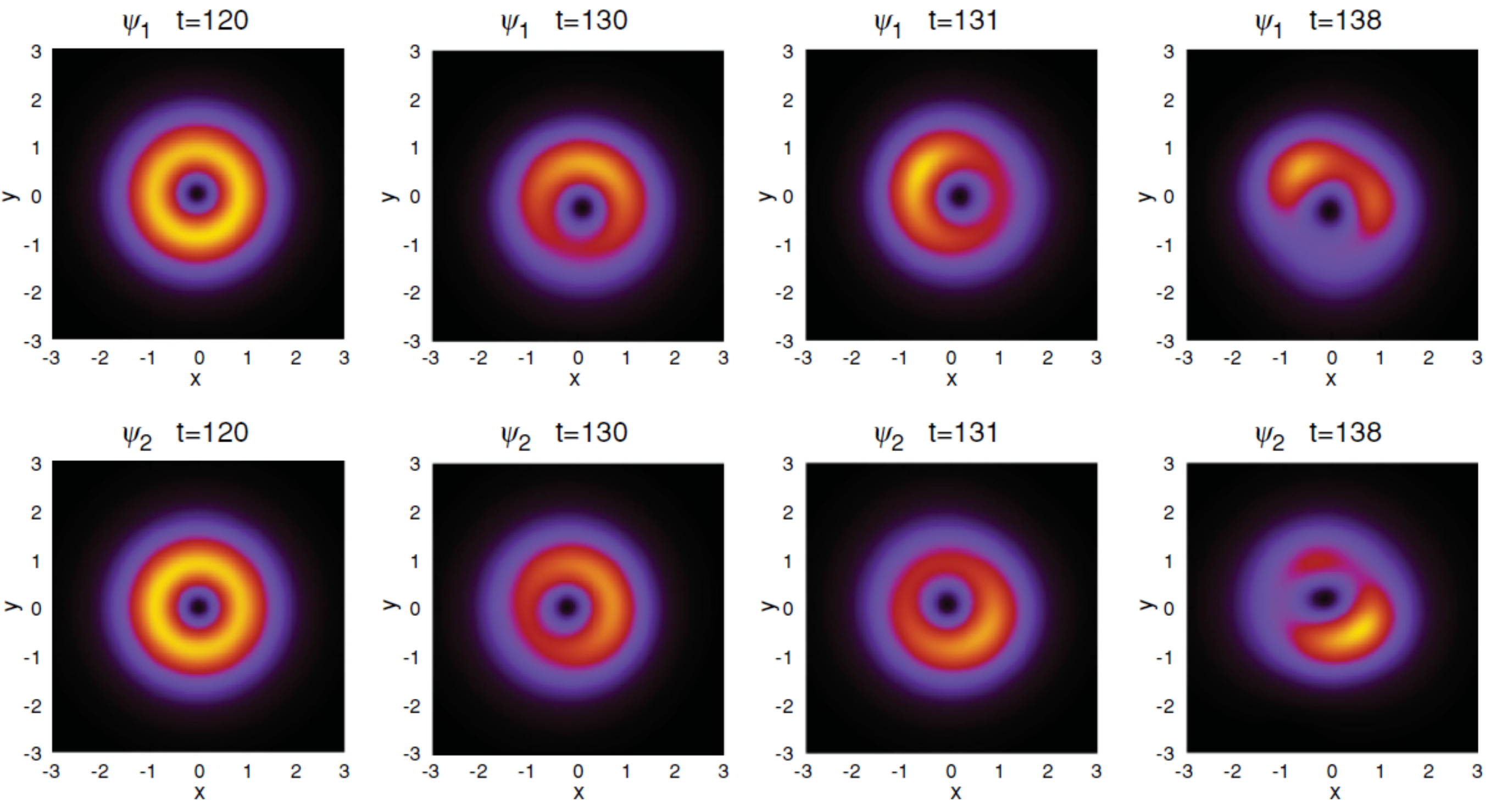}
\end{center}
\caption{The slowly developing instability of the HV soliton in the case
when the dominant perturbation eigenmode has azimuthal index $L=1$ (see Fig.
\protect\ref{fig5.6}). The top and bottom rows of panels display the
distribution of the densities in two components of the HV state at indicated
moments of time, as produced by simulations of Eqs. (5.24) with $\protect%
\beta =0.5$. The unperturbed HV soliton has the norm (in each component) $%
N=7.5$ and chemical potential $\protect\mu =0.83$ (source: Brtka, Gammal,
and Malomed\textit{\ }2010). }
\label{fig5.7}
\end{figure}

\begin{figure}[tbp]
\begin{center}
\includegraphics[width=0.90\textwidth]{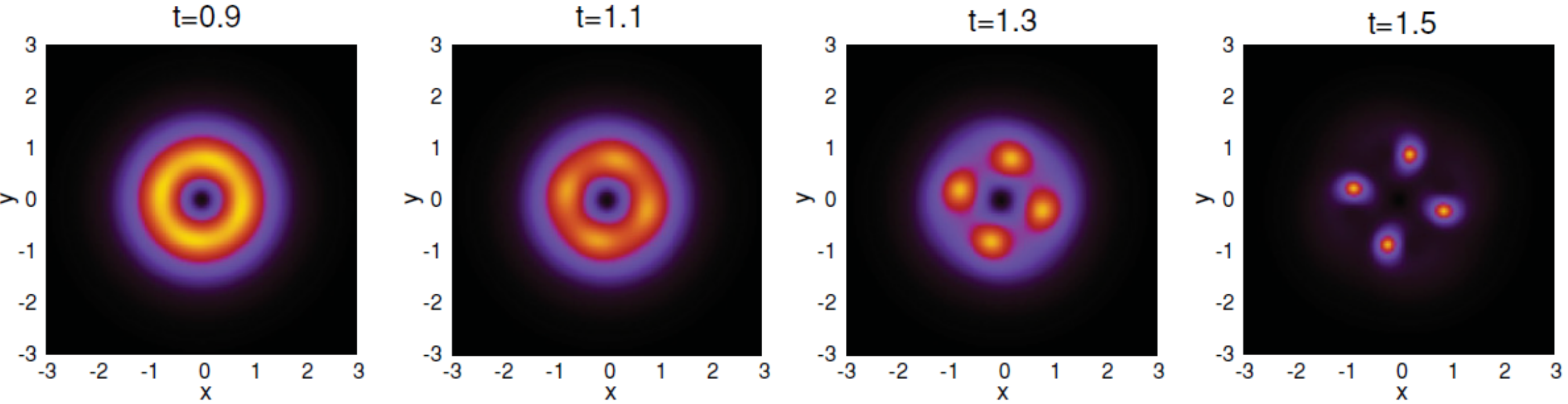}
\end{center}
\caption{The rapidly developing instability of the HV soliton in the case
when the dominant perturbation eigenmode has azimuthal index $L=4$ (see Fig.
\protect\ref{fig5.6}). The set of panels display identical distributions of
the density in both components at indicated moments of time. The results
were produced by simulations of Eq. (5.24) with $\protect\beta =-0.5$. The
unperturbed HV soliton has the norm (in each component) $N=34.1$ and
chemical potential $\protect\mu =-0.2$ (source: Brtka, Gammal, and Malomed%
\textit{\ }2010). }
\label{fig5.8}
\end{figure}

On the other hand, the instability corresponding to the dominant eigenmodes
with $L=3$ or $4$ splits the HV soliton into three or four fragments,
respectively (later, the fragments are destroyed by the intrinsic collapse).
In particular, the instability dominated by $L=4$ is very strong, leading to
fast splitting of the HV soliton, as shown in Fig. \ref{fig5.8}.

\section{Nonlinear trapping potentials}


The trapping potentials considered in the previous chapter provide the
traditional method for maintaining and stabilizing localized states, which,
by itself, applies independently of the presence of nonlinearity in the
system. \textit{Nonlinear potentials}, induced by spatial modulation of the
local strength of the cubic or other nonlinearity, offer a completely
different method for the creation of self-trapped states (quasi-solitons).
An efficient implementation of the latter method was proposed by Borovkova
\textit{et al}. (2011a,b), in the form of the \textit{self-repulsive} cubic
term with the coefficient growing fast enough from the center to periphery,
as per Eqs. (4.28) and (4.31) or (4.32). This scheme offers options for the
creation of various localized states that would not exist or would be
unstable without the use of nonlinear potentials. Although experimental
realization of the scheme has not yet been reported, many possibilities of
its use have been explored theoretically. In particular, an essential asset
of the theoretical work in this direction is that, although it is naturally
based on numerical methods, many important results may be obtained in an
analytical form, approximately or exactly.

\subsection{The basic setting}

The basic model providing self-trapping of various 2D and 3D modes in the
effective nonlinear trapping potential is (Borovkova \textit{et al}.
(2011a,b)),%
\begin{equation}
i\psi _{t}+\frac{1}{2}\nabla ^{2}\psi -\sigma (r)|\psi |^{2}\psi =0,
\tag{6.1}
\end{equation}%
where $r$ is the radial coordinate in the 2D or 3D space in which Eq. (6.1)
is written, and the local self-defocusing coefficient, $\sigma (r)>0$, grows
from the center ($r=0$) to periphery ($r\rightarrow \infty $). This equation
conserves the total norm defined as per Eq. (4.13),%
\begin{equation}
N_{\mathrm{3D}}=\int \int \int dxdydz\left\vert \psi \right\vert ^{2},N_{%
\mathrm{2D}}=\int \int dxdy\left\vert \psi \right\vert ^{2},  \tag{6.2}
\end{equation}%
the Hamiltonian (cf. Eq. (4.15)),%
\begin{equation}
H_{\mathrm{3D}}=\frac{1}{2}\int \int \int dxdydz\left[ \left\vert \nabla
\psi \right\vert ^{2}+\sigma (r)|\psi |^{4}\right]  \tag{6.3}
\end{equation}%
(or its counterpart in the 2D case), and the vectorial angular momentum,
given above by Eq. (4.16). In the 2D case, the norm, Hamiltonian, and $z$%
-component of the angular momentum are produced by obvious reduction of the
3D expressions.

Stationary states with real chemical potential $\mu $ are sought for as
solutions to Eq. (6.1) in the form of%
\begin{equation}
\psi \left( r,t\right) =\exp \left( -i\mu t\right) w(\mathbf{r}),  \tag{6.4}
\end{equation}%
where $\mathbf{r}$ is the set of coordinates in the respective 2D or 3D
space, and function $w$ (generally, a complex one) satisfies the stationary
equation,%
\begin{equation}
\mu w+\frac{1}{2}\nabla ^{2}w-\sigma (\mathbf{r})|w|^{2}w=0.  \tag{6.5}
\end{equation}%
The fundamental condition necessary for the existence of physically relevant
self-trapped states produced by Eq. (6.5) is the \emph{convergence} at $%
r\rightarrow \infty $ of the integral expression (6.2) which defines the
norm of the state. In particular, if $\sigma (r)$ grows asymptotically at $%
r\rightarrow \infty $ as
\begin{equation}
\sigma _{\mathrm{asympt}}(r)=\sigma _{0}r^{\alpha }  \tag{6.6}
\end{equation}%
with $\sigma _{0}>0$ and $\alpha >0$, it is easy to identify the convergence
condition by means of the TF approximation, which neglects term $(1/2)\nabla
^{2}w$ in Eq. (6.4) (a rigorous justification of the application of the TF
to models of the present type was developed by Malomed and Pelinovsky
(2015)). In this approximation, the solution is obvious,
\begin{equation}
\left( |w|^{2}\right) _{\mathrm{TF}}=\mu /\sigma (r)  \tag{6.7}
\end{equation}%
(it exists for $\mu >0$), and the substitution of this in Eq. (6.2)
immediately shows that the integral converges under the condition%
\begin{equation}
\alpha >D.  \tag{6.8}
\end{equation}%
In fact, condition (6.8) secures the self-trapping for localized modes of
all types, i.e., not only fundamental ones, but also states with an
intrinsic structure, such as embedded vorticity (Borovkova \textit{et al}.
(2011a)). It is relevant to mention that the same condition also secures the
convergence of the integral which defines Hamiltonian (6.3).

In the model with the asymptotic form (6.6) of the spatially modulated local
nonlinearity, the asymptotic solution including the term given by Eq. (6.7)
and the first post-TF correction, produced by term $(1/2)\nabla ^{2}w$ at $%
r\rightarrow \infty $, is%
\begin{equation}
w_{\mathrm{TF-corrected}}(r)\approx \sqrt{\frac{\mu }{\sigma _{0}}}%
r^{-\alpha /2}+\frac{1}{4\sqrt{\mu \sigma _{0}}}\left[ \frac{\alpha }{2}%
\left( \frac{\alpha }{2}-\left( D-2\right) \right) -m^{2}\right] r^{-\alpha
/2-2}  \tag{6.9}
\end{equation}%
(the term $\sim m^{2}$ appears in the 2D case, representing the effect of
the vorticity with integer winding number $m$, see Eq. (6.11) below).

Expression (6.9) shows that the TF approximation yields the \emph{%
asymptotically exact} expression for the soliton's tail at $r\rightarrow
\infty $, as the first correction to it is negligibly small, hence the
convergence condition, given by Eq. (6.8), is always correct. Further, it is
relevant to stress that stationary equation (6.5) with the
nonlinearity-modulation profile (6.6) is \emph{nonlinearizable}: while the
solution's tail decays at $r\rightarrow \infty $, its asymptotic form,
determined by the TF approximation (6.7) and (6.9), cannot be found from the
linearization of Eq. (6.5), unlike the usual situation for nonlinear
equations in the free space.

\subsection{Typical examples of the self-trapped modes}

Basic results illustrating the realization of the concept of the effective
nonlinear trapping potential were reported by Borovkova \textit{et al}.
(2011a), who considered 2D version of Eq. (6.1) with%
\begin{equation}
\sigma (r)=1+r^{\alpha },  \tag{6.10}
\end{equation}%
whose asymptotic form complies with expression (6.6) (the 1D version of the
model was also considered in that work). Along with the fundamental 2D
solitons, their counterparts with embedded vorticity $m$ were considered
too, with Eqs. (6.4) and (6.5) replaced by%
\begin{equation}
\psi \left( r,t\right) =\exp \left( -i\mu t+im\theta \right) w(r),
\tag{6.11}
\end{equation}%
\begin{equation}
\mu w+\frac{1}{2}\left( \frac{d^{2}w}{dr^{2}}+\frac{1}{r}\frac{dw}{dr}-\frac{%
m^{2}}{r^{2}}w\right) -\left( 1+r^{\alpha }\right) w^{3}=0,  \tag{6.12}
\end{equation}%
where $\theta $ is the angular coordinate. In this case, the substitution of
the TF approximation (6.6), with $\sigma (r)$ taken as per Eq. (6.10),
yields the prediction for the total norm of the fundamental ($m=0$)
solitons, which is relevant for all dimensions, $D=1$, $2$, and $3$:%
\begin{equation}
N_{\mathrm{TF}}=\frac{2\pi ^{D}\mu }{\alpha \sin \left( \pi D/\alpha \right)
}.  \tag{6.13}
\end{equation}%
In particular, this expression provides the convergent norm precisely under
condition (6.8).

\begin{figure}[tbp]
\begin{center}
\includegraphics[width=0.60\textwidth]{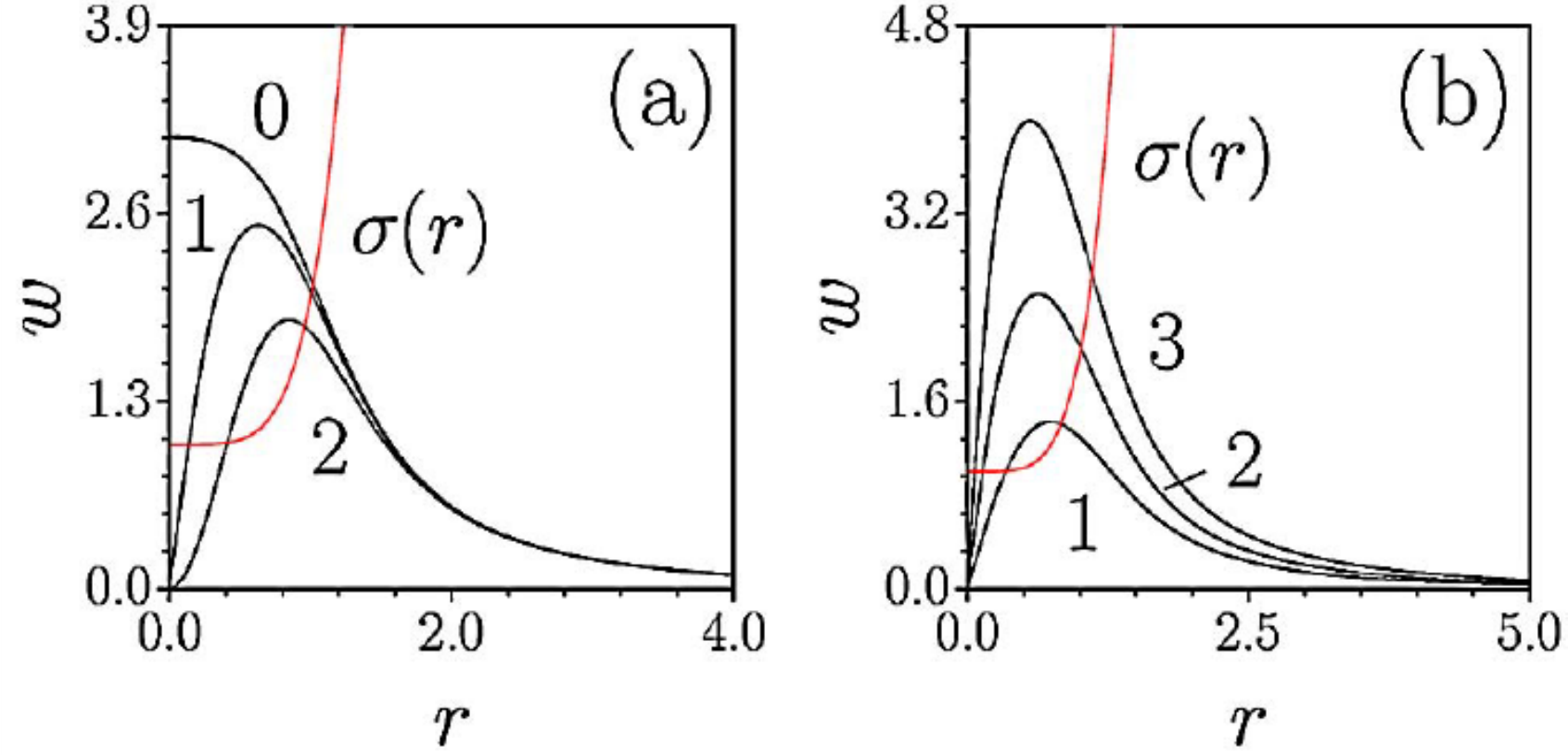}
\end{center}
\caption{(a) Radial profiles of self-trapped solutions produced by the
numerical solution of Eq. (6.12) with $\protect\alpha =5$ and a fixed
chemical potential, $\protect\mu =10$, for vorticities $m=0$, $m=1$, and $%
m=2 $. (b) Vortex states for $\protect\alpha =5$, $m=1$ and $\protect\mu =5$%
, $10 $, and $20$ (curves 1, 2, and 3, respectively). The red curve shows
the respective nonlinearity-modulation profile, defined as per Eq. (6.10)
(source: Borovkova \textit{et al}.,\textit{\ }2011a). }
\label{fig6.1}
\end{figure}

\begin{figure}[tbp]
\begin{center}
\includegraphics[width=0.66\textwidth]{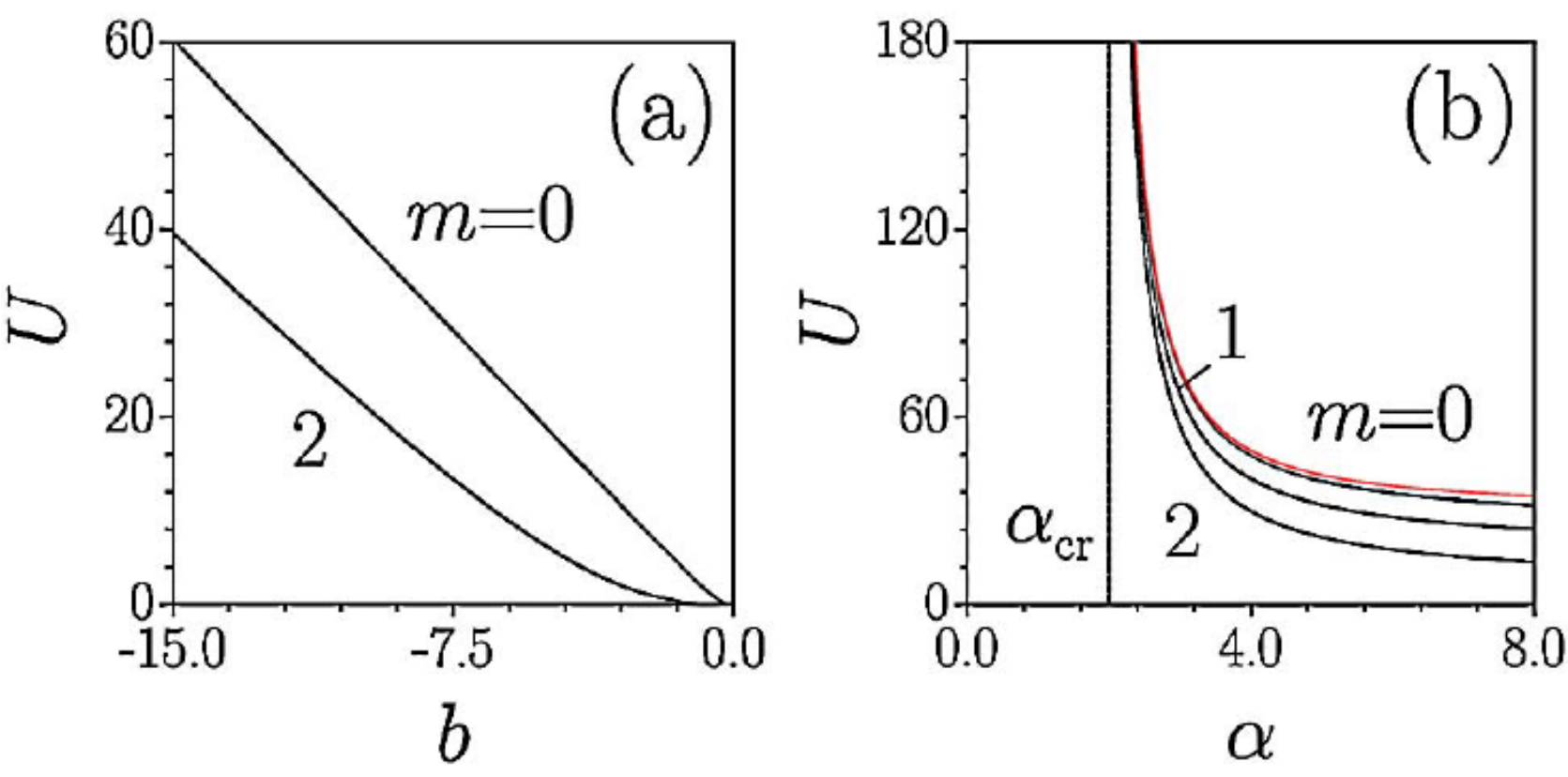}
\end{center}
\caption{(a) The norm of 2D self-trapped states (denoted $U$, instead of $N$%
) produced by the numerical solution of Eq. (6.12), vs. the chemical
potential (denoted $-b$, instead of $\protect\mu $), for $\protect\alpha =5$
and two values of the vorticity, $m=0$ and $2$. (b) Dependences of the 2D
norm on power $\protect\alpha $ from Eq. (6.10) for a fixed chemical
potential, $\protect\mu =10$, and three values of the vorticity: $m=0$, $1$,
and $2$. Value $\protect\alpha _{\mathrm{cr}}=2$ is determined by Eq. (6.7),
with the norm diverging at $\protect\alpha \geq \protect\alpha _{\mathrm{cr}%
} $. The red curve shows the analytical prediction for $m=0$ produced by the
TF approximation in the form of Eq. (6.13) (source: Borovkova \textit{et al}%
., 2011a). }
\label{fig6.2}
\end{figure}

Typical examples of radial profiles of stationary states, produced by the
numerical solution of Eq. (6.12) with $m=0$, $1$, and $2$, are displayed in
Fig. \ref{fig6.1}, and families of the respective solution families,
characterized by dependences $N(\mu )$ and $N(\alpha $), are shown in Fig.
55. The latter plot includes the TF prediction (6.13) for $m=0$, showing
that the TF approximation is quite accurate at all values of $\alpha $. Note
also that the linear dependence of $N_{\mathrm{TF}}$ on $\mu $, predicted by
Eq. (6.13), is in perfect agreement with the linear line for $m=0$ in Fig. %
\ref{fig6.2}(a).

\subsection{Stability of the self-trapped states}

Stability of the 2D\ self-trapped states produced by Eq. (6.12) against
small perturbations was investigated by Borovkova \textit{et al}. (2011(a))
by means of the numerical solution of the respective BdG equations, cf. Eqs.
(5.8). As a result, it was concluded that the entire family of fundamental
solitons, with $m=0$, is stable, in accordance with the fact that the
respective $N(\mu )$ dependence in Fig. \ref{fig6.2}(a) obeys the \textit{%
anti-VK criterion}, $dN/d\mu >0$, which is a necessary stability condition
for families of localized states supported by the self-repulsive
nonlinearity (on the contrary to the opposite VK criterion per se, in the
case of self-attraction), as shown by Sakaguchi and Malomed (2010). In the
notation adopted Fig. 6.2(a), the anti-VK criterion means $dU/db<0$.

Unlike the fundamental self-trapped states, ones with embedded vorticity
tend to be vulnerable to instability, their stabilization being possible for
sufficiently steep nonlinearity-modulation profiles in Eq. (6.10); in
particular, the entire family of the vortices with $m=1$ was found by
Borovkova et al. (2011a) to be completely stable at $\alpha \geq 8$, being
only partly stable at smaller $\alpha $ (e.g., for $-\mu >30$ at $\alpha =5$%
).

\begin{figure}[tbp]
\begin{center}
\includegraphics[width=0.58\textwidth]{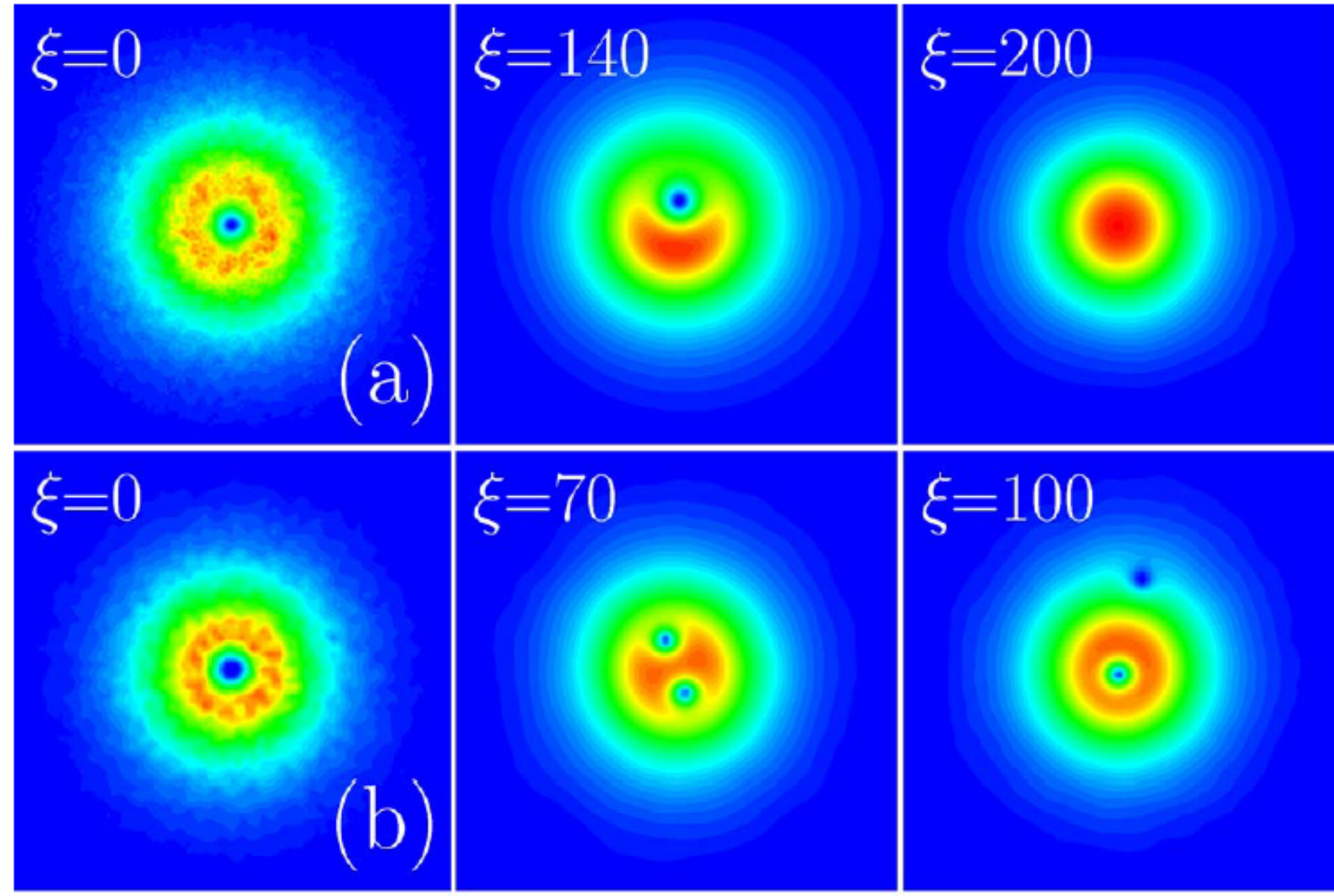}
\end{center}
\caption{(a) The development of the drift instability of the vortex state
with $m=1$ and $\protect\mu =3$, produced by simulations of Eq. (6.1) with
the nonlinearity-modulation profile (6.10), where $\protect\alpha =3.5$.
Colors denote values of the local density, $\left\vert \protect\psi \left(
x,y,t\right) \right\vert ^{2}$, with variable $t$ replaced by $\protect\xi $%
. (b) The same for the splitting-drift instability of the double vortex,
with $m=2$, $\protect\mu =20$, and $\protect\alpha =4$ (source: Borovkova
\textit{et al}., 2011a). }
\label{fig6.3}
\end{figure}

Direct simulations demonstrate that instability of the vortex mode with $m=1$
expels the vortex' pivot from the central position, and drives its drift
towards periphery. As shown in Fig. \ref{fig6.3}(a), the pivot eventually
vanishes in the peripheral area. The angular momentum of the original vortex
state is also ousted to the periphery where, in actual numerical
simulations, it is eliminated by a numerical absorber providing stability of
the numerical scheme. In the case when a double vortex with $m=2$ is
unstable, small perturbations may split it in two unitary vortices, one
staying as a stable one with $m=1$ at the center, while the other one
spontaneously drifting to the periphery, where it vanishes eventually. The
latter scenario of the instability development is shown in Fig. \ref{fig6.3}%
(b).

\subsection{Exact, approximate, and numerical results for the anti-Gaussian
spatial-modulation profile}

\subsubsection{The formulation and analytical findings}

Stronger results for self-trapped states supported by the spatial modulation
of the cubic repulsive term were reported by Borovkova \textit{et al}.
(2011b) in the framework of Eq. (6.1) with a very steep (\textit{%
anti-Gaussian}) modulation profile (cf. the algebraic one in Eq. (6.10)):%
\begin{equation}
\sigma (r)=\exp \left( \alpha r^{2}\right) .  \tag{6.14}
\end{equation}%
Note that the scaling invariance of Eq. (6.1) with $\sigma (r)$ taken as per
Eq. (6.14) implies that the solutions actually depend not separately on $%
\alpha $ and $\mu $, but solely on the ratio, $\mu /\alpha $.

A remarkable finding is that the 2D version of Eqs. (6.1) and (6.5) with $%
\sigma (r)$ taken as per Eq. (6.14) admits a particular \emph{exact solution}
for the self-trapped mode with embedded vorticity $m=1$:%
\begin{equation}
\psi =\frac{\alpha }{\sqrt{2}}r\exp \left( -i\mu t+i\theta -\frac{\alpha }{2}%
r^{2}\right) ,\mu =2\alpha  \tag{6.15}
\end{equation}%
(the exact solution is available for the single value of $\mu $, as written
in Eq. (6.15)). As concerns the asymptotic form of the solution's tail at $%
r\rightarrow \infty $, it is determined by the balance of the cubic and
gradient terms in Eq. (6.5):
\begin{equation}
w(r)=\left[ \frac{\alpha }{\sqrt{2}}r+\frac{2\mu -\alpha \left( D+2\right) }{%
2\sqrt{2}\alpha r}+\mathcal{O}\left( \frac{1}{r^{3}}\right) \right] \exp
\left( -\alpha r^{2}/2\right) .  \tag{6.16}
\end{equation}%
This asymptotic solution is a \emph{universal} one, as, apart from the
correction $\mathcal{O}\left( 1/r^{3}\right) $, it does not depend on $\mu $%
, nor on the presence of embedded vorticity $m$, and only the correction $%
\sim 1/r$ depends on the spatial dimension, $D$. Vanishing of the latter
term in the case of $D=2$, $\mu =2\alpha ~$agrees with the existence of the
exact solution given by Eq. (6.15).

Similar to the case of modulation profile (6.10), the one represented by Eq.
(6.14) makes equation (6.5) nonlinearizable, as the tail solution (6.16),
although it vanishes at $r\rightarrow \infty $, cannot be found in the
correct form, unless the nonlinear term is kept in the analysis. On the
other hand, the approximation leading to Eq. (6.16) is different from the TF
approximation, which was used above to derive expression (6.9), as the
gradient term is negligible in the latter case.

In any spatial dimension $D$, fundamental solitons (with $m=0$) produced by
Eqs. (6.5) and (6.14) can be very accurately approximated by the variational
method. To this end, one notes that the 2D ($D=2$) or 3D ($D=3$) Lagrangian
of Eq. (6.5) with real field $w(r)$ is
\begin{equation}
L_{w}=\frac{1}{2}\int_{0}^{\infty }\left[ \mu w^{2}-\frac{1}{2}\left( \frac{%
dw}{dr}\right) ^{2}-\frac{1}{2}\sigma (r)w^{4}\right] r^{D-1}dr.  \tag{6.17}
\end{equation}%
A natural ansatz approximating the fundamental soliton is%
\begin{equation}
w(r)=A\exp \left( -\frac{\alpha }{2}r^{2}\right) ,  \tag{6.18}
\end{equation}%
with the norm%
\begin{equation}
N=\left( \pi /\alpha \right) ^{D/2}A^{2},  \tag{6.19}
\end{equation}%
where amplitude $A$ is a free variational parameter.

The substitution of ansatz (6.18) in Lagrangian (6.29) yields the following
effective Lagrangian, in which $A^{2}$ is expressed in terms of $N$, by
means of relation (6.19):%
\begin{equation}
L_{\mathrm{2D}}=\frac{1}{4\pi }\left( \mu -\frac{\alpha }{2}\right) N-\frac{%
\alpha }{8\pi ^{2}}N^{2},  \tag{6.20}
\end{equation}%
\begin{equation}
L_{\mathrm{3D}}=\frac{1}{8\pi }\left( \mu -\frac{3\alpha }{4}\right) N-\frac{%
\alpha ^{3/2}}{16\pi ^{5/2}}N^{2}.  \tag{6.20}
\end{equation}%
Finally, the Euler-Lagrange equation, $\partial L/\partial N=0$, predicts
the following relation between the norm and chemical potential:%
\begin{equation}
N=\left( \pi /\alpha \right) ^{D/2}\left( \mu -D\alpha /4\right) .
\tag{6.21}
\end{equation}%
Numerically found dependences $N(\alpha )$ for the fundamental solitons are
indistinguishable from the VA prediction (6.21), in both cases $D=2$ and $3$
(Borovkova \textit{et al}. 2011b).

A two-component extension of Eqs. (6.1), in the form of%
\begin{equation}
i\left( \psi _{1,2}\right) _{t}+\frac{1}{2}\nabla ^{2}\psi _{1,2}-\sigma
(r)\left( |\psi _{1,2}|^{2}+C\left\vert \psi _{2,1}^{2}\right\vert \right)
\psi =0,  \tag{6.22}
\end{equation}%
with an XPM coefficient $C>0$, was elaborated by Kartashov \textit{et al}.
(2011).

\subsubsection{Vortices and rotary structures}

\begin{figure}[tbp]
\begin{center}
\includegraphics[width=0.40\textwidth]{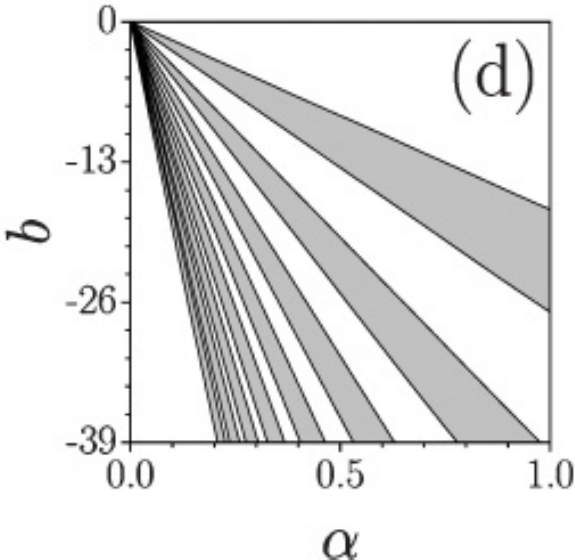}
\end{center}
\caption{Self-trapped vortex modes with $m=2$, generated by Eqs. (6.1) and
(6.14), are stable and unstable in white and shaded sectors of the
parametric plane $\left( \protect\alpha ,b\equiv b\equiv -\protect\mu %
\right) $. Note that the scaling invariance of the underlying equation
implies that all boundaries of the stability sectors are strict straight
lines, $\protect\mu /\protect\alpha =\mathrm{const}$, as seen in the figure
(source: Borovkova \textit{et al}., 2011b). }
\label{fig6.4}
\end{figure}

\begin{figure}[tbp]
\begin{center}
\includegraphics[width=0.62\textwidth]{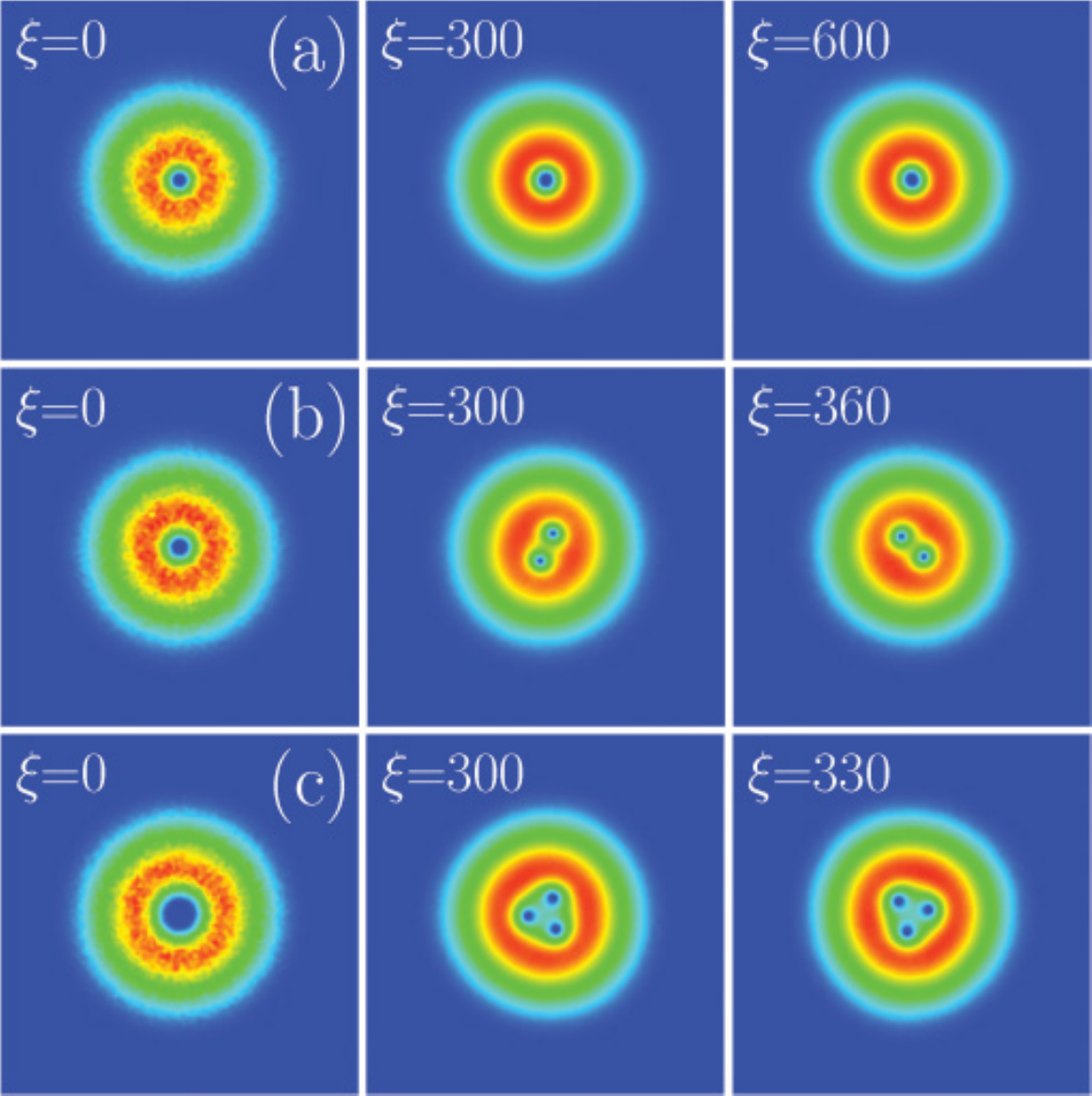}
\end{center}
\caption{(a) Relaxation to the stationary state of a perturbed stable vortex
solution of Eqs. (6.1) and (6.17) with $m=2$ and $\protect\mu =17$. (b)
Spontaneous splitting of an unstable vortex with $m=2$ and $\protect\mu =11$
into a steadily rotating pair of unitary vortices. (c) The splitting of the
unstable vortex with $m=3$ and $\protect\mu =9$ into a rotating triplet of
unitary vortices. In this figure $\protect\xi \equiv t$ (source: Borovkova
\textit{et al}., 2011b). }
\label{fig6.5}
\end{figure}

Shapes of fundamental and vortical self-trapped modes created by the
anti-Gaussian modulation profile (6.14) are quite similar to those shown
above in Fig. \ref{fig6.1} for profile (6.10). However, results for the
stability of vortices are different: unlike the above model, where vortices
even with the lowest winding number, $m=1$, tend to be unstable, see Fig.
56(a), all the 2D vortex states supported by the anti-Gaussian profile
(6.14) are \emph{completely stable}, as well as all the fundamental modes in
all dimensions (Borovkova \textit{et al}., 2011b). Alternation of stability
and instability commences from $m=2$, as shown in Fig. \ref{fig6.4}. In case
the vortex modes with $m=3$ and $3$ are unstable, they do not start drift to
$r\rightarrow \infty $, unlike what is shown above in Fig. \ref{fig6.3}(a);
instead, they spontaneously split in robust rotating sets of two or three
unitary vortices, as seen in Fig. \ref{fig6.5}.

The same 2D model, based on Eq. (6.1) with the modulation profile given by
Eq. (6.14), also supports stable rotational motion of a single vortex with
the pivot shifted off the center, as well of clusters composed of several
vortices (Kartashov \textit{et al.}, 2017). The respective GP equation,
written in the reference frame rotating with angular velocity $\omega $, is
written as%
\begin{equation}
i\psi _{t}+\frac{1}{2}\nabla ^{2}\psi +i\omega \left( x\frac{\partial }{%
\partial y}-y\frac{\partial }{\partial x}\right) \psi -\sigma (r)|\psi
|^{2}\psi =0.  \tag{6.23}
\end{equation}%
($\omega >0$ corresponds to the counter-clockwise rotation).

\begin{figure}[tbp]
\begin{center}
\includegraphics[width=0.43\textwidth]{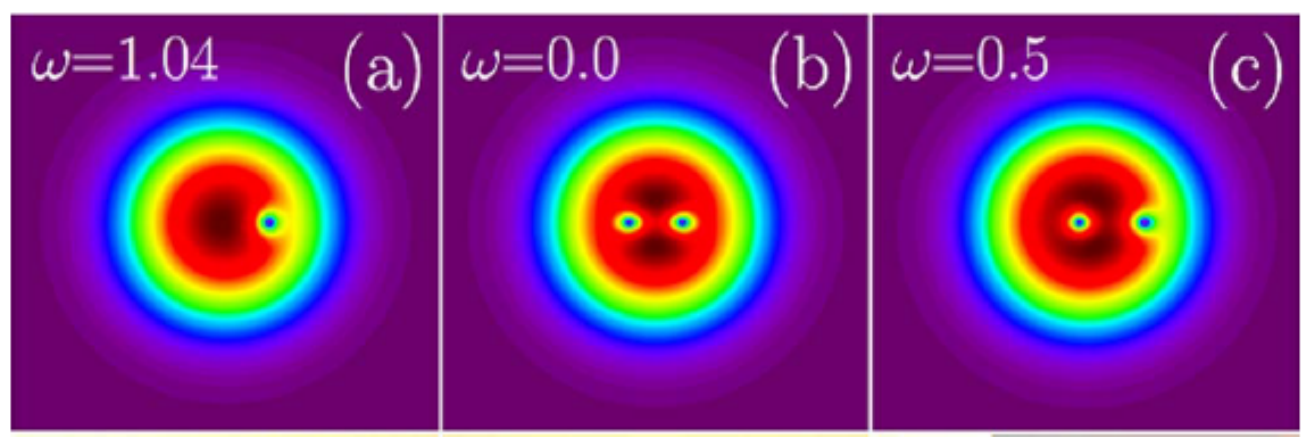} %
\includegraphics[width=0.43\textwidth]{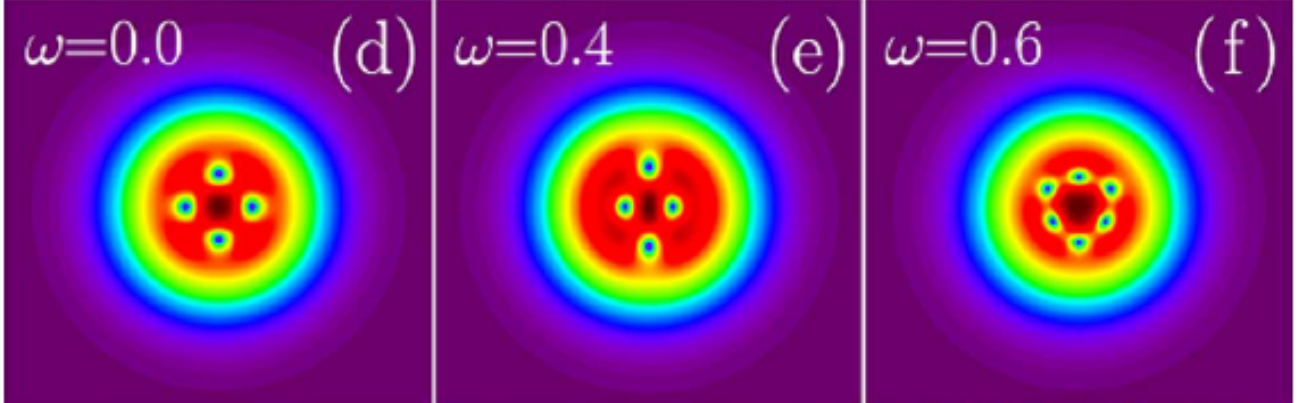}
\end{center}
\caption{(a): A snapshot of the density distribution in a 2D\ vortex with $%
m=1$ and norm $N=58.4$, which performs orbiting motion in the framework of
Eqs. (6.23) and (6.14) with $\protect\alpha =0.5$ and $\protect\omega =1.04$%
. Panels (b) and (d): the density distribution in stationary spatially
symmetric dipole and quadrupole vortex-antivortex clusters, produced by Eqs.
Eqs. (6.23) and (6.14) with $\protect\alpha =0.5$, $\protect\omega =0$, and $%
\protect\mu =10$. (c) and (e): The same clusters stabilized and
simultaneously deformed by the rotation. (f) The density distribution in a
stable rotating sextupole with $\protect\alpha =0.5$, $\protect\mu =20$,
built as a cluster of three vortices alternating with three antivortices.
Each panel shows an area in the $\left( x,y\right) $ plane of size $\left(
-5,+5\right) $ in each direction (source: Kartashov\textit{\ et al}., 2017).
}
\label{fig6.6}
\end{figure}

A snapshot of the density distribution in a stable vortex moving along a
circular trajectory is displayed in Fig. \ref{fig6.6}(a). Further, separated
vortex and antivortex, with winding numbers $\pm 1$, may form a non-rotating
dipole, as shown in Fig. \ref{fig6.6}(b), and a non-rotating quadrupole
built of alternating vortices and antivortices is shown in Fig. \ref{fig6.6}%
(d) (generally speaking, vortex dipoles and quadrupoles are subject to weak
instability in the absence of rotation (M\"{o}tt\"{o}nen \textit{et al}.,
2005)). In the usual confined BEC, vortex-antivortex dipoles were
experimentally created by Freilich \textit{et al}. (2010).

When the vortex dipole is unstable, its evolution may expel one vortex to
periphery, where it eventually disappears, while the remaining antivortex
falls onto the center, thus converting the unstable dipole into a stable
stationary vortex with the unitary topological charge. This
instability-development scenario resembles the one displayed above in Fig. %
\ref{fig6.3}(b). The application of the rotation ($\omega \neq 0$ in Eq.
(6.23)) helps to stabilize vortex-antivortex clusters. Simultaneously, Figs. %
\ref{fig6.6}(c), (e), and (f) demonstrate that dipoles, quadrupoles, and
sextupoles (a set of three vortices alternating with three antivortices),
which are symmetric complexes in the absence of the rotation, are deformed
by the rotation. The deformation is caused by an effective Coriolis force
acting with opposite signs on individual vortices and antivortices. In
particular, Fig. \ref{fig6.6}(c) shows that this force pushes the vortex and
antivortex, belonging to the dipole, towards $r=0$ and $r\rightarrow \infty $%
, respectively, and Fig. \ref{fig6.6}(f) shows that the sextupole, which has
a symmetric hexagonal shape at $\omega =0$, is deformed into a triangular
structure. Finally, if $\omega $ is too large, the fast rotation destroys
the clusters.

Additional results for 2D and 3D fundamental and vortex solitons were
produced by means of numerical methods, in similar models with more
sophisticated (specially devised) profiles of the spatial modulation of the
strength of the self-repulsive cubic nonlinearity (Tian \textit{et al}.
2012; Wu \textit{et al}. 2013). Another direction of the work was the
analysis of a similar model with the local quintic self-repulsive
nonlinearity (Zeng and Malomed, 2012).

\subsubsection{A discrete 2D\ model with the exponential
nonlinearity-modulation profile}

It is relevant to mention that a discrete version of the models considered
here was elaborated by Kevrekidis \textit{et al}. (2015). The corresponding
lattice (discrete) NLS equation, with the 2D discrete coordinates $\left(
m,n\right) $ is%
\begin{equation}
i\frac{d\psi _{mn}}{dt}=-\varepsilon \left( \psi _{m+1,n}+\psi _{m-1.n}+\psi
_{m,n+1}+\psi _{m,n-1}-4\psi _{m,n}\right) +\sigma \left( m,n\right)
\left\vert \psi _{m,n}\right\vert ^{2}\psi _{m,n}~,  \tag{6.24}
\end{equation}%
cf. Eqs. (2.163) and (6.1), with the exponential profile of the spatial
modulation of local nonlinearity,%
\begin{equation}
\sigma \left( m,n\right) =\exp \left( 2\left( |m|+|n|\right) \right) ,
\tag{6.25}
\end{equation}%
cf. Eq. (6.14). Various species of 2D discrete solitons supported by Eqs.
(6.24) and (6.25) have been constructed, and their stability has been
investigated. Among them were two types of discrete vortex solitons, \textit{%
viz}., OS-centered \textit{vortex crosses}, whose shape is similar to that
displayed above in Fig. 24, and IS-centered \textit{vortex squares}).

\subsection{Gyroscope solitons (vortex tori) in 3D}

Full consideration of the structure and dynamics of fundamental and vortical
solitons in the 3D version of Eq. (6.1) with the modulation profile defined
by Eq. (6.14), where $\alpha =1/2$ is fixed by means of rescaling, was
presented by Driben \textit{et al}. (2014a). The respective stationary
equation (6.5) for modes with embedded vorticity $S$, introduced as per Eq.
(6.11) with real function $w\left( r,z\right) $, takes the following form,
in cylindrical coordinates $\left( r,z\right) $:%
\begin{equation}
\mu w=-\frac{1}{2}\left( \frac{\partial ^{2}}{\partial r^{2}}+\frac{1}{r}%
\frac{\partial }{\partial r}-\frac{S^{2}}{r^{2}}\right) w+\exp \left( \frac{1%
}{2}\left( r^{2}+z^{2}\right) \right) w^{3}.  \tag{6.26}
\end{equation}%
As shown above, the VA based on Eqs. (6.18), (6.19), and (6.21) with $D=3$
produces a very accurate prediction for the family of fundamental solitons,
with $S=0$. For 3D solitons with embedded vorticity $m\geq 1$, the VA is
less accurate. However, in this case quite relevant is the TF approximation.
Indeed, neglecting the derivative terms in Eq. (6.26), one obtains%
\begin{equation}
w_{\mathrm{TF}}^{2}=%
\begin{array}{c}
0,~\mathrm{at}~r^{2}<m^{2}/\left( 2\mu \right) , \\
\left[ \mu -S^{2}/\left( 2r^{2}\right) \right] \exp \left( -\frac{1}{2}%
\left( r^{2}+z^{2}\right) \right) ,~\mathrm{at}~r^{2}>S^{2}/\left( 2\mu
\right) .%
\end{array}
\tag{6.27}
\end{equation}%
The TF approximation (6.27) produces the norm of the vortex state in the
form of%
\begin{equation}
N_{\mathrm{TF}}(\mu )=4\left( 2\pi \right) ^{3/2}\mu \exp \left( -\frac{S^{2}%
}{4\mu }\right) \int_{0}^{\infty }\frac{e^{-R}RdR}{4R+S^{2}/\mu }  \tag{6.28}
\end{equation}%
(here, the integral factor should be computed numerically).

\begin{figure}[tbp]
\begin{center}
\includegraphics[width=0.72\textwidth]{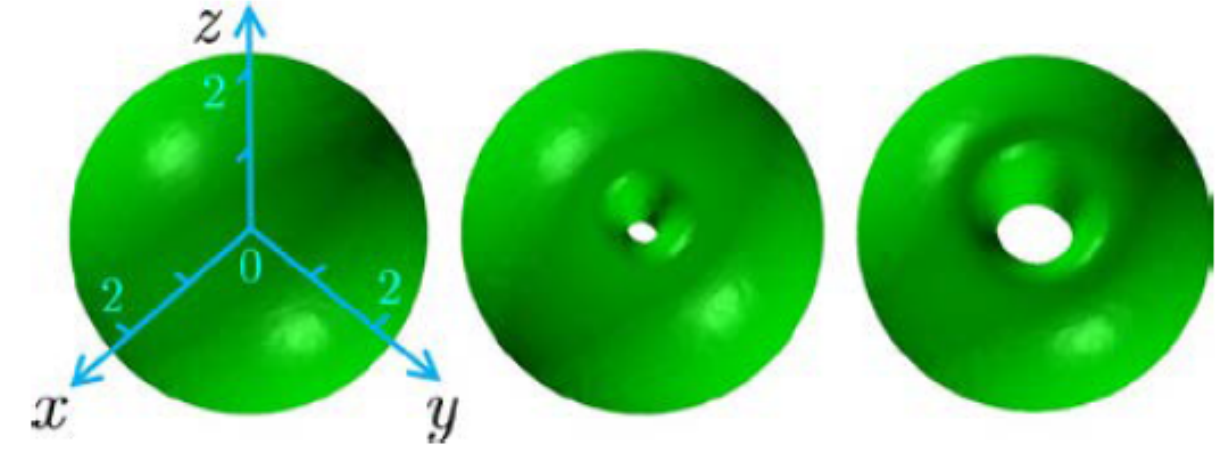}
\end{center}
\caption{Typical examples of stable 3D self-trapped modes produced by the
numerical solution of Eq. (6.26) with $\protect\mu =16$ and $S=0$, norm $%
N=246.1$ (left), $S=1$, $N=225.2$ (central), and $S=2$, $N=191.7$. The modes
are displayed by means of isosuraces, $\left\vert w\left( x,y,z\right)
\right\vert ^{2}=3$ (source: Driben \textit{et al}., 2014a). }
\label{fig6.7}
\end{figure}

\begin{figure}[tbp]
\begin{center}
\includegraphics[width=0.60\textwidth]{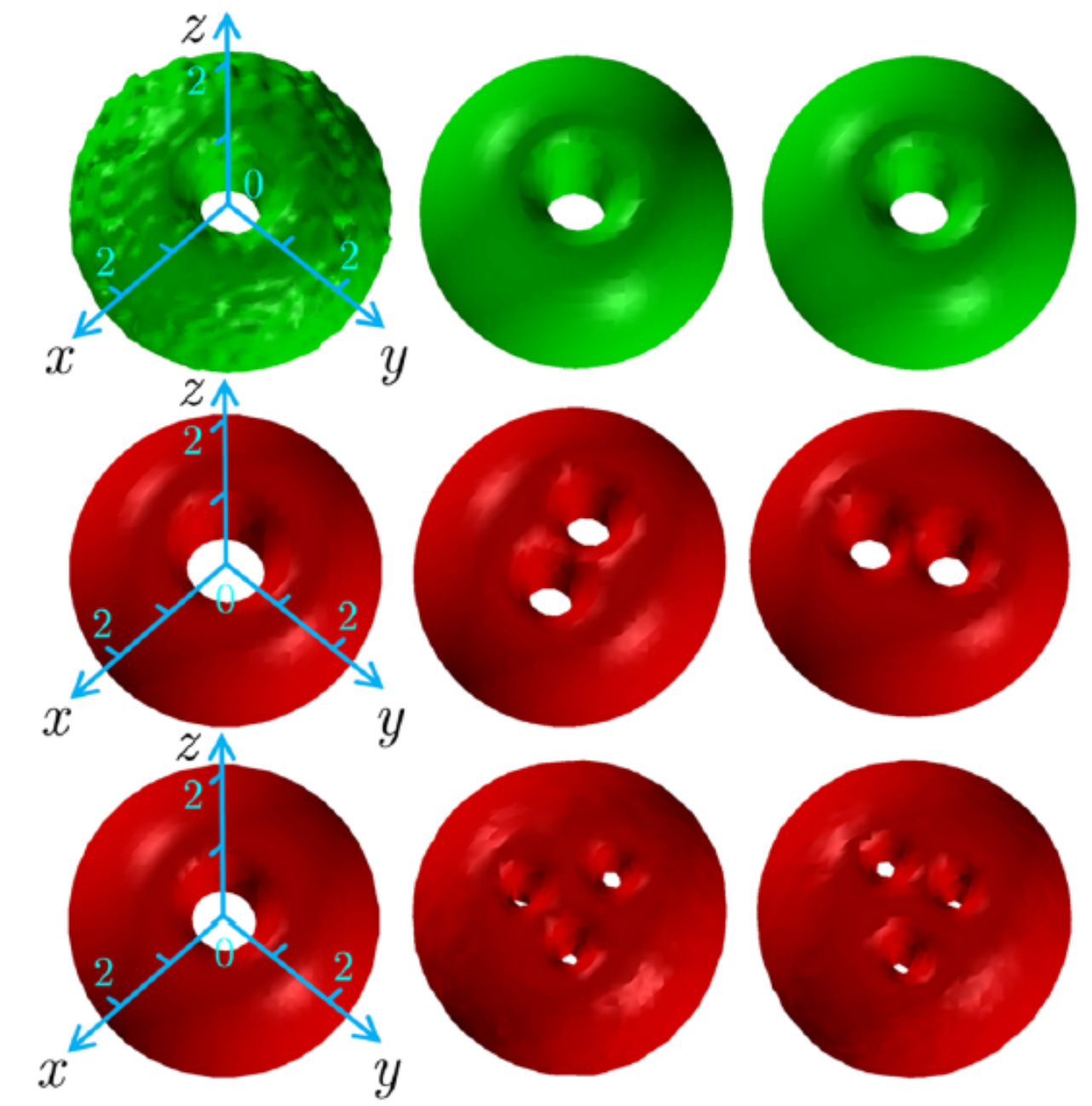}
\end{center}
\caption{Generic examples of the perturbed dynamics of 3D solitons with
embedded vorticity $S=1$ in the top row (a stable vortex torus with $\protect%
\mu =16$; the plots are shown at $t=0$, $t=230$, $t=310$); $S=2$ in the
middle row (an unstable vortex torus with $\protect\mu =13$, shown at $t=0$,
$t=230$, $t=310$); and $S=3$ (an unstable vortex torus with $\protect\mu =25$%
, shown at $t=0$, $t=165$, $t=180$). $N=191.7$. The top and middle rows
display density isosurfaces $\left\vert \protect\psi \left( x,y,z\right)
\right\vert =2.5$, and bottom one displays $\left\vert \protect\psi \left(
x,y,z\right) \right\vert =3$ (source: Driben \textit{et al}., 2014a). }
\label{fig6.8}
\end{figure}

Numerically found shapes of typical \emph{stable} 3D solitons with
vorticities $S=0$, $1$, and $2$ are displayed in Fig. \ref{fig6.7} (because
of their shape, the solitons with $S\neq 0$ are often called \textit{vortex
tori}, alias 3D vortex rings). It was found that all the states with $S=0$
and $S=1$ are stable, while ones with $S=2$ and $3$ feature alternating
stability and instability zones. Generic examples of the dynamics of stable
and unstable solitons are presented in Fig. \ref{fig6.8}. It is seen that a
stable soliton cleans itself of a strong initial perturbation, while
unstable ones with $S=2$ and $3$ spontaneously split in rotating clusters of
two and three unitary vortices, respectively.

\begin{figure}[tbp]
\begin{center}
\includegraphics[width=0.60\textwidth]{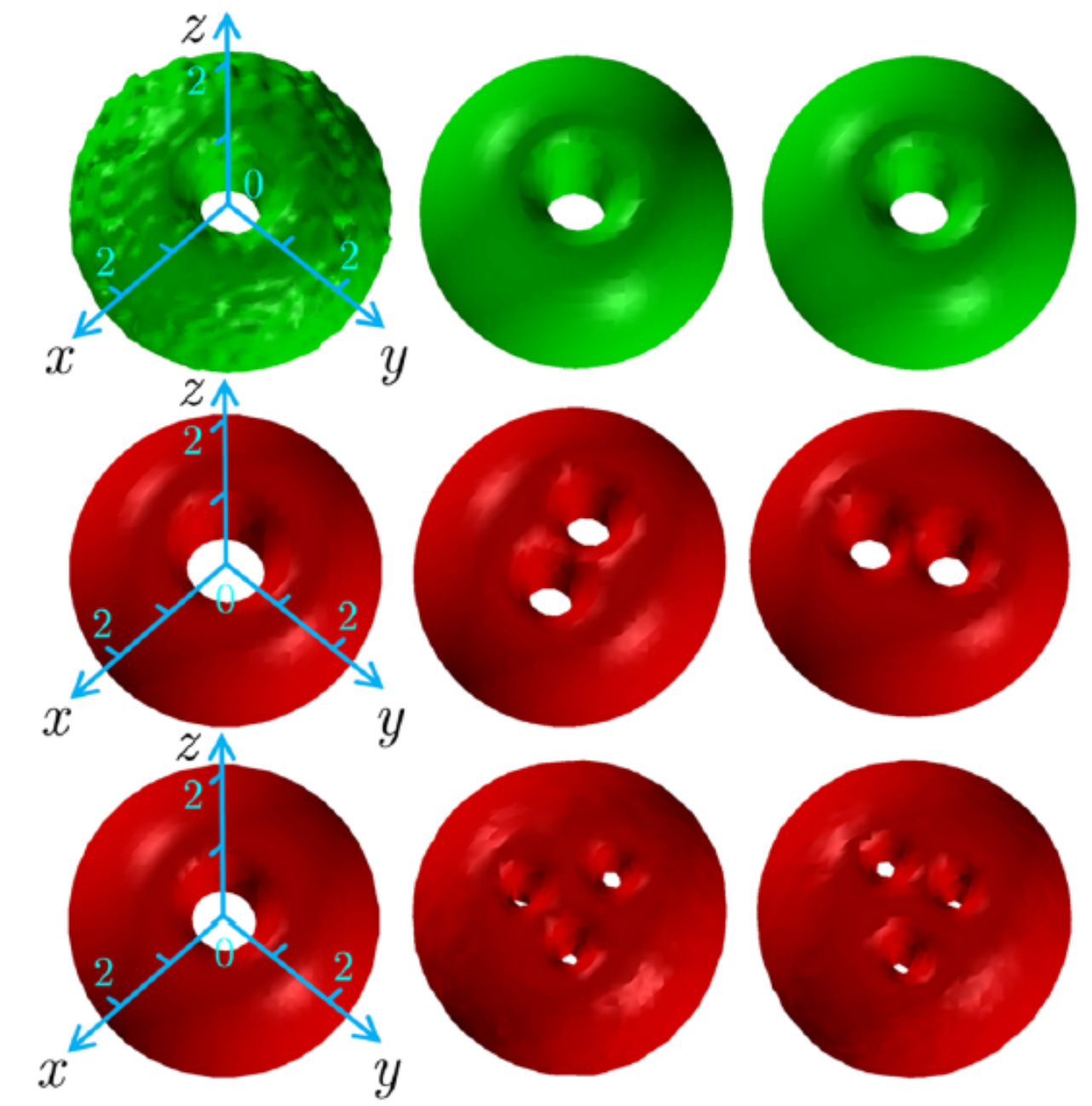}
\end{center}
\caption{Black lines show the norm of the fundamental solitons (a) and ones
with $S=1$ (b) vs. their chemical potential. Chains of green dots represent
the VA prediction for the fundamental solitons, given by Eq. (6.21) with $%
D=3 $, and the TF prediction for $S=1$, given by Eq. (6.28). The empty
circle in (a) and bold red dot in (b) correspond, respectively, to the left
plot in Fig. 60, and the top row in Fig. \protect\ref{fig6.8} (source:
Driben \textit{et al}., 2014a). }
\label{fig6.9}
\end{figure}

Families of the completely stable solitons with $S=0$ and $1$ are
characterized by dependences $N(\mu )$ shown in Fig. \ref{fig6.9}. It is
clearly seen that the VA for the fundamental solitons with $S=0$ and the TF
approximation for ones with $S=1$ produce the $N(\mu )$ lines which are
indistinguishable from their numerically generated counterparts.

The stable 3D solitons with embedded vorticity demonstrate macroscopic
dynamics resembling that of mechanical gyroscopes. An example if displayed
in Fig. 63. To generate it, Eqs. (6.1) and (6.14) with $\alpha =1/2$ was
simulated, with the input taken as the stable soliton with $S=1$, shown in
Figs. 60 and 61, to which a \textit{torque} is applied at $t=0$, multiplying
the soliton's wave function by the phase factor%
\begin{equation}
T=\exp \left( i\beta z\tanh \left( x/x_{0}\right) \right) ,  \tag{6.29}
\end{equation}%
with real $\beta $. This factor adds a nonzero component $M_{y}$ of the
angular momentum (see Eq. (4.16)). In particular, if the unperturbed vortex
is taken in the form defined by the TF approximation, as per Eqs. (6.11),
(6.27), and (6.28), the momentum produced by torque (6.29) is%
\begin{equation}
M_{y}=\frac{\beta }{2x_{0}}\left[ N-\left( 2\pi \right) ^{3/2}\left( \mu -%
\frac{S^{2}}{4}\right) \right] ,  \tag{6.30}
\end{equation}%
while the the momentum of the unperturbed vortex soliton is%
\begin{equation}
M_{z}=SN.  \tag{6.31}
\end{equation}%
Then, as it may be expected, the application of the torque to the gyroscope
sets it in the state of precession, as shown in Fig. 63. This dynamical
regime corresponds to the total angular momentum with components (6.31) and
(6.30).
\begin{figure}[tbp]
\begin{center}
\includegraphics[width=0.60\textwidth]{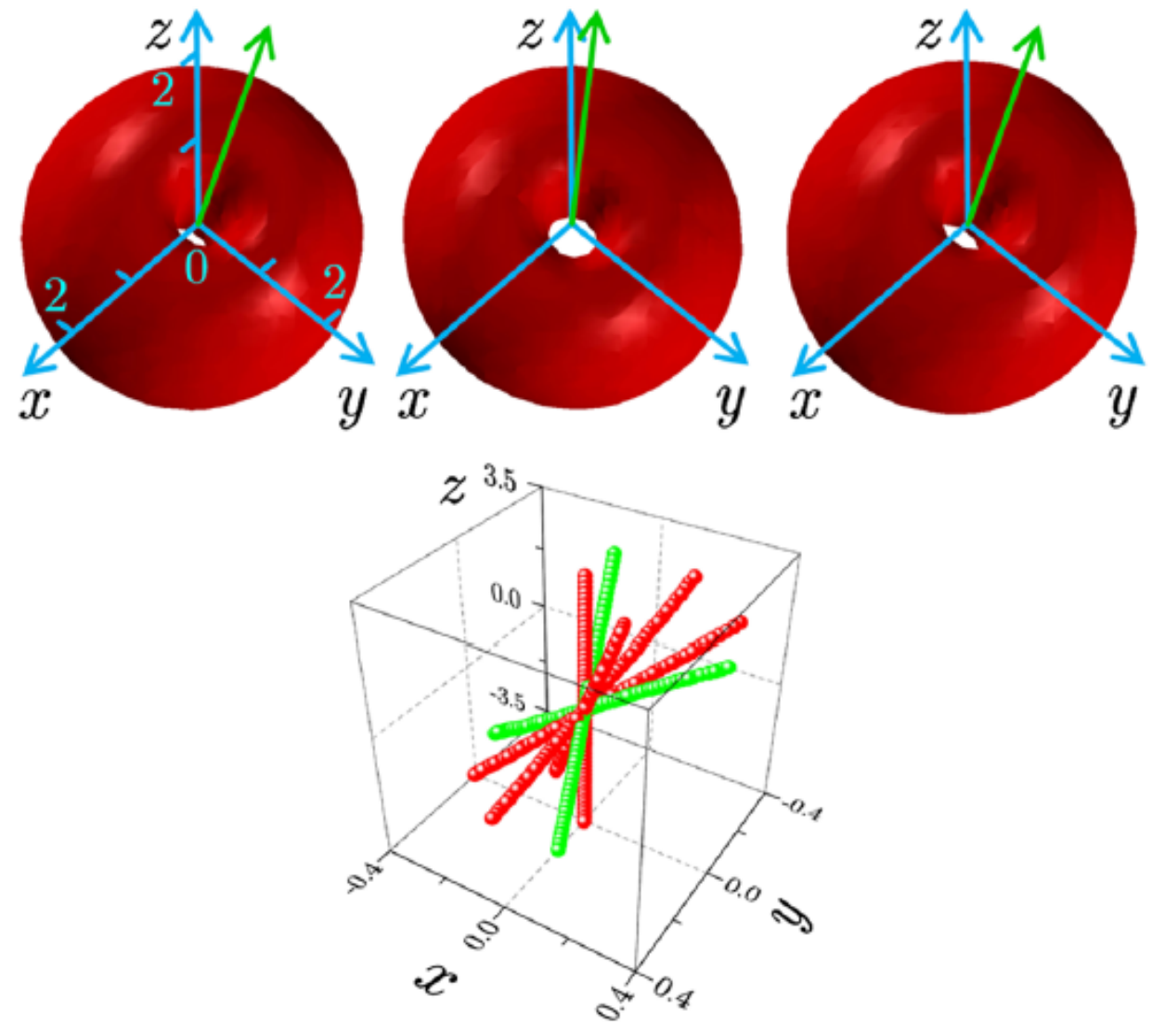}
\end{center}
\caption{The gyroscopic precession of the 3D vortex soliton (torus) with $S=1
$ and $\protect\mu =16$, initiated by the application of torque (6.29) with $%
\protect\beta =4$ and $x_{0}=5$. The plots are produced by simulations of
Eqs. (61.) and (6.14) with $\protect\alpha =1/2$. Three images in the top
row correspond to times $t=90$, $90.8$, and $91.5$. Time $\Delta t=1.5$
between the first and third images is equal to the precession period. Green
arrows indicate the corresponding orientations of the axle of the precessing
gyroscope. The bottom plot shows a set of instantaneous orientations of the
axle, the green ones corresponding to those displayed in the top row
(source: Driben \textit{et al}., 2014a).}
\label{fig6.10}
\end{figure}

\subsection{Hopfions: 3D solitons with two independent topological charges}

As shown above in Fig. \ref{fig6.7}, the density distribution in 3D solitons
with embedded vorticity $m$ is shaped as a torus (\textquotedblleft donut").
An additional possibility is to impose intrinsic twist, characterized by
integer number $s$, onto the torus, in an attempts to create self-trapped
modes carrying two independent topological charges. States of such a type
were originally predicted in the Faddeev-Skyrme model (Faddeev, 1976;
Faddeev and Niemi, 1997), based on a set of three real scalar fields.
Soliton-like states in that system realize the \textit{Hopf map}, $%
R^{3}\rightarrow S^{2}$, therefore they are called \textit{hopfions}.

\begin{figure}[tbp]
\begin{center}
\includegraphics[width=0.60\textwidth]{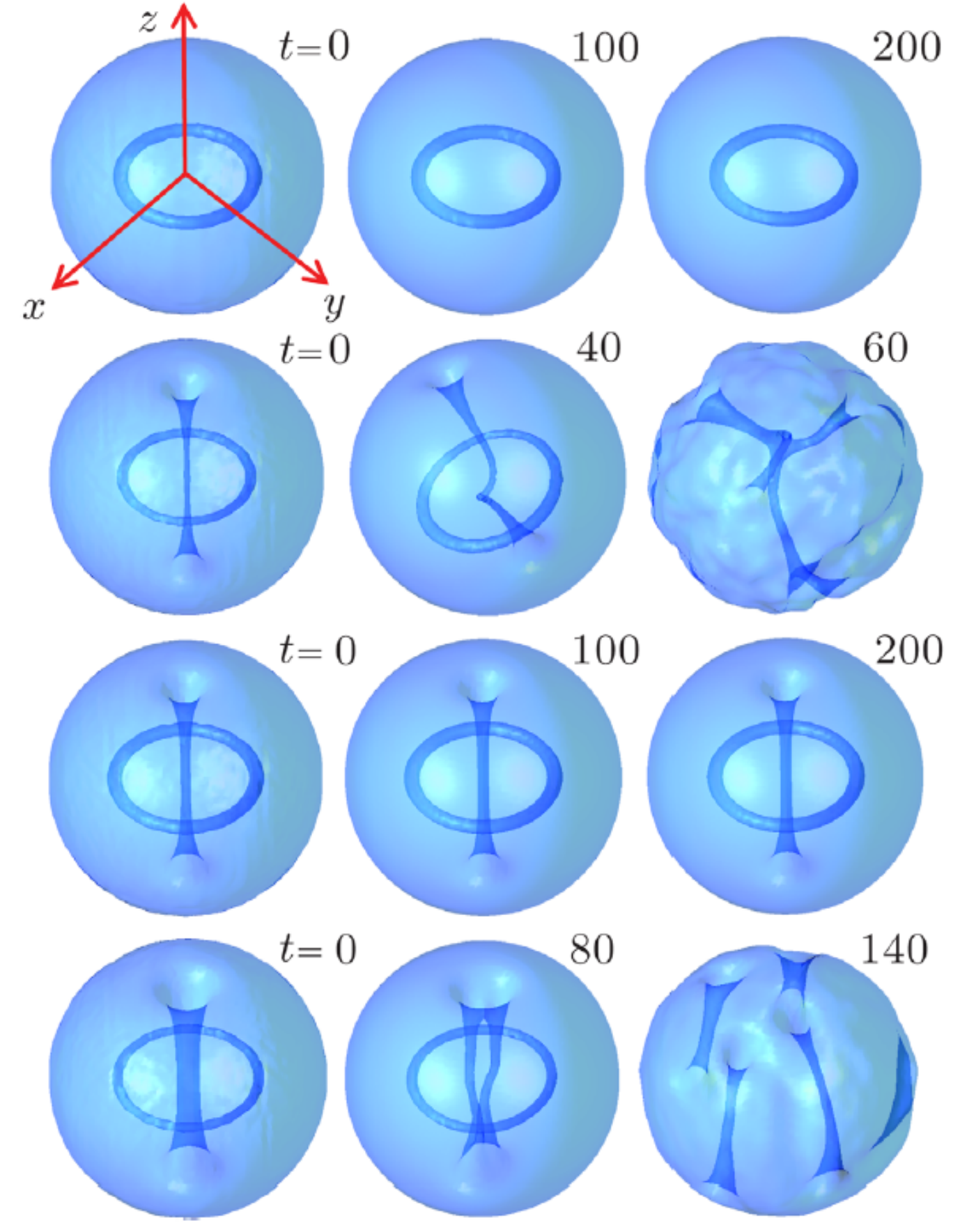}
\end{center}
\caption{Four rows display, from left to right, the distribution of the
density in hopfion stationary solutions of Eq. (6.33), and their evolution,
governed by the underlying equations (6.1) and (6.14) (with $\protect\alpha %
=1/2$). Parameters (twist number $s$ of the intrinsic torus, overall
vorticity $m$, and chemical potential $\protect\mu $) are, from the top to
bottom: $\left( s=1,m=0,\protect\mu =15\right) $, $\left( s=1,m=1,\protect%
\mu =16\right) $, $\left( s=1,m=1,\protect\mu =22\right) $, and $\left(
s=1,m=2,\protect\mu =22\right) $. Snapshots of the evolving hopfions are
taken at moments of time indicated in the figure. In the second row, the
density isosurface is drawn as $\left\vert \protect\psi \left( x,y,z\right)
\right\vert ^{2}=0.2$, and in all others is it $\left\vert \protect\psi %
\left( x,y,z\right) \right\vert ^{2}=1$ (source: Kartashov \textit{et al}.,
2014).}
\label{fig6.11}
\end{figure}

In usual settings, the creation of hopfions requires, at the minimum level,
two complex components (Sutcliffe and Manton, 2004; Sutcliffe, 2007). A
unique possibility offered by the 3D equation (6.1) with the modulation
profile (6.14) is to predict single-component stable hopfions (Kartashov
\textit{et al}., 2014). The respective stationary solutions were looked for,
in the cylindrical coordinates $\left( r,\theta ,z\right) $, as%
\begin{equation}
\psi =\exp \left( -i\mu t+im\theta \right) \,w\left( r,z\right) ,  \tag{6.32}
\end{equation}%
with complex field $w$ satisfying equation%
\begin{equation}
\mu w=-\frac{1}{2}\left( \frac{\partial ^{2}}{\partial r^{2}}+\frac{1}{r}%
\frac{\partial }{\partial r}-\frac{m^{2}}{r^{2}}\right) w+\exp \left( \frac{1%
}{2}\left( r^{2}+z^{2}\right) \right) |w|^{2}w.  \tag{6.33}
\end{equation}%
Examples of hopfions with sets of topological numbers $\left( s=1,m=0\right)
$, $\left( s=1,m=1\right) ,$ $\left( s=1,m=1\right) $, and $\left(
s=1,m=2\right) $, found as numerical stationary solutions of Eq. (6.1), are
shown, from the top to bottom, in the left images of Fig. \ref{fig6.11}.
Solutions with $s>1$ do not exist (note that the solution with $S=0$ and $%
s=1 $, i.e., zero overall vorticity, exists, as shown in the top row of Fig. %
\ref{fig6.11}). Further, the evolution of the hopfions, displayed in Fig. %
\ref{fig6.11}, demonstrates that all the hopfions with $\left(
s=1,m=0\right) $ are stable, ones with $\left( s=m=1\right) $ may be stable
or unstable, depending on their chemical potential (or norm), and all the
states with $\left( s=1,m\geq 2\right) $ are completely unstable against
spontaneous breakup (the results for the hopfions with $s=-1$ are the same).

Shapes of the hopfions displayed in Fig. \ref{fig6.11} suggest that the
respective solution to Eq. (6.26) may be approximated by the following
ansatz, proposed by Kartashov \textit{et al}. (2014) (assuming $s\geq 0$ and
$m\geq 0$):%
\begin{equation}
w=A\left[ \left( r-R\right) +iz\right] ^{s}r^{m}\exp \left[ -\frac{1}{2}%
\left( r^{2}+z^{2}\right) \right] .  \tag{6.34}
\end{equation}%
Here variational parameter $R$ is the radius of the coiled core of the
hopfion, at which the field must vanish, in the midplane drawn through $z=0$%
, due to the presence of the vorticity with winding number $s$ in the cross
section of the torus. The norm of this ansatz is%
\begin{equation}
N=\left( 2\pi \right) ^{3/2}A^{2}\left( R^{2}-\sqrt{2\pi }R+3\right) .
\tag{6.35}
\end{equation}

The Lagrangian of Eq. (6.33) is (cf. Lagrangian (6.17) for the spherically
symmetric real field):%
\begin{equation}
L_{w}=\int_{0}^{\infty }rdr\int_{-\infty }^{+\infty }dz\left[ \left( \mu -%
\frac{m^{2}}{2r^{2}}\right) |w|^{2}-\frac{1}{2}\left( \left\vert \frac{%
\partial w}{\partial r}\right\vert ^{2}+\left\vert \frac{\partial w}{%
\partial z}\right\vert ^{2}\right) -\frac{1}{2}\exp \left( \frac{1}{2}\left(
r^{2}+z^{2}\right) \right) |w|^{4}\right] .  \tag{6.36}
\end{equation}%
The substitution of ansatz (6.33) in expression (6.36) yields the respective
effective Lagrangian, in which $A^{2}$ is replaced by $N$ according to Eq.
(6.35). The ensuing Euler-Lagrange equation, $\partial \left( L_{w}\right) _{%
\mathrm{eff}}/\partial R=0$, leads to a cumbersome but usable equation which
determines $R$ as a function of $N$. It takes a rather simple form for
\textquotedblleft heavy" hopfions, in the limit of $N\rightarrow \infty $
(which may be considered as the TF limit):%
\begin{equation}
2\left( 4-\pi \right) R_{\infty }^{3}-3\sqrt{2\pi }R_{\infty }^{2}+4(2\pi
-3)R_{\infty }=3\sqrt{2\pi }.  \tag{6.37}
\end{equation}%
A relevant root of Eq. (6.37) is%
\begin{equation}
R_{\infty }\approx 1.06.  \tag{6.38}
\end{equation}%
$\allowbreak $

\begin{figure}[tbp]
\begin{center}
\includegraphics[width=0.62\textwidth]{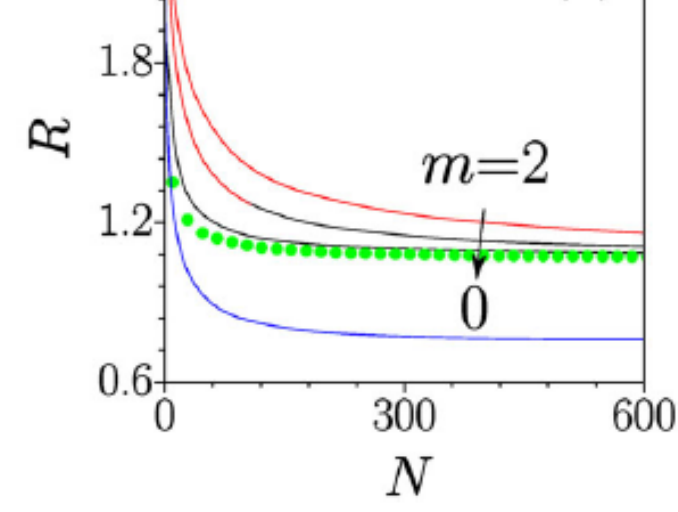}
\end{center}
\caption{Continuous black and red lines represent dependences of radius $R$
of the coiled axis of the hopfion, along which the wave function vanishes,
on its norm, as obtained form the numerical solution of Eq. (6.33). Black
and red segments correspond, respectively, to stable and unstable hopfions
(the stability boundary splits the solution branch with $s=m=1$ as per Eq.
(6.40)). The chain of green dots represents the VA prediction based on
ansatz (6.34). The lower blue curve shows the $R(N)$ dependence for a stable
hopfion family with $s=1$ and $m=0$, produced by Eq. (6.1) with a smoother
spherically isotropic modulation profile, $\protect\sigma (r)=1+r^{6}$,
instead of one defined by Eq. (6.14) (source: Kartashov \textit{et al}.,
2014).}
\label{fig6.12}
\end{figure}

Numerical and analytical findings for hopfion families are presented in Fig. %
\ref{fig6.12} by means of $R(N)$ dependences, which represent the most
relevant property of the states. The figure includes results for the
stability, produced by means of both numerical solution of the respective
BgG equations and direct simulations of the perturbed evolution in the
framework of Eq. (6.1) (in particular, some results of the simulations are
displayed in Fig. \ref{fig6.12}). It is seen that the VA prediction for $%
R(N) $ becomes very close to its numerical counterpart at $N\gtrsim 100$.
Actually, Eq. (6.38) yields an \emph{asymptotically exact} value of the
radius of the coiled core of the hopfion in the limit of $N\rightarrow
\infty $. An additional prediction of the VA, which is also very close to
the numerical findings, is that the radius of the cross section of the
hopfion's toroidal core shrinks with the increase of $N$ (unlike the
asymptotically constant overall radius of the core, given by Eq. (6.38)):
\begin{equation}
\rho _{\mathrm{core}}\approx \left( 2\pi ^{3}\right) ^{1/4}/\sqrt{N}.
\tag{6.39}
\end{equation}%

As concerns the stability, Fig. \ref{fig6.12} demonstrates, as mentioned
above, that the hopfions with $s=1,m=0$ are completely stable, the ones with
$s=m=1$ are stable at
\begin{equation}
\mu \geq \mu _{\min }\approx 11.2,  \tag{6.40}
\end{equation}%
and the modes with $s=1,m\geq 2$ are completely unstable.

\begin{figure}[tbp]
\begin{center}
\includegraphics[width=0.80\textwidth]{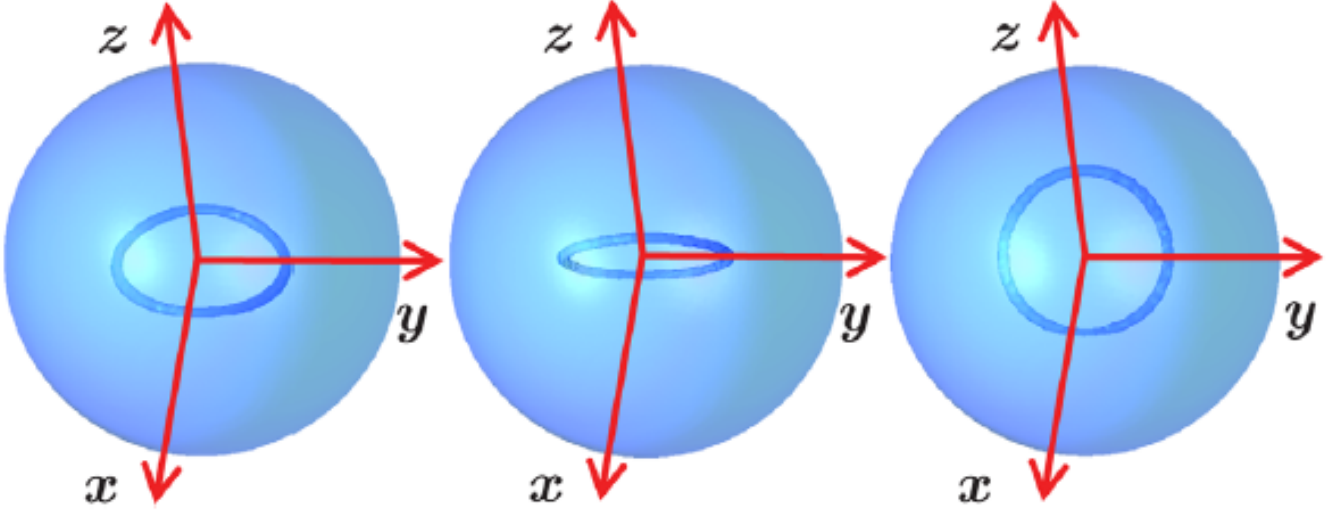}
\end{center}
\caption{Stable rotation of the hopfion with $s=1,m=0$, initiated by
multiplying the stationary wave function by torque factor (6.30) with $%
\protect\beta =x_{0}=5$. Densities isosurfaces, $\left\vert \protect\psi %
\left( x,y,z\right) \right\vert ^{2}=0.2$, are plotted at times $t=141$, $156
$, and $177,$ the rotation period being $T\approx 48$ (source: Kartashov
\textit{et al}., 2014).}
\label{fig6.13}
\end{figure}

Finally, the application of torque (6.30), with the axis belonging to the
plane of the coiled axis of the hopfion, may readily set the stable hopfion
in mechanical rotation around the torque's axis. An example is displayed in
Fig. \ref{fig6.13} for the hopfion with $s=1,m=0$.

\subsection{Vortex-antivortex hybrids in the peanut-shaped modulation profile%
}

\subsubsection{Composite modes with opposite vorticities}

Another species of unusual 3D self-trapped states is supported by Eq. (6.1)
in which the anti-Gaussian isotropic modulation profile (6.14) is replaced
by a \textquotedblleft peanut"-shaped one, which is stretched along axis $z$
by distance $d$. It was introduced by Driben \textit{et al}. (1914b) in
terms of the cylindrical coordinates:%
\begin{equation}
\sigma \left( r,z\right) =\exp \left[ \frac{1}{2}\left( r^{2}+\left( |z|-%
\frac{d}{2}\right) ^{2}\right) \right] .  \tag{6.41}
\end{equation}%
If $d$ is large enough, this modulation profile makes it possible to create
\textquotedblleft hybrid states", alias composite ones, in the form of
vortex tori with \emph{opposite winding numbers}, $S=\pm 1$, trapped in the
top and bottom lobes of the \textquotedblleft peanut". Examples of both the
usual vortex mode, with winding numbers $S=1$ in both lobes, i.e., with%
\begin{equation}
\psi =e^{iS\theta }w\left( r,z)\right) ,  \tag{6.42}
\end{equation}%
and hybrids, with vorticities $\pm 1$ and $\pm 2$, are presented in Fig. \ref%
{fig6.14}.
\begin{figure}[tbp]
\begin{center}
\includegraphics[width=0.46\textwidth]{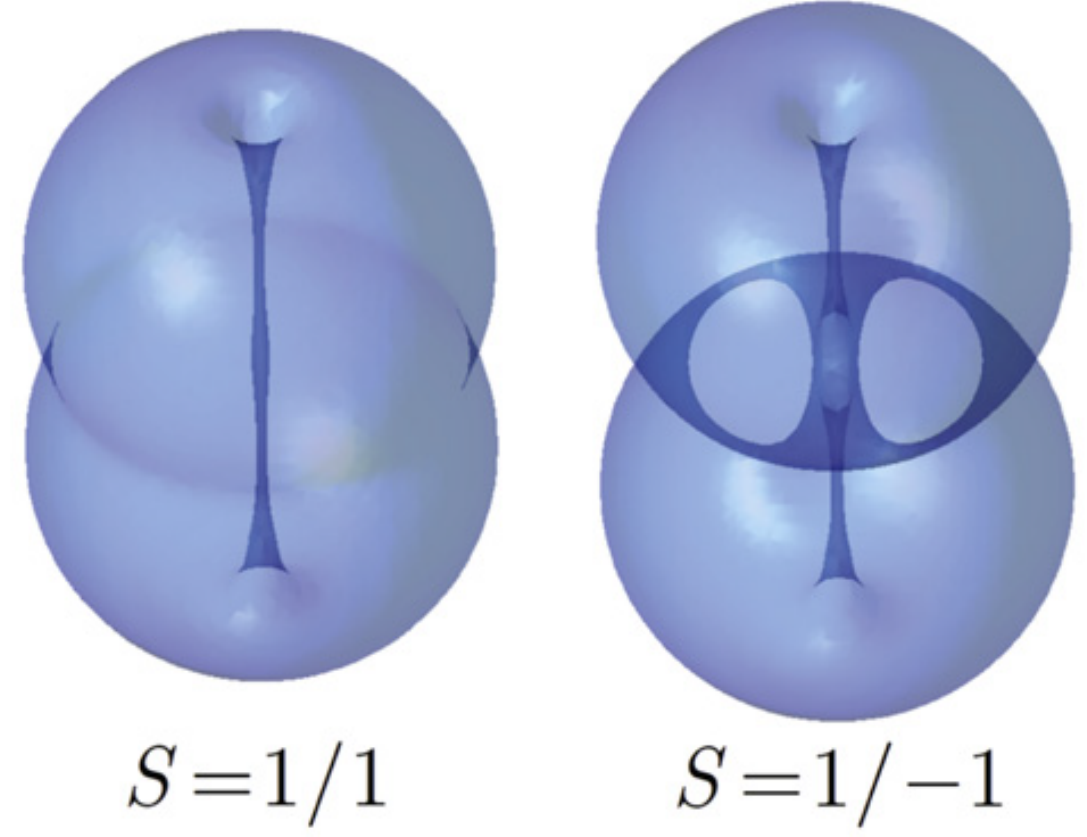} %
\includegraphics[width=0.25\textwidth]{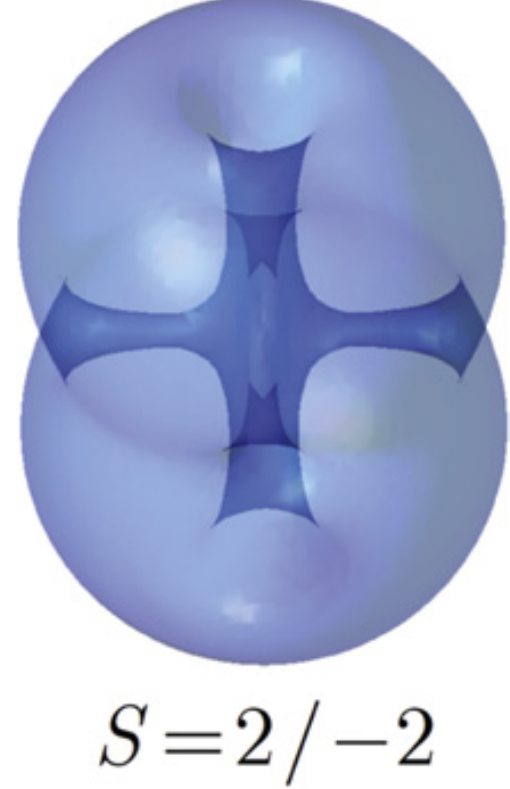}
\end{center}
\caption{Density profiles of 3D vortex modes supported by Eq. (6.1) with the
\textquotedblleft peanut"-shaped modulation profile of the local
nonlinearity strength, defined as per Eq. (6.41). The profiles are shown by
means of isosurfaces $\left\vert \protect\psi \left( x,y,z\right)
\right\vert ^{2}=0.2$. The left panel displays a usual stable vortex state,
taken in the form of Eq. (6.42), with equal winding numbers $S=1$ in both
lobes of the peanut. The central and right panels represent
\textquotedblleft hybrids" (composite modes), with opposite vorticities, $%
S/-S$, in the top and bottom lobes (note the strong breaking of the axial
symmetry exhibited by the hybrids). The left and right panels correspond to $%
d=3$ in Eq. (6.41) and chemical potential $\protect\mu =10$. The central
panel corresponds to $d=5$ and $\protect\mu =7$. The usual and hybrid vortex
modes with vorticities $1$ and $\pm 1$, shown in the left and central
panels, are stable. The hybrid with vorticities $\pm 2$, shown in the right
panel, is unstable (source: Driben\textit{\ et al}., 2014b). }
\label{fig6.14}
\end{figure}

The hybrid states may be represented by an approximate stationary solution
in the following form:%
\begin{equation}
\psi =e^{-i\mu t}\left[ w_{+}\left( r,z\right) e^{iS\theta }+w_{-}\left(
r,z\right) e^{-iS\theta }\right] .  \tag{6.42}
\end{equation}%
The substitution of this ansatz in Eq. (6.1) leads, in the approximation
which neglects angular harmonics $\exp \left( \pm 3iS\theta \right) $, to a
system of two coupled stationary equations for real functions $w_{\pm }$:%
\begin{equation}
\left[ \mu +\frac{1}{2}\left( \frac{\partial ^{2}}{\partial r^{2}}+\frac{1}{r%
}\frac{\partial }{\partial r}-\frac{S^{2}}{r^{2}}+\frac{\partial ^{2}}{%
\partial z^{2}}\right) \right] \phi _{\pm }=\sigma \left( r,z\right) \left[
2\left( \phi _{\mp }\right) ^{2}+\left( \phi _{\pm }\right) ^{2}\right] \phi
_{\pm },  \tag{6.44}
\end{equation}%
where $\sigma \left( r,z\right) $ is taken as per Eq. (6.41). Numerical
solution of Eqs. (6.44) predicts shapes of the hybrids which are quite close
to those produced, as stationary solutions, by full equation (6.1) with the
modulation profile (6.41).

The hybrid modes, combining opposite azimuthal harmonics, break the axial
symmetry of the resulting states. This can be easily seen looking at ansatz
(6.42) in the midplane, $z=0$, where $\phi _{\pm }(r,z=0)$ amount to the
single real function, $\phi _{0}(r)$:%
\begin{equation}
\left\vert \psi \left( r,\theta ,z=0\right) \right\vert ^{2}=4\phi
_{0}^{2}(r)\cos ^{2}\left( S\theta \right) .  \tag{6.45}
\end{equation}%
The strong azimuthal pattern exhibited by Eq. (6.45) (it resembles the
so-called \textit{azimuthon} states known in 2D systems (Desyatnikov,
Sukhorukov, and Kivshar, 2005)) models the actual structure seen in the
middle panel of Fig. 67.

The study of stability of the hybrids was performed by Driben \textit{et al}%
. (2014b) by solving the respective BdG equations for small perturbations,
and also by means of direct simulations of the perturbed evolution. As a
result, it was found that the hybrids with vorticities $\pm 1$ may be stable
provided that the stretching parameter $d$ in Eq. (6.41) exceeds a minimum
value,%
\begin{equation}
d>\left( d_{\min }\right) _{S=+1,-1}\approx 4.  \tag{6.46}
\end{equation}%
In particular, the hybrids are stable at $\mu <15.8$ (alias $N<394.9$) at $%
d=4.5$, and at $\mu <13.5$ ($N<329.3$) at $d=5$. The instability of the
hybrids at larger values of $\mu $ is explained by the fact that (as shown,
in particular, by the TF approximation -- see Eq. (6.37)) the increase of $%
\mu $ makes the localized modes broader, hence the stretching parameter $d$
becomes relatively smaller, failing to produce the stabilization. In direct
simulations, those hybrids with $S=\pm 1$ which are unstable develop the
instability in the form of corrugation of the mode's central axis. All
hybrids with vorticities $\pm S$ are unstable for $S\geq 2$.

\subsubsection{Fundamental-vortex complexes}

The same model, based on Eqs. (6.1) and (6.41), admits dynamical composite
modes, in the form of fundamental and vortex states juxtaposed in the bottom
and top lobes of the peanut-shaped trapping profile. As shown in Fig. \ref%
{fig6.15}, this complex may readily form a state in which the vortex
component with $S_{1}=1$ performs completely stable periodic precession on
top of the fundamental \textquotedblleft base", with $S_{2}=0$. A broad
stability area for the precessing composite states was found provided that $%
d $ exceeds a minimum value,
\begin{equation}
d>\left( d_{\min }\right) _{S_{1,2}=1,0}\approx 4.8,  \tag{6.47}
\end{equation}%
cf. Eq. (6.46).
\begin{figure}[tbp]
\begin{center}
\includegraphics[width=0.62\textwidth]{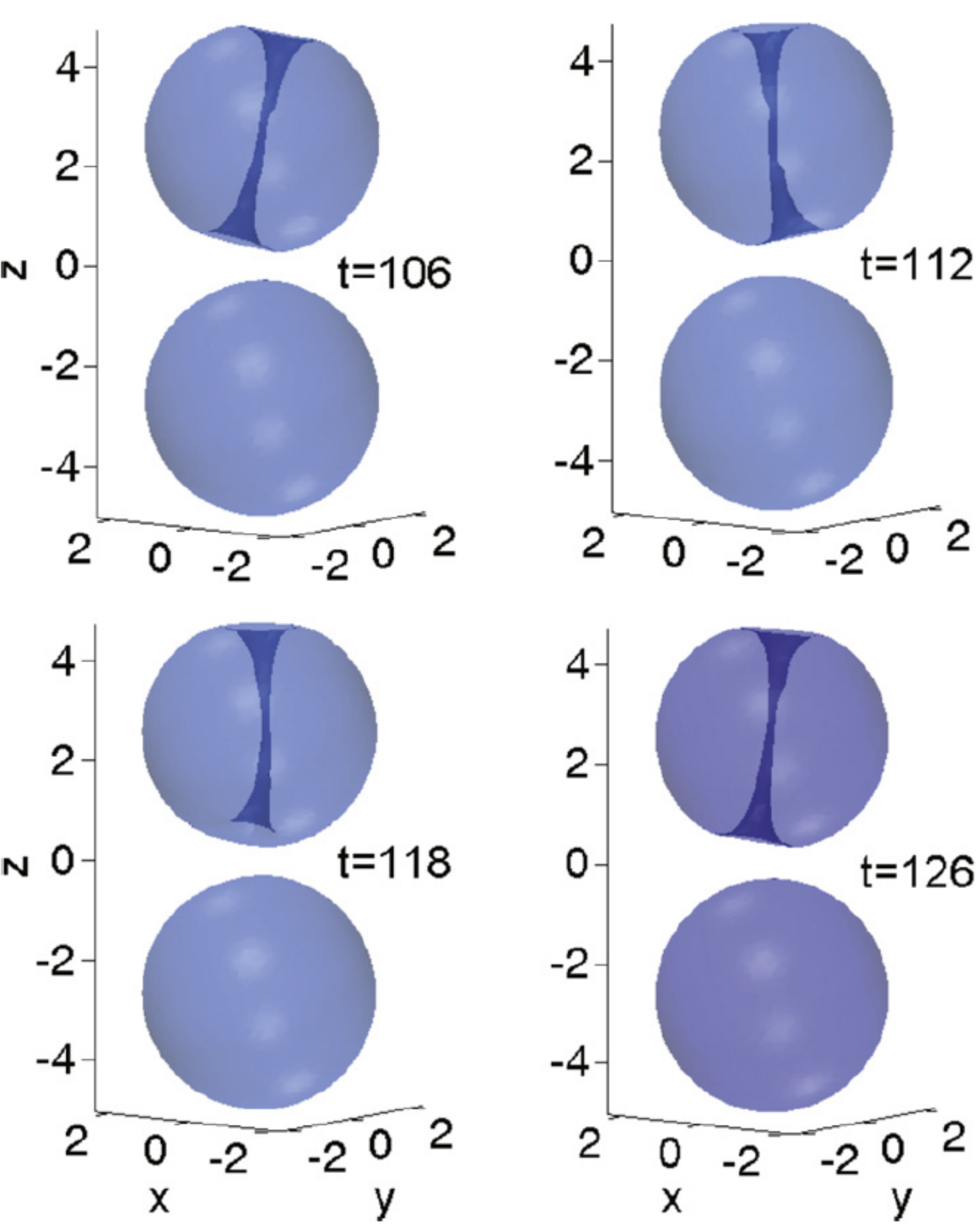}
\end{center}
\caption{An example of robust periodic precession of a vortex mode with $S=1$
on top of the immobile zero-vorticity one, which are trapped, respectively,
in the top and bottom lobes of the \textquotedblleft peanut-shaped"
structure (6.41) with $d=5$. The initial state was built taking halves of
the respective stationary states with $\protect\mu =15$. The motion is
displayed by means of density isosurfaces $\left\vert \protect\psi \left(
x,y,z\right) \right\vert ^{2}=1$, produced by simulations of Eq. (6.1). In
the present case, the period of the precession is $\approx 20$ (source:
Driben \textit{et al}., 2014b).}
\label{fig6.15}
\end{figure}

\subsection{Weakly localized dark vortices}

As mentioned above, the fundamental condition which secures the convergence
of the norm of localized modes produced by Eqs. (6.1) and (6.6) is given by
Eq. (6.8). If this condition does not hold, Eq. (6) gives rise to weakly
localized states with the divergent norm, which may be considered as \textit{%
localized dark solitons} (Zeng and Malomed, 2017), as they share the
property of the divergence of the norm with dark solitons in usual
(free-space) models.

Localized dark solitons were considered by Zeng and Malomed (2017) in the 2D
version of Eq. (6.1) with the modulation profile%
\begin{equation}
\sigma (r)=r^{\alpha },~\alpha \leq 2,  \tag{6.48}
\end{equation}%
cf. Eq. (6.10). The weakly localized solutions to Eq. (6.1) with vorticity $%
S $ were looked for, in the polar coordinates, as%
\begin{equation}
\psi =\exp \left( -i\mu t+iS\theta \right) w(r),  \tag{6.49}
\end{equation}%
(essentially the same as Eq. (6.11)), where real function $w(r)$ satisfies
the equation%
\begin{equation}
\mu w+\frac{1}{2}\left( \frac{d^{2}w}{dr^{2}}+\frac{1}{r}\frac{dw}{dr}-\frac{%
S^{2}}{r^{2}}w\right) -r^{\alpha }w^{3}=0,  \tag{6.50}
\end{equation}%
cf. Eq. (6.12). The asymptotic form of the solution at $r\rightarrow
\infty $ is%
\begin{equation}
w=\sqrt{\mu }r^{-\alpha /2}+\frac{1}{4\sqrt{\mu }}\left( \frac{\alpha ^{2}}{4%
}-S^{2}\right) r^{-\alpha /2-2}+\mathcal{O}\left( r^{-\alpha /2-4}\right) ,
\tag{6.51}
\end{equation}%
where the first term is tantamount to the TF approximation, cf. Eq. (6.9).
Solutions (6.49) with $S=0$ and $S\neq 0$ may be called, respectively,
\textit{localized CW states} and \textit{localized dark vortices}.

First, in the case of $\alpha =2$, which corresponds to the boundary between
localized dark and bright states, Eq. (6.50) with $S=1$ gives rise to
simple exact solutions, for all values $\mu >0$ (although these dark-vortex
solutions are unstable, see Fig. \ref{fig6.17} below):
\begin{equation}
w(r;\alpha =2;S=1)=\sqrt{\mu }r^{-1}.  \tag{6.52}
\end{equation}%
Obviously, the divergence of the integral norm of solutions to Eq. (6.50) is
logarithmic at $r\rightarrow \infty $.

\begin{figure}[tbp]
\begin{center}
\includegraphics[width=0.70\textwidth]{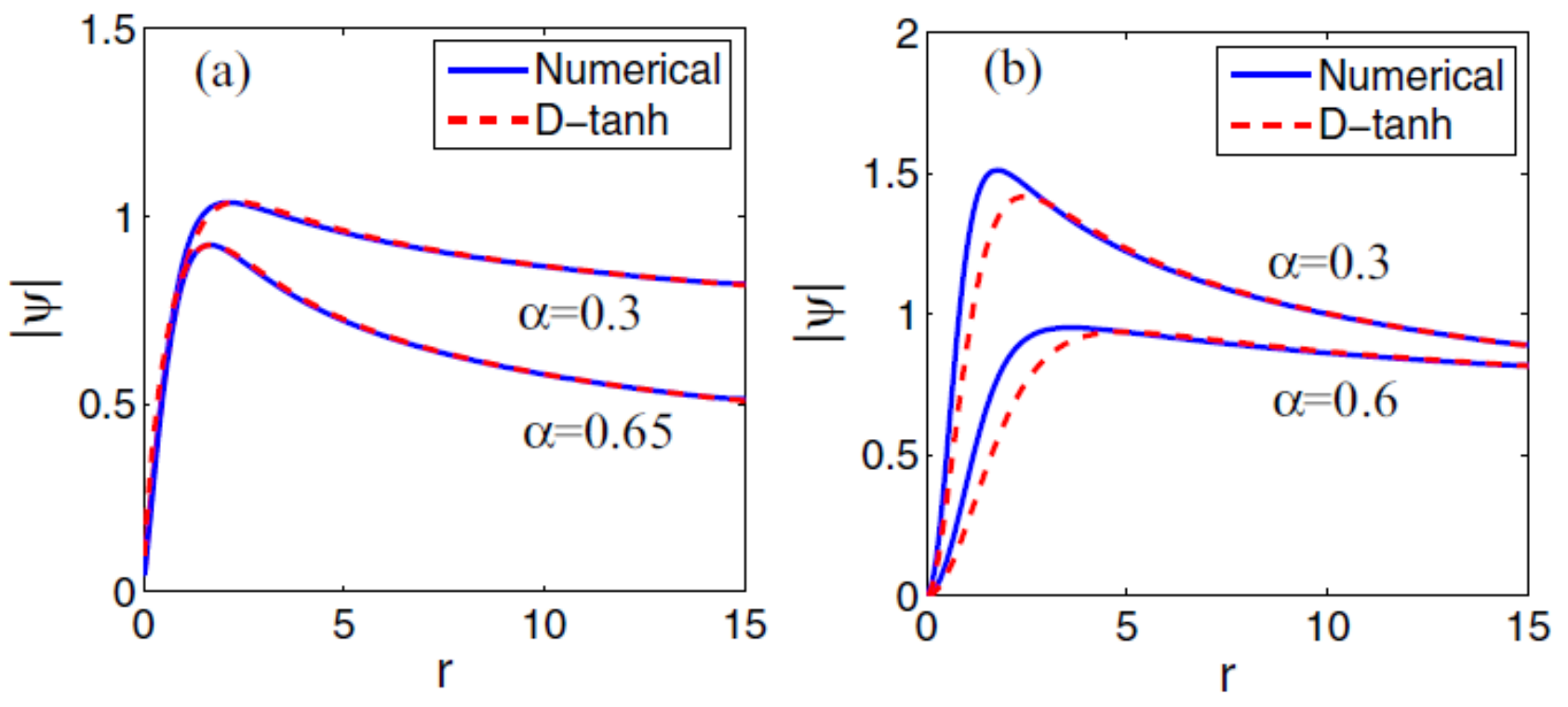}
\end{center}
\caption{Blue lines show examples of localized dark vortices, produced by
the numerical solution of Eq. (6.50) with values of $\protect\alpha $
indicated in the figure, and vorticities $S=1$ (a) and $S=2$ (b). The
chemical potential is $\protect\mu =4$ for $\protect\alpha =0.6$ in (b), and
$\protect\mu =1.5$ for other solutions. Red dashed lines show the analytical
approximation given by Eq. (6.53), with fitting factors $\protect\lambda %
=1.1 $ in (a), and $\protect\lambda =1.5$ in (b). The vortices with $S=1$
shown in (a) are stable, while the double vortices in (b) are weakly
unstable (source: Zeng and Malomed, 2017).}
\label{fig6.16}
\end{figure}

The dark CW states may be quite accurately approximated by the TF
approximation (which corresponds to the first term on the right-hand side of
Eq. (6.51)) multiplied by a factor which accounts for vorticity $S\geq 1$:
\begin{equation}
w(r)=\sqrt{\mu }r^{-\alpha /2}\left[ \tanh (\lambda r)\right] ^{S},
\tag{6.53}
\end{equation}%
where $\lambda $ is a fitting parameter. Typical radial profiles of the
localized dark vortices, along with approximation (6.53), are displayed, for
$S=1$ and $2$, in Fig. \ref{fig6.16}.

Stability of the localized CW states and dark vortices was identified by
means of systematic simulations of their perturbed evolution in the
framework of the Eq. (6.1). The conclusion is that the CW states are
completely stable (actually, they realize the system's ground state), while
all the localized dark vortices with $S\geq 2$ are unstable. The analysis
for vortices with $S=1$ produces a nontrivial stability boundary, which is
shown in Fig. \ref{fig6.17}. It shows that the increase of both $\mu $ and $%
\alpha $ leads to destabilization of the dark vortex. Actually, they may be
stable only at
\begin{equation}
\alpha <\alpha _{\max }\approx 0.55,  \tag{6.54}
\end{equation}%
i.e., if the the dark vortex is sufficiently stretched in the radial
direction (as seen from the approximate expression (6.53)).
\begin{figure}[tbp]
\begin{center}
\includegraphics[width=0.40\textwidth]{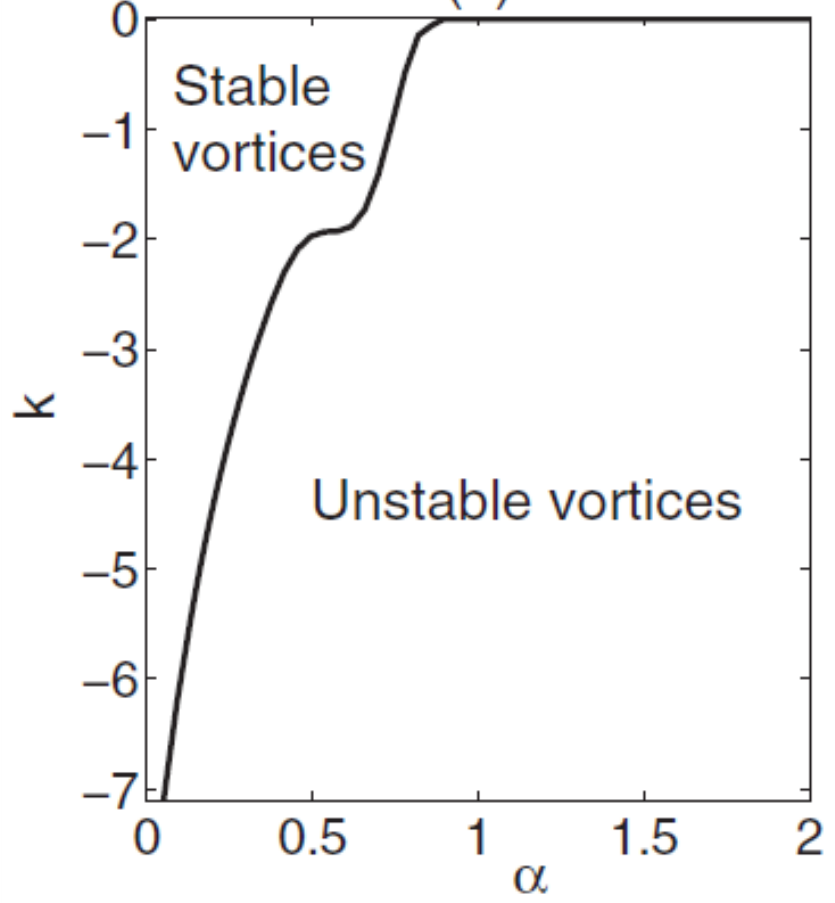}
\end{center}
\caption{The stability chart for the localized dark vortices with $S=1$,
produced by Eq. (6.50). In this plot, notation $k\equiv -\protect\mu $ is
used. The vortices are completely unstable at $\protect\alpha >0.55,$ see
Eq. (6.54). All the localized dark states with $S=0$ (CW) are completely
stable, while all vortices with $S\geq 2$ are unstable (source: Zeng and
Malomed, 2017).}
\label{fig6.17}
\end{figure}

\section{Conclusion}


This chapter produces a systematic summary of theoretical results, both
numerical and analytical ones, obtained in numerous works aimed at the
search for physically relevant settings which allow one to create stable 2D
and 3D solitons, including topological modes with embedded vorticity, as
well as complex 3D modes in the form of \textit{hopfions}, which carry two
independent topological charges. Stability is the main challenge in the
studies of multidimensional solitons because, on the contrary to diverse 1D
settings, in which solitons typically emerge as stable objects, they tend to
be strongly unstable in basic 2D and 3D situations. In particular, the
ubiquitous NLS (nonlinear Schr\"{o}dinger) equation with the cubic
self-focusing term (essentially the same PDE is known as the GP
(Gross-Pitaevskii) equation in the BEC theory) creates only unstable
solitons in 2D and 3D spaces alike, because exactly the same equation gives
rise to the destructive effects in the form of the critical and
supercritical wave collapse in 2D and 3D cases, respectively.

The present chapter is focused on physically relevant multidimensional
models of the NLS/GP type. The chapter offers a relatively detailed review
of two \ basic topics. First, it is a general survey of various schemes
which have been elaborated to secure stability of 2D and 3D solitons,
including those with embedded vorticity. The main stabilization schemes
outlined here are: (i) competing (e.g., cubic-quintic) and saturable
nonlinearities; (2) linear and nonlinear trapping potentials; (3) the
stabilizing effect provided by the LHY (Lee-Huang-Yang) corrections to the
mean-field BEC dynamics, leading to the formation of stable QDs (quantum
droplets), including ones with embedded vorticity; (4) SOC
(spin-orbit-coupling) effects in two- and three-dimensional binary BEC; and
(5) emulation of SOC in nonlinear optical waveguides, including $\mathcal{PT}
$-symmetric ones.

The second topic addressed in the chapter is a detailed summary of results
which demonstrate the creation of stable 2D and 3D solitons by means of
schemes based on the usual linear trapping potentials or effective nonlinear
ones. Nonlinear potentials may be induced by means of spatial modulation of
the local strength of the nonlinear term. The latter option is especially
promising, making it possible to use self-defocusing media, with the local
nonlinearity strength growing fast enough from the center to periphery, for
the creation of a great variety of multidimensional self-trapped modes. In
addition to fundamental quasi-solitons and vortex rings, the 3D modes may be
hopfions, i.e., twisted vortex rings. A remarkable fact is that many
essential results for the multidimensional solitons can be obtained, in such
settings, not only in a numerical form, but also by means of efficient
analytical methods, exact and approximate ones. In particular, basic
properties of such complex objects as hopfions are accurately predicted by
the analytical variational approximation.

The work on the vast topic surveyed in this chapter is far from being
completed. Among challenging open problems is, for instance, a possibility
to create stable complex 3D structures in the form of two linked rings, each
one being a vortex torus or, possibly, a hopfion. The greatest challenge is
to implement many theoretical predictions, outlined in the present review,
in real physical experiments -- especially in BEC\ and nonlinear optics.%
\newline

\section*{Acknowledgments}

I would like to thank my coauthors, in collaboration with whom many results
included in this review have been produced: Gregory Astrakharchik, Bakhtiyor
Baizakov, Milivoj Beli\'{c}, Marijana Brtka, Olga Borovkova, George Boudebs,
Gennadiy Burlak, Cid de Ara\'{u}jo, Zhaopin Chen, Guangjiong Dong, Shenhe
Fu, Arnaldo Gammal, Yaroslav Kartashov, Vladimir Konotop, Herv\'{e} Leblond,
Falk Lederer, Ben Li, Yongyao Li, Valery Lobanov, Dumitru Mazilu, Torsten
Meier, Dumitru Mihalache, Michele Modugno, Wei Pang, Dmitry Pelinovsky, Han
Pu, Jieli Qin, Albert Reyna, Hidetsugu Sakaguchi, Mario Salerno, Elad
Shamriz, Evgeny Sherman, Yasha Shnir, Leticia Tarruell, Lluis Torner, Viktor
Vysloukh, Li Wang, Zhenya Yan, and Jianhua Zeng.

My thanks are to Prof. Weizhu Bao who has invited me to produce this
chapter. My work in this research area was supported, in part, by the Israel
Science Foundation through grant No. 1286/17.


\begin{thebibliography}{999}
\bibitem{1D Townes} Abdullaev F. Kh. and M. Salerno, \textquotedblleft
Gap-Townes solitons and localized excitations in low-dimensional
Bose-Einstein condensates in optical lattices", Phys. Rev. A 72, 033617
(2005).

\bibitem{Ablowitz-Segur} Ablowitz M. and H. Segur, \textit{Solitons and the
Inverse Scattering Transform} (SIAM, Philadelphia, 1981).

\bibitem{Adhikari-3D trapped vortex} Adhikari S. K., \textquotedblleft
Collapse of attractive Bose-Einstein condensed vortex states in a
cylindrical trap", Phys. Rev. E \textbf{65}, 016703 (2001).

\bibitem{Adhikari spin-1 SOC} Adhikari S. K., Multiring, stripe, and
superlattice solitons in a spin-orbit-coupled spin-1 condensate, Phys. Rev.
A \textbf{103}, L011301 (2021).

\bibitem{Alexander Berge} Alexander T. J. and L. Berg\'{e},
\textquotedblleft Ground states and vortices of matter-wave condensates and
optical guided waves", Phys. Rev. E \textbf{65}, 026611 (2002).

\bibitem{Anderson-saturable} Anderson D., and M. Bonnedal, \textquotedblleft
Variational approach to nonlinear self-focusing of Gaussian laser beams",
Phys. Fluids \textbf{22}, 105-109 (1979)

\bibitem{Anderson-1983} Anderson D., ``Variational approach to nonlinear
pulse propagation in optical fibers", Phys. Rev. A \textbf{27}, 3135-3144
(1983).

\bibitem{BBB-EPL} Baizakov B. B., B. A. Malomed, and M. Salerno,
\textquotedblleft Multidimensional solitons in periodic potentials".
Europhys. Lett. \textbf{63}, 642-648 (2003).

\bibitem{BBB low-dim} Baizakov B. B., B. A. Malomed, and M. Salerno,
\textquotedblleft Multidimensional solitons in a low-dimensional periodic
potential", Phys. Rev. A \textbf{70}, 053613 (2004).

\bibitem{BBB radial} Baizakov B. B., B. A. Malomed, and M. Salerno,
\textquotedblleft Matter-wave solitons in radially periodic potentials",
Phys. Rev. E \textbf{74}, 066615 (2006).

\bibitem{Townes sol 2-comp Rb-87} Bakkali-Hassani B., C. Maury, Y.-Q. Zou,
E. Le Cerf, R. Saint-Jalm, P. C. M. Castilho, S. Nascimbene, J. Dalibard,
and J. Beugnon, \textquotedblleft Realization of a Townes Soliton in a
Two-Component Planar Bose Gas", Phys. Rev. Lett. \textbf{127}, 023603 (2021).

\bibitem{Weizhu3} Bao W. Z. and Y. Y. Cai, "Mathematical theory and
numerical methods for Bose-Einstein condensation", Kinetic and Related
Models \textbf{6}, 1-135 (2013).

\bibitem{Weizhu2} Bao W. Z. and Q. Du, \textquotedblleft Computing the
ground state solution of Bose-Einstein condensates by a normalized gradient
flow", SIAM J. Sci. Comp. \textbf{25}, 1674-1697 (2004).

\bibitem{Weizhu} Bao W. Z., D. Jaksch, and P. A. Markowich,
\textquotedblleft Numerical solution of the Gross-Pitaevskii equation for
Bose-Einstein condensation", J. Comp. Phys. \textbf{187}, 318-342 (2003).

\bibitem{magn FR light contr} Bauer D. M., M. Lettner, C. Vo, G. Rempe, and
S. D\"{u}rr, \textquotedblleft Control of a magnetic Feshbach resonance with
laser light", Nature Phys. \textbf{5}, 339-342 (2009)

\bibitem{Bender PT} Bender C. M., \textquotedblleft Making sense of
non-Hermitian Hamiltonians", Rep. Prog. Phys. \textbf{70}, 947-1018 (2007).

\bibitem{Berge'} Berg\'{e} L., \textquotedblleft Wave collapse in physics:
principles and applications to light and plasma waves", Phys. Rep. \textbf{%
303}, 259-370 (1998).

\bibitem{Borovkova1} Borovkova O. V., Y. V. Kartashov, B. A. Malomed, and L.
Torner, Algebraic bright and vortex solitons in defocusing media, Opt. Lett.
\textbf{36}, 3088-3090 (2011a).

\bibitem{Borovkova2} Borovkova O. V., Y. V. Kartashov, L. Torner, and B. A.
Malomed, Bright solitons from defocusing nonlinearities, Phys. Rev. E
\textbf{84}, 035602 (R) (2011b).

\bibitem{chalcogenide-Angers} G. Boudebs, S. Cherukulappurath, H. Leblond,
J. Troles, F. Smektala, and F. Sanchez, \textquotedblleft Experimental and
theoretical study of higher-order nonlinearities in chalcogenide glasses",
Opt. Commun. \textbf{219}, 427-432 (2003).

\bibitem{Huler first} Bradley C. C., C. A. Sackett, J. J. Tollett, and R. G.
Hulet, \textquotedblleft Evidence of Bose-Einstein condensation in an atomic
gas with attractive interactions", Phys. Rev. Lett. 75, 1687-1690 (1995);
Erratum: Phys. Rev. Lett. \textbf{79}, 1170 (1997).

\bibitem{Brazhnyi Konotop} Brazhnyi V. A. and V. V. Konotop, Theory of
nonlinear matter waves in optical lattices, Mod. Phys. Lett. B \textbf{18},
627-651 (2004).

\bibitem{Brtka} Brtka M., A. Gammal, and B. A. Malomed, \textquotedblleft
Hidden vorticity in binary Bose-Einstein condensates", Phys. Rev. A \textbf{%
82}, 053610 (2010).

\bibitem{Burlak 2D PT coupler} Burlak G. and B. A. Malomed,
\textquotedblleft Stability boundary and collisions of two-dimensional
solitons in $\mathcal{PT}$-symmetric couplers with the cubic-quintic
nonlinearity", Phys. Rev. E \textbf{88}, 062904 (2013).

\bibitem{Buryak} Buryak A. V., P. Di Trapani, D. V. Skryabin, and S. Trillo,
\textquotedblleft Optical solitons due to quadratic nonlinearities: from
basic physics to futuristic applications", Phys. Rep. \textbf{370}, 63-235
(2002).

\bibitem{Rashba} Bychkov Yu. A. and E. I. Rashba, \textquotedblleft
Oscillatory effects and the magnetic susceptibility of carriers in inversion
layers", J. Phys. C: Solid State Phys. \textbf{17}, 6039-6045 (1984).

\bibitem{Leticia1} Cabrera C., L. Tanzi, J. Sanz, B. Naylor, P. Thomas, P.
Cheiney, and L. Tarruell, \textquotedblleft Quantum liquid droplets in a
mixture of Bose-Einstein condensates", Science \textbf{359}, 301-304 (2018).

\bibitem{Calogero-Degasperis} Calogero F, and A. Degasperis, \textit{%
Spectral Transform and Solitons: Tools to Solve and Investigate Nonlinear
Evolution Equations} (North-Holland, Amsterdam, 1982).

\bibitem{Clark and Carr} Carr L. D. and C. W. Clark, \textquotedblleft
Vortices in attractive Bose-Einstein condensates in two dimensions", Phys.
Rev. Lett. \textbf{97}, 010403 (2006).

\bibitem{Leticia2} Cheiney P., C. R. Cabrera, J. Sanz, B. Naylor, L. Tanzi,
and L. Tarruell, \textquotedblleft Bright soliton to quantum droplet
transition in a mixture of Bose-Einstein condensates", Phys. Rev. Lett.
\textbf{120}, 135301 (2018).

\bibitem{Chen Hung} Chen C.-A. and C.-L. Hung, ``Observation of Universal
Quench Dynamics and Townes Soliton Formation from Modulational Instability
in Two-Dimensional Bose Gases", Phys. Rev. Lett. 125, 250401 (2020).

\bibitem{Chen Hung2} Chen C.-A. and C.-L. Hung, \textquotedblleft
Observation of scale invariance in two-dimensional matter-wave Townes
solitons", Phys. Rev. Lett. \textbf{127}, 023604 (2021).

\bibitem{Chen-Liu old} Chen H.-H. and C.-H. Liu, \textquotedblleft Solitons
in nonuniform media", Phys. Rev. Lett. \textbf{37}, 693-697 (1976).

\bibitem{Townes} Chiao R. Y., E. Garmire, and C. H. Townes, ``Self-trapping
of optical beams", Phys. Rev. Lett. \textbf{13}, 479-482 (1964).

\bibitem{Chin } Chin C., R. Grimm, P. Julienne, and E. Tiesinga, ``Feshbach
resonances in ultracold gases", Rev. Mod. Phys. \textbf{82}, 1225-1286
(2010).

\bibitem{Clark opt FR} Clark L. W., L.-C. Ha, C.-Y. Xu, and C. Chin,
\textquotedblleft Quantum dynamics with spatiotemporal control of
interactions in a stable Bose-Einstein condensate", Phys. Rev. Lett. \textbf{%
115}, 155301 (2015).

\bibitem{semicond doped glass CQ 2} Coutaz J. L. and M. Kull,
\textquotedblleft Saturation of the nonlinear index of refraction in
semiconductor-doped glass", J. Opt. Soc. Am. B \textbf{8}, 95-98 (1991).

\bibitem{Dauxois-Peyrard} Dauxois T. and M. Peyrard, \textit{Physics of
Solitons} (Cambridge University Press, Cambridge, 2006), ISBN 0-521-85421-0.

\bibitem{Davydova-Yakimenko} Davydova T. A. and A. I. Yakimenko,
\textquotedblleft Stable multicharged localized optical vortices in
cubic--quintic nonlinear media", J. Optics A: Pure Appl. Opt. \textbf{6},
S197-S201 (2004).

\bibitem{hetero QD} D'Errico, C., A. Burchianti, M. Prevedelli, L.
Salasnich, F. Ancilotto, M. Modugno, F. Minardi, and C. Fort,
\textquotedblleft Observation of quantum droplets in a heteronuclear bosonic
mixture", Phys. Rev. Res. \textbf{1}, 033155 (2019).

\bibitem{Desaix} Desaix M., D. Anderson, and M. Lisak, ``Variational
approach to collapse of optical pulses", J. Opt. Soc. Am. B \textbf{8},
2082-2086 (1991).

\bibitem{azimuthon} Desyatnikov, A. S., A. A. Sukhorukov, and Y. S. Kivshar,
\textquotedblleft Azimuthons: Spatially modulated vortex solitons", Phys.
Rev. Lett. \textbf{95}, 203904 (2005).

\bibitem{cascaded CQ} Dolgaleva K., H. Shin, and R. W. Boyd,
\textquotedblleft Observation of a microscopic cascaded contribution to the
fifth-order nonlinear susceptibility", Phys. Rev. Lett. \textbf{103}, 113902
(2009).

\bibitem{Dresselhaus} Dresselhaus G., \textquotedblleft Spin-orbit coupling
effects in zinc blende structures", Phys. Rev. \textbf{100}, 580-586 (1955).

\bibitem{Radik PT coupler} Driben R. and B. A. Malomed, \textquotedblleft
Stability of solitons in parity-time-symmetric couplers", Opt. Lett. \textbf{%
36}, 4323-4325 (2011).

\bibitem{gyroscopes} Driben R., Y. V. Kartashov, B. A. Malomed, T. Meier,
and L. Torner, \textquotedblleft Soliton gyroscopes in media with spatially
growing repulsive nonlinearity", Phys. Rev. Lett. \textbf{112}, 020404
(2014a).

\bibitem{hybrids} Driben R., Y. Kartashov, B. A. Malomed, T. Meier, and L.
Torner, \textquotedblleft Three-dimensional hybrid vortex solitons", New J.
Phys. \textbf{16}, 063035 (2014b).

\bibitem{Dror spatiotemp vort} Dror N. and B. A. Malomed, \textquotedblleft
Symmetric and asymmetric solitons and vortices in linearly coupled
two-dimensional waveguides with the cubic-quintic nonlinearity", Physica D
\textbf{240,} 526-541 (2011).

\bibitem{Feddersen} Duncan D. B., J. C. Eilbeck, H. Feddersen, and J. A. D.
Wattis, ``Solitons on lattices", Physica D \textbf{68}, 1-11 (1993).

\bibitem{Dutta} Dutta O., M. Gajda, P. Hauke, M. Lewenstein, D.-S. Luhmann,
B. Malomed, T. Sowinski, and J. Zakrzewski, ``Non-standard Hubbard models in
optical lattices: a review", Rep. Prog. Phys. \textbf{78}, 066001 (2015).

\bibitem{Edilson} Edilson L., L. Falc\~{a}o Filho, C. B. de Ara\'{u}jo, G.
Boudebs, H. Leblond, and V. Skarka, \textquotedblleft Robust two-dimensional
spatial solitons in liquid carbon disulfide", Phys. Rev. Lett. \textbf{110},
013901 (2013).

\bibitem{Efremidis-Segev 2D lattice solitons} Efremidis, N. K., J. Hudock,
D. N. Christodoulides, J. W. Fleischer, O. Cohen, and M. Segev
\textquotedblleft Two-dimensional optical lattice
solitons,\textquotedblright\ Phys. Rev. Lett. \textbf{91}, 213905 (2003).

\bibitem{lattice in photorefr} Efremidis N. K., Sears S., Christodoulides D.
N., Fleischer J. W., and M. Segev, \textquotedblleft Discrete solitons in
photorefractive optically induced photonic lattices", Phys. Rev. E \textbf{66%
}, 046602 (2002).

\bibitem{Faddeev} Faddeev L. D., ``Some comments on the many-dimensional
solitons", Lett. Math. Phys. \textbf{1}, 289-293 (1976).

\bibitem{Faddeev Niemi} Faddeev L. D and Niemi A. J., \textquotedblleft
Knots and particles", Nature \textbf{387}, 58-61 (1997).

\bibitem{Fetter-RMP} Fetter A. L., \textquotedblleft Rotating trapped
Bose-Einstein condensates", Rev. Mod. Phys. \textbf{81}, 657-691 (2009).

\bibitem{Fibich} Fibich G., \textit{The Nonlinear Schr\"{o}dinger Equation:
Singular Solutions and Optical Collapse} (Springer, Heidelberg, 2015).

\bibitem{Skryabin} Firth W. J. and D. V. Skryabin, \textquotedblleft Optical
solitons carrying orbital angular momentum", Phys. Rev. Lett. \textbf{79},
2450-2453 (1997).

\bibitem{Hall vort-antivort dipole} Freilich D. V., D. M. Bianchi, A. M.
Kaufman, T. K. Langin, and D. S. Hall, \textquotedblleft Real-time dynamics
of single vortex lines and vortex dipoles in a Bose-Einstein condensate",
Science \textbf{329}, 1182-1185 (2010).

\bibitem{SOC in BEC} Galitski V. and I. B. Spielman, \textquotedblleft
Spin-orbit coupling in quantum gases", Nature \textbf{494}, 49-54 (2013).

\bibitem{3D infinitely deep box} Gaunt A. L., T. F. Schmidutz, I.
Gotlibovych, R. P. Smith, and Z. Hadzibabic, \textquotedblleft Bose-Einstein
condensation of atoms in a uniform potential", Phys. Rev. Lett. \textbf{110}%
, 200406 (2013).

\bibitem{Adhikari spin 1 SOC 3D} Gautam S. and S. K. Adhikari,
\textquotedblleft Three-dimensional vortex-bright solitons in a
spin-orbit-coupled spin-1 condensate", Phys. Rev. A \textbf{97}, 013629
(2018).

\bibitem{Hannaford} Ghanbari, S., T. D. Kieu, A. Sidorov, and P. Hannaford,
\textquotedblleft Permanent magnetic lattices for ultracold atoms and
quantum degenerate gases,\textquotedblright\ J. Phys. B: At. Mol. Opt. Phys.
\textbf{39}, 847-860 (2006).

\bibitem{Rb-97-gas-CQ} Greenberg J. A. and D. J. Gauthier, \textquotedblleft
High-order optical nonlinearity at low light levels", EPL \textbf{98}, 24001
(2012).

\bibitem{Harrison} Harrison, W. A., \textit{Pseudopotentials in the Theory
of Metals} (Benjamin, New York, 1966).

\bibitem{} Hauke P., F. M. Cucchietti, L. Tagliacozzo, I. Deutsch, and M.
Lewenstein, \textquotedblleft Can one trust quantum simulators?", Rep. Prog.
Phys. \textbf{75}, 082401 (2012).

\bibitem{HHu derivation} Hu H. and X.-J. Liu, \textquotedblleft Microscopic
derivation of the extended Gross-Pitaevskii equation for quantum droplets in
binary Bose mixtures", Phys. Rev. A \textbf{102}, 043302 (2020).

\bibitem{2D infinite box} Hueck K., N. Luick, L. Sobirey, J. Siegl, T.
Lompe, and H. Moritz, \textquotedblleft Two-dimensional homogeneous Fermi
gases", Phys. Rev. Lett. \textbf{120}, 060402 (2018).

\bibitem{Kip} Hukriede, J., D. Runde, and D. Kip, \textquotedblleft
Fabrication and application of holographic Bragg gratings in lithium niobate
channel waveguides,\textquotedblright\ J. Phys. D \textbf{36}, R1 (2003).

\bibitem{Ilg dim reduction} Ilg T., J. Kumlin, L. Santos, and D. S. Petrov,
and, H. P. B\"{u}chler, \textquotedblleft Dimensional crossover for the
beyond-mean-field correction in Bose gases", Phys. Rev. A \textbf{98},
051604 (2018).

\bibitem{Joannopoulos} Joannopoulos J. D., S. G. Johnson, J. N. Winn, and R.
D. Meade, \textit{Photonic Crystals: Molding the Flow of Light} (Princeton
University Press, Princeton, 2008).

\bibitem{nonlin latt RMP} Kartashov Y. V., B. A. Malomed, and L. Torner,
\textquotedblleft Solitons in nonlinear lattices", Rev. Mod. Phys. \textbf{83%
}, 247-306 (2011).

\bibitem{two-comp SDF} Kartashov Y. V., V. A. Vysloukh, L. Torner, and B. A.
Malomed, \textquotedblleft Self-trapping and splitting of bright vector
solitons under inhomogeneous defocusing nonlinearities", Opt. Lett. 36,
4587-4589 (2011).

\bibitem{hopfions} Kartashov Y. V., B. A. Malomed, Y. Shnir, and L. Torner,
\textquotedblleft Twisted toroidal vortex-solitons in inhomogeneous media
with repulsive nonlinearity", Phys. Rev. Lett. \textbf{113}, 264101 (2014).

\bibitem{SOC emulation Barcelona} Kartashov Y. V., B. A. Malomed, V. V.
Konotop, V. E. Lobanov, and L. Torner, \textquotedblleft Stabilization of
solitons in bulk Kerr media by dispersive coupling", Opt. Lett. \textbf{40},
1045-1048 (2015).

\bibitem{rotation} Kartashov Y. V., B. A. Malomed, V. A. Vysloukh, M. R. Beli%
\'{c}, and L. Torner, Rotating vortex clusters in media with inhomogeneous
defocusing nonlinearity, Opt. Lett. \textbf{42}, 446-449 (2017).

\bibitem{swirling} Kartashov Y. V., B. A. Malomed, L. Tarruell, and L.
Torner, \textquotedblleft Three-dimensional droplets of swirling
superfluids", Phys. Rev. A \textbf{98}, 013612 (2018).

\bibitem{Nature Phys Rev} Kartashov Y., G. Astrakharchik, B. Malomed, and L.
Torner, \textquotedblleft Frontiers in multidimensional self-trapping of
nonlinear fields and matter", Nature Reviews Phys. \textbf{1}, 185-197 (2019)

\bibitem{low-dim SOC} Kartashov Y. V., L. Torner, M. Modugno, E. Ya.
Sherman, B. A. Malomed, and V. V. Konotop, \textquotedblleft
Multidimensional hybrid Bose-Einstein condensates stabilized by
lower-dimensional spin-orbit coupling", Phys. Rev. Research \textbf{2},
013036 (2020).

\bibitem{KiAgr} Kivshar Y. S and G. P. Agrawal, \textit{Optical Solitons:
From Fibers to Photonic Crystals} (Academic Press, San Diego, 2003).

\bibitem{CS2-chi5-exper2} Kong D. G., Q. Chang, H. A. Ye, Y. C. Gao, Y. X.
Wang, X. R. Zhang, K. Yang, W. Z. Wu, and Y. L. Song, \textquotedblleft The
fifth-order nonlinearity of CS$_{2}$", J. Phys. B: At. Mol. Opt. Phys.
\textbf{42}, 065401 (2009).

\bibitem{PT sol Zezyulin} Konotop V. V., J. Yang, and D. A. Zezyulin,
\textquotedblleft Nonlinear waves in $\mathcal{PT}$-symmetric systems", Rev.
Mod. Phys. \textbf{88}, 035002 (2016).

\bibitem{Kruglov Vlasov} Kruglov V. I. and R. A. Vlasov, \textquotedblleft
Spiral self-trapping propagation of optical beams", Phys. Lett. A \textbf{111%
}, 401-404 (1985).

\bibitem{TS with S > 0} Kruglov V. I., V. M. Volkov, R. A. Vlasov, and V. V.
Drits, \textquotedblleft Auto-waveguide propagation and the collapse of
spiral light beams in non-linear media," J. Phys. A: Math. Gen. \textbf{21},
4381--4395 (1988).

\bibitem{in-advance-5} Lavoine L. and T. Bourdel, Beyond-mean-field
crossover from one dimension to three dimensions in quantum droplets of
binary mixtures, Phys. Rev. A \textbf{103}, 033312 (2021).

\bibitem{Leblond quasi-2D} Leblond H., B. A. Malomed, and D. Mihalache,
\textquotedblleft Three-dimensional vortex solitons in quasi-two-dimensional
lattices", Phys. Rev. E \textbf{76}, 026604 (2007).

\bibitem{Lederer} Lederer F., G. I. Stegeman, D. N. Christodoulides, G.
Assanto, M. Segev, and Y. Silberberg, \textquotedblleft Discrete solitons in
optics", Phys. Rep. \textbf{463}, 1-126 (2008).

\bibitem{LHY} Lee T. D., K. Huang, and C. N. Yang, \textquotedblleft
Eigenvalues and eigenfunctions of a Bose system of hard spheres and its
low-temperature properties", Phys. Rev. 1\textbf{06}, 1135-1145 (1957).

\bibitem{Lewenstein emulating book} Lewenstein M., A. Sanpera, and V.
Ahufinger, \textit{Ultracold Atoms in Optical Lattices: Simulating Quantum
Many-Body Systems} (Oxford: Oxford University Press, 2012)

\bibitem{2D SOC gap sol Raymond} Li Y., Y. Liu, Z. Fan, W. Pang, S. Fu, and
B. A. Malomed, \textquotedblleft Two-dimensional dipolar gap solitons in
free space with spin-orbit coupling", Phys. Rev. A \textbf{95}, 063613
(2017).

\bibitem{vortex 2D QD} Li Y., Z. Chen, Z. Luo, C. Huang, H. Tan, W. Pang,
and B. A. Malomed, Two-dimensional vortex quantum droplets, Phys. Rev. A
\textbf{98}, 063602 (2018).

\bibitem{eps-near-0 2017} Liberal I. and N. Engheta, \textquotedblleft
Near-zero refractive index photonics,\textquotedblright\ Nature Photonics
\textbf{11}, 149-158 (2017).

\bibitem{Spielman-Nature} Lin Y.-J., K. Jim\'{e}nez-Garc\'{\i}a, and I. B.
Spielman, \textquotedblleft Spin-orbit-coupled Bose-Einstein condensates",
Nature \textbf{471}, 83-86 (2011).

\bibitem{N=3 breather} Luo D., Y. Jin, J. H. V. Nguyen, B. A. Malomed, O. V.
Marchukov, V. A. Yurovsky, V. Dunjko, M. Olshanii, and R. G. Hulet,
\textquotedblleft Creation and characterization of matter-wave breathers",
Phys. Rev. Lett. \textbf{125}, 183902 (2020).

\bibitem{PT review} Makris, K. G., R. El-Ganainy, D. N. Christodoulides, and
Z. H. Musslimani, \textquotedblleft $\mathcal{PT}$ symmetric periodic
optical potentials", Int. J. Theor. Phys. \textbf{50}, 1019-1041 (2011).

\bibitem{Malomed-progress} Malomed B. A., \textquotedblleft Variational
methods in nonlinear fiber optics and related fields", Prog. Optics \textbf{%
43}, 71-193 (2002);
http://www.sciencedirect.com/science/article/pii/S0079663802800269.

\bibitem{Special Topics} Malomed B. A., \textquotedblleft Multidimensional
solitons: Well-established results and novel findings", Eur. Phys. J.
Special Topics \textbf{225}, 2507-2532 (2016).

\bibitem{PhysD-vortices} Malomed B. A., (INVITED)\ Vortex solitons: Old
results and new perspectives, Physica D \textbf{399}, 108-137 (2019).

\bibitem{Physica-D CQ vortex} Malomed B. A., L.-C. Crasovan, and D.
Mihalache, \textquotedblleft Stability of vortex solitons in the
cubic-quintic model", Physica D \textbf{161}, 187-201 (2002).

\bibitem{3D vort PLA} Malomed, B. A., F. Lederer, D. Mazilu, and D.
Mihalache, \textquotedblleft On stability of vortices in three-dimensional
self-attractive Bose-Einstein condensates", Phys. Lett. A \textbf{361},
336-340 (2007).

\bibitem{Malomed 2005} Malomed B. A., D. Mihalache, F. Wise, and L. Torner,
``Spatiotemporal optical solitons", J. Optics B: Quant. Semicl. Opt. \textbf{%
7}, R53-R72 (2005).

\bibitem{Malomed 2016} Malomed B. A., D. Mihalache, F. Wise, and L. Torner,
\textquotedblleft Viewpoint: On multidimensional solitons and their legacy
in contemporary Atomic, Molecular and Optical physics", J. Phys. B: At. Mol.
Opt. Phys. \textbf{49}, 170502 (2016).

\bibitem{Pelin} Malomed B. A. and D. E. Pelinovsky, \textquotedblleft
Persistence of the Thomas-Fermi approximation for ground states of the
Gross-Pitaevskii equation supported by the nonlinear confinement", Appl.
Math. Lett. \textbf{40}, 45-48 (2015).

\bibitem{Manton-book} Manton N. and P. Sutcliffe, \textit{Topological
Solitons} (Cambridge University Press, Cambridge, 2004).

\bibitem{Mihalache 3D cluster} Mihalache D., D. Mazilu, L.-C. Crasovan, B.
A. Malomed, F. Lederer, and L. Torner, \textquotedblleft Soliton clusters in
three-dimensional media with competing cubic and quintic nonlinearities", J.
Optics B 6, S333-S340 (2004a).

\bibitem{Barcelona quasi2D} Mihalache, D. D. Mazilu, F. Lederer, Y. V.
Kartashov, L.-C. Crasovan, and L. Torner, \textquotedblleft Stable
three-dimensional spatiotemporal solitons in a two-dimensional photonic
lattice", Phys. Rev. E \textbf{70}, 055603(R) (2004b).

\bibitem{Mihalache stability vort in OH} Mihalache D., D. Mazilu, B. A.
Malomed, and F. Lederer. Vortex stability in nearly-two-dimensional
Bose-Einstein condensates with attraction. Phys. Rev. A 73, 043615 (2006).

\bibitem{Mihalache multidim sol review} Mihalache D., ``Multidimensional
localized structures in optical and matter-wave media: A topical survey of
recent literature", Romanian Reports in Physics \textbf{69}, 403 (2017).

\bibitem{Mineev} Mineev V. P., \textquotedblleft The theory of the solution
of two near-ideal Bose gases", Zh. Eksp. Teor. Fiz. \textbf{67}, 263-272
(1974) [English translation: Sov. Phys. -- JETP \textbf{40}, 132-136 (1974)].

\bibitem{Moiseyev} Moiseyev N., \textit{Non-Hermitian Quantum Mechanics}
(Cambridge University Press, Cambridge, 2011).

\bibitem{Morsch-Oberthaler} O. Morsch and M. Oberthaler, \textquotedblleft
Dynamics of Bose-Einstein condensates in optical lattices", Rev. Mod. Phys.
\textbf{78}, 179-212 (2006).

\bibitem{vortex dipole} M\"{o}tt\"{o}nen M., S. M. M. Virtanen, T. Isoshima,
and M. M. Salomaa, \textquotedblleft Stationary vortex clusters in
nonrotating Bose-Einstein condensates", Phys. Rev. A \textbf{71}, 033626
(2005).

\bibitem{Fortran} Muruganandam P. and S. K. Adhikari, \textquotedblleft
Fortran programs for the time-dependent Gross-Pitaevskii equation in a fully
anisotropic trap", Comp. Phys. Comm. \textbf{180}, 1888-1912 (2009).

\bibitem{Navon 3D pot box} Navon N., A. L. Gaunt, R. P. Smith, and Z.
Hadzibabic, \textquotedblleft Critical dynamics of spontaneous symmetry
breaking in a homogeneous Bose gas", Science \textbf{347}, 167-170 (2015).

\bibitem{Newell} Newell A., \textit{Solitons in Mathematics and Physics}
(SIAM, Philadelphia, 1985).

\bibitem{eps-near-0 2018} Niu X., X. Hu, S. Chu, and Q. Gong,
\textquotedblleft Epsilon-near-zero photonics: a new platform for integrated
devices,\textquotedblright\ Adv. Opt. Mater. \textbf{6}, 1701292 (2018).

\bibitem{Pego} Pego R. L. and H. A. Warchall, \textquotedblleft Spectrally
stable encapsulated vortices for nonlinear Schr\"{o}dinger equations, J.
Nonlinear Sci. \textbf{12}, 347-394 (2002).

\bibitem{Petrov} Petrov D. S., \textquotedblleft Quantum mechanical
stabilization of a collapsing Bose-Bose mixture. Phys. Rev. Lett. \textbf{115%
}, 155302 (2015).

\bibitem{Petrov Astrakharchik} Petrov D. S. and G. E. Astrakharchik,
``Ultradilute low-dimensional liquids", Phys. Rev. Lett. \textbf{117},
100401 (2016).

\bibitem{Pit-Str} Pitaevskii L. P. and S. Stringari, \textit{Bose-Einstein
Condensation} (Oxford University Press, Oxford, 2003).

\bibitem{Hulet Feshbach} Pollack S. E., D. Dries, M. Junker, Y. P. Chen, T.
A. Corcovilos, and R. G. Hulet, Extreme Tunability of Interactions in a Li-7
Bose-Einstein Condensate, Phys. Rev. Lett. \textbf{102}, 090402 (2009).

\bibitem{Dong giant vort} Qin J., G. Dong, and B. A. Malomed,
\textquotedblleft Stable giant vortex annuli in microwave-coupled atomic
condensates," Phys. Rev. A \textbf{94}, 053611 (2016).

\bibitem{Quiroga} Quiroga-Teixeiro M. and H. Michinel, \textquotedblleft
Stable azimuthal stationary state in quintic nonlinear optical media", J.
Opt. Soc. Amer. B \textbf{14}, 2004-2009 (1997).

\bibitem{low-index-CQ} Reshef O., E. Giese, M. Z. Alam, I. de Leon, J.
Upham, and R. W. Boyd, \textquotedblleft Beyond the perturbative description
of the nonlinear optical response of low-index materials", Opt. Lett.
\textbf{42}, 3225-3228 (2017).

\bibitem{Cid-quint-sept} Reyna A. S. and C. B. de Ara\'{u}jo,
\textquotedblleft Spatial phase modulation due to quintic and septic
nonlinearities in metal colloids", Opt. Exp. \textbf{22}, 22456 (2014).

\bibitem{Cid-vortex} Reyna A. S., G. Boudebs, B. A. Malomed, and C. B. de Ara%
\'{u}jo, \textquotedblleft Robust self-trapping of vortex beams in a
saturable optical medium", Phys. Rev. A \textbf{93}, 013840 (2016).

\bibitem{Rogers Schieff} Rogers C. and W. K. Schief, \textit{B\"{a}cklund
and Darboux Transformations: Geometry and Modern Applications in Soliton
Theory} (Cambridge University Press, New York, 2002).

\bibitem{Saito and Ueda} Saito H. and M. Ueda, \textquotedblleft Split
instability of a vortex in an attractive Bose-Einstein condensate", Phys.
Rev. Lett. \textbf{89}, 190402 (2002).

\bibitem{Ben Li} Sakaguchi H., B. Li, and B. A. Malomed, \textquotedblleft
Creation of two-dimensional composite solitons in spin-orbit-coupled
self-attractive Bose-Einstein condensates in free space", Phys. Rev. E
\textbf{89}, 032920 (2014).

\bibitem{Sakaguchi gap vort} Sakaguchi H. and B. A. Malomed,
\textquotedblleft Two-dimensional loosely and tightly bound solitons in
optical lattices and inverted traps", J. Phys. B \textbf{37}, 2225-2239
(2004).

\bibitem{HS sharp annulus} Sakaguchi H. and B. A. Malomed, Two-dimensional
solitons in the Gross-Pitaevskii equation with spatially modulated
nonlinearity, Phys. Rev. E \textbf{73}, 026601 (2006).

\bibitem{anti-VK} Sakaguchi H. and B. A. Malomed, \textquotedblleft Solitons
in combined linear and nonlinear lattice potentials", Phys. Rev. A \textbf{81%
}, 013624 (2010).

\bibitem{collapse suppression} Sakaguchi H. and B. A. Malomed, Suppression
of the quantum-mechanical collapse by repulsive interactions in a quantum
gas, Phys. Rev. A \textbf{83}, 013607 (2011).

\bibitem{HS sharp edge OL} Sakaguchi H. and B. A. Malomed, \textquotedblleft
Stable two-dimensional solitons supported by radially inhomogeneous
self-focusing nonlinearity", Opt. Lett. \textbf{37}, 1035-1037 (2012).

\bibitem{SOC-PT Hidetsugu} Sakaguchi H. and B. A. Malomed, \textquotedblleft
One- and two-dimensional solitons in $\mathcal{PT}$-symmetric systems
emulating spin-orbit coupling", New J. Phys. \textbf{18}, 105005 (2016).

\bibitem{SOC 2D gap sol Hidetsugu} Sakaguchi H. and B. A, Malomed, One- and
two-dimensional gap solitons in spin-orbit-coupled systems with Zeeman
splitting, Phys. Rev. A \textbf{97}, 013607 (2018).

\bibitem{Romanian-Sherman} Sakaguchi H., B. Li, E. Ya. Sherman, and B. A.
Malomed, \textquotedblleft Composite solitons in two-dimensional spin-orbit
coupled self-attractive Bose-Einstein condensates in free space", Romanian
Reports in Physics \textbf{70}, 502 (2018).

\bibitem{tubular vortex} Salasnich L., B. A. Malomed, and F. Toigo,
\textquotedblleft Matter-wave vortices in cigar-shaped and toroidal
waveguides", Phys. Rev. A \textbf{76}, 063614 (2007).

\bibitem{Satsuma-Yajima} Satsuma J. and N. Yajima, ``Initial Value Problems
of One-Dimensional Self-Modulation of Nonlinear Waves in Dispersive Media",
Suppl. Prog. Theor. Phys. No. 55, 284-306 (1974).

\bibitem{Segev} Segev M., B. Crosignani, A. Yariv, and B. Fischer,
\textquotedblleft Spatial solitons in photorefractive media", Phys. Rev.
Lett. \textbf{68}, 923-926 (1992).

\bibitem{Inguscio} Semeghini G., G. Ferioli, L. Masi, C. Mazzinghi, L.
Wolswijk, F. Minardi, M. Modugno, G. Modugno, M. Inguscio, and M. Fattori,
\textquotedblleft Self-bound quantum droplets of atomic mixtures in free
space?", Phys. Rev. Lett. \textbf{120}, 235301 (2018).

\bibitem{Shamriz1} Shamriz E., Z. Chen, and B. A. Malomed, Suppression of
the quasi-two-dimensional quantum collapse in the attraction field by the
Lee-Huang-Yang effect, Phys. Rev. A \textbf{101}, 063628 (2020a).

\bibitem{Shamriz2} Shamriz E., Z. Chen, and B. A. Malomed, \textquotedblleft
Stabilization of one-dimensional Townes solitons by spin-orbit coupling in a
dual-core system", Comm. Nonlin. Sci. Numer. Simul. \textbf{91}, 105412
(2020b).

\bibitem{Silberberg OL} Silberberg Y., \textquotedblleft Collapse of optical
pulses,\textquotedblright\ Opt. Lett. \textbf{15}, 1282-1284 (1990).

\bibitem{Skoro} M. Skorobogatiy M. and J. Yang, \textit{Fundamentals of
Photonic Crystal Guiding} (Cambridge University Press, Cambridge, 2009).

\bibitem{Sutcliffe} P. M. Sutcliffe, \textquotedblleft Knots in the
Skyrme--Faddeev model", Proc. R. Soc. A \textbf{463}, 3001-3020 (2007).

\bibitem{Takhtadjian-Faddeev} Takhtadjian L. A. and L. D. Faddeev, \textit{%
The Hamiltonian Approach in the Theory of Solitons} (Nauka Publishers,
Moscow, 1986).

\bibitem{SDF concocted} Tian Q., L. Wu, Y. Zhang, and J.-F. Zhang,
\textquotedblleft Vortex solitons in defocusing media with spatially
inhomogeneous nonlinearity", Phys. Rev. E \textbf{85}, 056603 (2012).

\bibitem{FR control immiscibility} Tojo S., Y. Taguchi, Y. Masuyama, T.
Hayashi, H. Saito, and T. Hirano, \textquotedblleft Controlling phase
separation of binary Bose-Einstein condensates via mixed-spin-channel
Feshbach resonance", Phys. Rev. A \textbf{82}, 033609 (2010).

\bibitem{CS2-chi5} Tominaga K. and K. Yoshihara, Fifth order optical
response of liquid CS$_{2}$ observed by ultrafast nonresonant six-wave
mixing, Phys. Rev. Lett. \textbf{74}, 3061 (1995).

\bibitem{Vakh-Kol} Vakhitov N. G. and A. A. Kolokolov, \textquotedblleft
Stationary solutions of the wave equation in a medium with nonlinearity
saturation", Radiophys. Quantum Electron. \textbf{16}, 783-789 (1973);
https://doi.org/10.1007/BF01031343.

\bibitem{Vlasov} Vlasov S. N., V. A. Petrishchev, and V. I. Talanov, Izv.
Vyssh. Uchebn. Zaved. Radiofiz. \textbf{14}, 1353 (1971) [English
translation: Radiophys. Quantum Electron. \textbf{14}, 1062 (1971)].

\bibitem{C} Vudragovi\'{c} D, I. Vidanovi\'{c} A. Bala\v{z}, P.
Muruganandam, and S. K. Adhikari, \textquotedblleft C programs for solving
the time-dependent Gross-Pitaevskii equation in a fully anisotropic trap",
Comp. Phys. Comm. \textbf{183}, 2021-2025 (2012).

\bibitem{Li Wang} Wang L., B. A. Malomed, and Z. Yan, \textquotedblleft
Attraction centers and $\mathcal{PT}$-symmetric delta-functional dipoles in
critical and supercritical self-focusing media", Phys. Rev. E \textbf{99},
052206 (2019).

\bibitem{HHu QD at finite T} Wang J., X.-J. Liu, and H. Hu,
\textquotedblleft Ultradilute self-bound quantum droplets in Bose--Bose
mixtures at finite temperature", Chi. Phys. B \textbf{30}, 010306 (2021).

\bibitem{SDF-concocted-2} Wu Y., Q. Xie, H. Zhong, L. Wen, and W. Hai,
\textquotedblleft Algebraic bright and vortex solitons in self-defocusing
media with spatially inhomogeneous nonlinearity", Phys. Rev. A \textbf{87},
055801 (2013).

\bibitem{2D SOC} Wu Z., L. Zhang, W. Sun, X.-T. Xu, B.-Z. Wang, S.-C. Ji, Y.
Deng, S. Chen, X.-J. Liu, and J.-W. Pan, \textquotedblleft Realization of
two-dimensional spin-orbit coupling for Bose-Einstein condensates", Science
\textbf{354}, 83-86 (2016).

\bibitem{Yakimenko-Lashkin} Yakimenko A. I., Yu. A. Zaliznyak, and V. M.
Lashkin, Two-dimensional nonlinear vector states in Bose-Einstein
condensates, Phys. Rev. A \textbf{79}, 043629 (2009).

\bibitem{submicron FR} Yamazaki R., S. Taie, S. Sugawa, and Y. Takahashi,
\textquotedblleft Submicron spatial modulation of an interatomic interaction
in a Bose-Einstein condensate", Phys. Rev. Lett. \textbf{105}, 050405 (2010).

\bibitem{Yang J.} Yang J., \textit{Nonlinear Waves in Integrable and
Nonintegrable Systems} (SIAM, Philadelphia, 2010).

\bibitem{Yang-Musslimani} Yang J. and Z. H. Musslimani, \textquotedblleft
Fundamental and vortex solitons in a two-dimensional optical lattice", Opt.
Lett. \textbf{28}, 2094-2096 (2003).

\bibitem{ZakhKuz} Zakharov V. E. and E. A. Kuznetsov, ``Solitons and
collapses: two evolution scenarios of nonlinear wave systems", Physics -
Uspekhi \textbf{55}, 535-556 (2012).

\bibitem{Zakharov-et-al} Zakharov V. E., S. V. Manakov, S. P. Novikov, and
L. P. Pitaevskii, \textit{Theory of Solitons: The Inverse Problem Method}
(Nauka Publishers, Moscow, 1980) (English translation: Consultants Bureau,
New York, 1984).

\bibitem{Zakharov-Shabat} Zakharov V. E. and A. B. Shabat, \textquotedblleft
Exact Theory of Two-dimensional Self-focusing and One-dimensional
Self-modulation of Waves in Nonlinear Media", Zh. Eksp. Teor. Fiz. \textbf{61%
}, 118-134 (1971) [English translation: J. Exp. Theor. Phys. \textbf{34},
62-69 (1972)].

\bibitem{Seng-SDF-quintic} Zeng J. and B. A. Malomed, \textquotedblleft
Bright solitons in defocusing media with spatial modulation of the quintic
nonlinearity", Phys. Rev. E \textbf{86}, 036607 (2012).

\bibitem{Zeng-dark} Zeng J. and B. A. Malomed, \textquotedblleft Localized
dark solitons and vortices in defocusing media with spatially inhomogeneous
nonlinearity", Phys. Rev. E \textbf{95}, 052214 (2017).

\bibitem{organic-CQ} Zhan C., D. Zhang, D. Zhu, D. Wang, Y. Li, D. Li, Z.
Lu, L. Zhao, and Y. Nie, \textquotedblleft Third- and fifth-order optical
nonlinearities in a new stilbazolium derivative", J. Opt. Soc. Am. B 19,
369-375 (2002).

\bibitem{Han Pu 3D} Zhang Y.-C., Z.-W. Zhou, B. A. Malomed, and H. Pu,
\textquotedblleft Stable solitons in three dimensional free space without
the ground state: Self-trapped Bose-Einstein condensates with spin-orbit
coupling", Phys. Rev. Lett. \textbf{115}, 253902 (2015).

\bibitem{Zin dim reduction} Zin P., M. Pylak, T. Wasak, M. Gajda, and Z.
Idziaszek, \textquotedblleft Quantum Bose-Bose droplets at a dimensional
crossover?", Phys. Rev. A \textbf{98}, 051603(R) (2018).
\end{thebibliography}
\end{document}